\documentclass[12pt]{article}
\usepackage{amssymb}
\usepackage{tikz}
\usepackage{xr}

\usetikzlibrary{arrows, calc, decorations.pathmorphing, backgrounds, positioning, fit, petri, automata}
\definecolor{grey1}{rgb}{0.5,0.5,0.5}

\newcommand{\blind}{1}

\usepackage{algorithmicx}
\usepackage{algcompatible}
\usepackage{algorithm}

\usepackage{changebar}

\usepackage{hyperref}
\hypersetup{
  colorlinks,
  citecolor=blue,
  linkcolor=red,
  urlcolor=magenta}

\usepackage{mathtools}
\usepackage{graphicx,psfrag,epsf}
\usepackage{subfigure}
\usepackage{amsmath,amsthm,bm,bbm}
\usepackage{rotating}
\usepackage{mathrsfs}
\usepackage{chngcntr}
\usepackage{apptools}
\usepackage[titletoc,title]{appendix}
\usepackage{comment}
\renewcommand{\baselinestretch}{1.3}
\usepackage{xcolor}         
\definecolor{grau}{rgb}{0.8,0.8,0.8}

\newcommand{\chen}[1]{\color{orange}}
\usepackage{setspace,lscape,longtable}
\usepackage{amsmath,amsthm,bm}
\usepackage{fullpage}  

\usepackage{titlesec}
\usepackage[english]{babel}
\usepackage{xcolor}         
\definecolor{grau}{rgb}{0.8,0.8,0.8}

\theoremstyle{plain}

\newtheorem{theorem}{Theorem}
\newtheorem{lemma}{Lemma}
\newtheorem{corollary}{Corollary}

\theoremstyle{definition}

\theoremstyle{remark}
\newtheorem{remark}{Remark}

\usepackage[comma,authoryear]{natbib}

\bibpunct{(}{)}{;}{a}{,}{,}
\numberwithin{equation}{section}

\DeclareMathOperator*{\argmin}{arg\,min}

\newcommand{\iidsim}{{\overset{\mathrm{i.i.d.}}{\sim}}}
\newcommand{\transpose}{^{\mathrm{T}}}

\newcommand{\calA}{{\mathcal{A}}}
\newcommand{\calB}{{\mathcal{B}}}
\newcommand{\calC}{{\mathcal{C}}}
\newcommand{\calD}{{\mathcal{D}}}
\newcommand{\calE}{{\mathcal{E}}}
\newcommand{\calF}{{\mathcal{F}}}

\newcommand{\calH}{{\mathcal{H}}}

\newcommand{\calJ}{{\mathcal{J}}}

\newcommand{\calL}{{\mathcal{L}}}
\newcommand{\calM}{{\mathcal{M}}}
\newcommand{\calN}{{\mathcal{N}}}

\newcommand{\calP}{{\mathcal{P}}}

\newcommand{\calU}{{\mathcal{U}}}

\newcommand{\calW}{{\mathcal{W}}}
\newcommand{\calX}{{\mathcal{X}}}
\newcommand{\calY}{{\mathcal{Y}}}

\newcommand{\bdot}{{\boldsymbol{\cdot}}}

\newcommand{\bF}{{\mathbf{F}}}

\newcommand{\bv}{{\mathbf{v}}}
\newcommand{\bx}{{\mathbf{x}}}

\newcommand{\bA}{{\mathbf{A}}}
\newcommand{\bB}{{\mathbf{B}}}
\newcommand{\bW}{{\mathbf{W}}}
\newcommand{\bw}{{\mathbf{w}}}
\newcommand{\bk}{{\mathbf{k}}}

\newcommand{\bt}{{\mathbf{t}}}
\newcommand{\br}{{\mathbf{r}}}

\newcommand{\by}{{\mathbf{y}}}

\newcommand{\bV}{{\mathbf{V}}}

\newcommand{\bSigma}{{\bm{\Sigma}}}

\newcommand{\eye}{{\mathbf{I}}}
\newcommand{\btheta}{{\bm{\theta}}}

\newcommand{\zero}{{\bm{0}}}

\newcommand{\eps}{\epsilon}

\makeatletter
\g@addto@macro\normalsize{%
	\setlength\abovedisplayskip{1ex}
	\setlength\belowdisplayskip{1ex}
	\setlength\abovedisplayshortskip{1ex}
	\setlength\belowdisplayshortskip{1ex}
}
\makeatother

\usepackage{lipsum}

\usepackage{enumitem}
\setlist[enumerate]{itemsep=0mm}
\setlist[itemize]{itemsep=0mm}

\allowdisplaybreaks
\begin{document}

\def\spacingset#1{\renewcommand{\baselinestretch}%
{#1}\small\normalsize} \spacingset{1}


\if1\blind
{
  \title{\bf Bayesian Projected Calibration of Computer Models}
  \author{Fangzheng Xie
    and 
    Yanxun Xu\thanks{Correspondence should be addressed to Yanxun Xu (yanxun.xu@jhu.edu)}\\
    Department of Applied Mathematics and Statistics, Johns Hopkins University}
    \date{}
  \maketitle
} \fi

\if0\blind
{
  \bigskip
  \bigskip
  \bigskip
  \begin{center}
    {\LARGE\bf Bayesian Projected Calibration of Computer Models}
\end{center}
  \medskip
} \fi
\spacingset{1.45}
\bigskip
\begin{abstract}
We develop a Bayesian approach called Bayesian projected calibration to address the problem of calibrating an imperfect computer model using observational data from a complex physical system. The calibration parameter and the physical system are parametrized in an identifiable fashion via $L_2$-projection. The physical process is assigned a Gaussian process prior, which naturally induces a prior distribution on the calibration parameter through the $L_2$-projection constraint. The calibration parameter is estimated through its posterior distribution, which provides a natural and non-asymptotic way for the uncertainty quantification. We provide a rigorous large sample justification for the proposed approach by 
establishing the asymptotic normality of the posterior of the calibration parameter with the efficient covariance matrix. 
In addition, two efficient computational algorithms based on stochastic approximation are designed with theoretical guarantees. 
Through extensive simulation studies and two real-world datasets analyses, we show that the Bayesian projected calibration can accurately estimate the calibration parameters, appropriately calibrate the computer models, and compare favorably to alternative approaches.  
An R package implementing the Bayesian projected calibration is publicly available at \url{https://drive.google.com/open?id=1Sij0P-g5ocnTeL_qcQ386b-jfBfV-ww_}. 

\end{abstract}

\noindent%
\vfill

\newpage
\spacingset{1.45} 


\section{Introduction} 
\label{sec:introduction}

With the rapid development of computational techniques and mathematical tools, computer models have been widely adopted by researchers to study large and complex physical systems. One can think of computer models as complicated nonlinear functions designed by experts using scientific knowledge \citep{sacks1989design,fang2005design}.  Compared to physical experiments, computer models are typically much faster and cheaper to run. Furthermore, computer models can be used to generate data that are infeasible to collect in practice. For example, a public available computer model called TITAN2D \citep{sheridan2002visualization} was developed to simulate granular mass flows over digital elevation models of natural terrain, to better understand the loss of life and disruption of infrastructure due to volcanic phenomena, the data of which are impossible to collect in real life. For more applications of computer models, we refer to \cite{fang2005design}, \cite{santner2013design}, and the April 2018 issue of \emph{Statistica Sinica} (\url{http://www3.stat.sinica.edu.tw/statistica}), which is devoted to computer experiments and uncertainty quantification.

In this paper we consider the \emph{calibration} problem in computer models when they include not only variables that can be measured, often referred to as \emph{design}, but also unknown parameters that are not directly available in the physical system. These parameters are called \emph{calibration parameters} in the literature \citep{kennedy2001bayesian}. The goal of calibration is to estimate calibration parameters by combining observational data from the physical system and simulated data from the computer model, so that the computer model with the estimated calibration parameters plugged-in provides decent approximation to the underlying physical system. Formally, we model the data $(y_i)_{i=1}^n$ of the physical system $\eta$ at design $(\bx_i)_{i=1}^n$ through a nonparametric regression model
\[
y_i = \eta(\bx_i)+e_i,\quad i =1,\cdots,n,
\]
where $(e_i)_{i=1}^n$ are independent $\mathrm{N}(0,\sigma^2)$ residuals. The computer model $y^s(\cdot,\btheta)$, also referred to as the simulator, is a function designed by scientific experts to model the unknown physical system $\eta(\cdot)$ when the calibration parameter $\btheta$ is appropriately estimated.

Despite the success of computer models in many scientific studies, researchers often ask the following question: is the computer model a suitable surrogate for the real physical system? Compared to the physical system, traditional computer models are rarely perfect or exact due to their fixed parametric nature or simplifications of complex physical phenomenon \citep{tuo2015efficient}: \emph{i.e.}, there exists discrepancy between a physical system $\eta(\cdot)$ and its corresponding computer model $y^s(\cdot,\btheta)$ even if the computer model is well calibrated.  \cite{kennedy2001bayesian} first tackled this discrepancy issue under a Bayesian framework, which has been influential among many other statisticians and quality control engineers. For an incomplete list of references, we refer to \cite{higdon2004combining,bayarri2007framework,qian2008bayesian,joseph2009statistical,wang2009bayesian,chang2014model,brynjarsdottir2014learning,storlie2015calibration} among others. 

Theoretical properties of calibration problem were not well understood until \cite{tuo2015efficient, tuo2016theoretical}, who pointed out that the calibrated computer models estimated by \cite{kennedy2001bayesian} could lead to poor approximations to physical systems. 
Identifiability issue of the calibration parameter in \cite{kennedy2001bayesian} was also noticed by H. P. Wynn, among several other discussants, in their written discussion of \cite{kennedy2001bayesian}. In short, identifiability issue refers to the phenomenon that the distribution of the observed data from physical system does not uniquely determine the value of the corresponding calibration parameter given the computer model.
There have been several attempts to tackle the identifiability issue, many of which are Bayesian approaches. For example,  \cite{bayarri2007framework}  suggested to incorporate experts' information into the prior distribution of the calibration parameter $\btheta$ to reduce the confounding caused by nonidentifiability. 
\cite{brynjarsdottir2014learning} presented a concrete example, in which the derivative information of the model discrepancy was incorporated through a constrained Gaussian process prior to better estimate $\btheta$. These Bayesian approaches, however, lack theoretical guarantees and mathematical rigor. 
In contrast to the Bayesian methods, which are traditionally applied to computer model calibration problems, in \cite{tuo2015efficient, tuo2016theoretical} and \cite{wong2017frequentist} the authors addressed the identifiability issue rigorously in frequentist frameworks and provided corresponding theoretical justifications. 

We propose a Bayesian method for computer model calibration called Bayesian projected calibration. To the best of our knowledge, our work is the first one in the literature that simultaneously achieves the following objectives:
\begin{itemize}
	\item[a)]\textbf{Identifiability: }The proposed approach is formulated in a rigorously identifiable fashion. \cite{tuo2015efficient,tuo2016theoretical} and \cite{wong2017frequentist} defined the ``true'' value of the calibration parameter to be the one minimizing the $L_2$ distance between the computer model $y^s(\cdot,\btheta)$ and the physical system $\eta(\cdot)$. Following their work, the proposed Bayesian projected calibration provides a Bayesian method to estimate this ``true'' value of the calibration parameter.
	\item[b)]\textbf{Uncertainty quantification: }The proposed Bayesian projected calibration provides a natural way for the uncertainty quantification of the calibration parameter through its full posterior distribution. 
  \cite{tuo2015efficient} showed the asymptotic normality of the $L_2$-projected calibration estimator for quantifying the uncertainty of the calibration parameter, which may not work in practice, since the amount of physical data is usually very limited \citep{tuo2017adjustments}. Hence a Bayesian approach is desired, especially when the observational data are scarce.  
	\item[c)]\textbf{Theoretical guarantee: }We show that the full posterior distribution of the calibration parameter is asymptotically normal with efficient covariance matrix. 
	Earlier literature either only provides asymptotic results of specific point estimators \citep{tuo2015efficient,tuo2016theoretical,wong2017frequentist,tuo2017adjustments}, or formulates a Bayesian methodology for calibration problem without large sample evaluation \citep{plumlee2017bayesian}. Our method represents the first effort in providing  theoretical guarantees for the full posterior distribution of Bayesian methods in computer model calibration. 
  \item[d)]\textbf{Efficient computational algorithms: } We design two stochastic approximation algorithms for posterior inference of the calibration parameter. Compared to the orthogonal Gaussian process approach in \cite{plumlee2017bayesian}, the proposed algorithms are computationally cheaper. This is illustrated in Section \ref{sec:numerical_examples}. Furthermore,  theoretical properties of these algorithms such as convergence   are discussed. 
\end{itemize}

The rest of the paper is organized as follows. In Section \ref{sec:problem_formulation} we formulate the calibration problem rigorously in an identifiable fashion and introduce the Bayesian projected calibration method. Section \ref{sec:asymptotic_properties} elaborates on the asymptotic properties of the posterior distribution of the calibration parameter. 
We discuss computational strategies for  
 the Bayesian projected calibration and its approximation in Section \ref{sec:algorithmic_issue}, in which two algorithms based on stochastic approximation are designed with strong theoretical support. 
In Section \ref{sec:numerical_examples},  we demonstrate advantages of the Bayesian projected calibration in terms of estimation accuracy and the uncertainty quantification via simulation studies and two real-world data examples. 
Conclusion and further discussion are  in Section \ref{sec:conclusion_and_discussion}. 

\section{Problem Formulation} 
\label{sec:problem_formulation}

\subsection{Background} 
\label{sub:background_on_computer_model_calibration}


We first briefly review the frequentist \emph{$L_2$-projected calibration} approach proposed in \cite{tuo2015efficient}
before introducing the Bayesian projected calibration method, which can be regarded as the Bayesian version of the $L_2$-projected calibration. 
 
Suppose one has collected responses $(y_i)_{i=1}^n$ from a physical system $\eta$ on a set of design points $(\bx_i)_{i=1}^n\subset\Omega\subset\mathbb{R}^p$, 
where $\eta:\Omega\to\mathbb{R}$ is a deterministic function, and the design space $\Omega$ is the closure of a connect bounded convex open set in $\mathbb{R}^p$. The physical responses $(y_i)_{i=1}^n$ are noisy due to measurement or observational errors, and hence can be modeled by the following nonparametric regression model:
\begin{align}
\label{eqn:physical_system}
y_i = \eta(\bx_i)+e_i,\quad i=1,\ldots,n,
\end{align}
where $(e_i)_{i = 1}^n$ are independent $\mathrm{N}(0,\sigma^2)$ residuals. Such a model has been widely adopted in the literature of calibration \citep{kennedy2001bayesian,tuo2015efficient,tuo2017adjustments,wong2017frequentist}.

Let $\Theta$ be the parameter space of the calibration parameter $\btheta$. We assume that $\Theta\subset\mathbb{R}^q$ is compact. A computer model is a deterministic function $y^s:\Omega\times\Theta\to\mathbb{R}$ that produces an output $y^s(\bx,\btheta)$ given a controllable input $\bx\in\Omega$ and  $\btheta\in\Theta$. The goal of calibration is to estimate  $\btheta$ given the computer model $y^s$ and the physical data $(y_i)_{i=1}^n$, such that the calibrated computer model approximates the physical system well. 
However, as pointed out by \cite{tuo2016theoretical} and \cite{wong2017frequentist}, the calibration parameter $\btheta$ cannot be identified without further restriction, in the sense that $\btheta$ cannot be uniquely determined by the distribution of $(\bx_i,y_i)_{i = 1}^n$. More precisely, by expressing the physical system $\eta$ in terms of the computer model $y^s(\bx,\btheta)$ and a discrepancy $\delta(\bx)$ as follows \citep{kennedy2001bayesian,wong2017frequentist,plumlee2017bayesian,tuo2017adjustments}:
\[
\eta(\bx)=y^s(\bx,\btheta)+\delta(\bx),
\]
where $\delta$ is completely nonparametric, it is clear that $(\btheta,\delta)$ cannot be uniquely identified by  $\eta$. Therefore, the ``true'' value of the calibration parameter that gives rise to the physical data is not well-defined.

In \cite{kennedy2001bayesian} the authors first studied 
the computer model calibration problem by assigning a Gaussian process prior to the discrepancy function $\delta(\cdot)$, and then obtaining its posterior distribution. Although the Kennedy and O'Hagan (abbreviated as KO) approach did not tackle the identifiability issue directly, \cite{tuo2016theoretical} discussed its potential in a simplified setting. Specifically,   
if the discrepancy function follows a mean-zero Gaussian process prior with covariance function $\Psi$, $\btheta$ follows a uniform prior, and the physical data are noise-free (\emph{i.e.}, $e_i$'s are zeros) in the KO approach, then the posterior density of $\btheta$ is  
\[
\pi(\btheta\mid(\bx_1,y_1),\ldots,(\bx_n,y_n))\propto \exp\left[-\frac{1}{2}(\by - \by^s_\btheta)\transpose\boldsymbol{\Psi}(\bx_{1:n},\bx_{1:n})^{-1}(\by - \by^s_\btheta)\right],
\]
where $\by = [y_1,\cdots,y_n]\transpose{}$,  $\by^s_\btheta = [y^s(\bx_1,\btheta),\ldots,y^s(\bx_n,\btheta)]\transpose$, and $\boldsymbol{\Psi}(\bx_{1:n},\bx_{1:n}) = [\Psi(\bx_i,\bx_j)]_{n\times n}$. 
Instead of a fully Bayesian approach, \cite{tuo2016theoretical} computed the maximum \emph{a posteriori} estimator $\widehat\btheta$ and proved that $\widehat\btheta$ asymptotically  converged to a point $\btheta^*$ that minimized the \emph{reproducing kernel Hilbert space} (RKHS) 
 norm of $\delta$ associated with the covariance function $\Psi$ under certain regularity conditions. 
Therefore, in this simplified KO approach the ``true'' value of $\btheta$ can be defined to be $\btheta^*$. However, when the physical data are noisy,  \cite{tuo2015efficient,tuo2016theoretical} showed that such an approach was no longer valid for defining $\btheta^*$, and the resulting  $\widehat\btheta$ did not converge to the desired $\btheta^*$. 

Alternatively, as pointed out in Section 4.2 of \cite{tuo2016theoretical}, 
it is more  straightforward to define the ``true'' value of $\btheta$ through the  $L_2$-projection: 
\begin{align}\label{eqn:L_2_projection}
\btheta^*=\argmin_{\btheta\in\Theta}\left\|\eta(\cdot)-y^s(\cdot,\btheta)\right\|^2_{L_2(\Omega)}=\argmin_{\btheta\in\Theta}\int_{\Omega}[\eta(\bx)-y^s(\bx,\btheta)]^2\mathrm{d}\bx.
\end{align}
The $L_2$-projected calibration method provides an estimate $\widehat{\btheta}_{L_2}$ for $\btheta^*$ 
using a two-step procedure. First, an estimator $\widehat{\eta}$ of the physical system $\eta$ is obtained via the \emph{kernel ridge regression} \citep{wahba1990spline} given $(\bx_i,y_i)_{i=1}^n$:
\begin{align}\label{eqn:kernel_ridge_regression}
\widehat{\eta}=\argmin_{f\in\mathbb{H}_\Psi(\Omega)}\frac{1}{n}\sum_{i=1}^n[y_i-f(\bx_i)]^2+\lambda\|f\|_{\mathbb{H}_\Psi(\Omega)},
\end{align}
where $\Psi:\Omega\times\Omega\to\mathbb{R}$ is a positive definite covariance function, and $\|\cdot\|_{\mathbb{H}_\Psi(\Omega)}$ is the RKHS norm associated with $\Psi$ \citep{wendland2004scattered}. 
Then, the $L_2$-projected calibration estimate $\widehat{\btheta}_{L_2}$ for $\btheta^*$ is given by
\begin{align}\label{eqn:L_2_calibration}
\widehat{\btheta}_{L_2}:=\argmin_{\btheta\in\Theta}\left\|\widehat{\eta}(\cdot)-y^s(\cdot,\btheta)\right\|_{L_2(\Omega)}^2.
\end{align}
The $L_2$-projected calibration has very nice theoretical properties: $\widehat{\btheta}_{L_2}$ is not only $\sqrt{n}$-consistent for $\btheta^*$, but also semiparametric efficient \citep{tuo2016theoretical}. In other words, it provides an optimal estimator to the ``true'' calibration parameter. More importantly, unlike the simplified KO approach, the $L_2$-calibration approach can directly deal with noisy physical data. 

\subsection{Bayesian Projected Calibration} 
\label{sub:bayesian_projected_calibration}

The $L_2$-projected calibration estimate $\widehat{\btheta}_{L_2}$ enjoys nice asymptotic properties. Nevertheless, its uncertainty quantification needs to be assessed via additional procedures (\emph{e.g.}, bootstrapping, see \citealp{wong2017frequentist}). Such a limitation motivates us to 
develop a Bayesian projected calibration method with carefully selected prior distributions, and assess the uncertainty   via the posterior distribution of the parameters of interest. 
Estimating deterministic parameters by Bayesian inference has been gaining popularities since it provides a natural and flexible way to quantify the uncertainty of the parameters. There has been extensive work on  frequentist justifications of the Bayesian estimation for deterministic parameters in nonparametric and high-dimensional problems.  The readers are referred to \cite{ghosal2017fundamentals} for a thorough review. 

We follow the definition of the ``true'' value $\btheta^*$ of $\btheta$ given in \eqref{eqn:L_2_projection}, as it minimizes the discrepancy between the computer model and the physical system in the $L_2$-sense. 
There are two unknown parameters: the physical system $\eta$, taking values in some function space $\calF$, and the calibration parameter $\btheta\in\Theta$. We define the statistical model for calibration as follows, 
\[
\calP = \left\{\phi_\sigma(y-\eta(\bx)):\eta\in\calF, \btheta^* = \argmin_{\btheta\in\Theta}\|\eta(\cdot)-y^s(\cdot,\btheta)\|_{L_2(\Omega)}^2\right\},
\]
where $\phi_\sigma(\cdot)$ is the density function of $\mathrm{N}(0,\sigma^2)$. Namely, the parameter $(\eta,\btheta^*)$ is constrained on a manifold in $\calF\times\Theta$ defined by 
\begin{align}\label{eqn:manifold_calibration}
\calM = \left\{(\eta,\btheta^*)\in\calF\times\Theta:\btheta^* = \argmin_{\btheta\in\Theta}\|\eta(\cdot)-y^s(\cdot,\btheta)\|_{L_2(\Omega)}^2\right\}.
\end{align}
We will rigorously show in Section \ref{sec:asymptotic_properties} that under certain regularity conditions, $\calM$ is a differentiable Banach manifold. 
It is therefore natural to treat the ``true'' calibration parameter  as a functional $\btheta^*:\calF\to\Theta$, $\eta\mapsto \argmin_\btheta\|\eta(\cdot)-y^s(\cdot,\btheta)\|_{L_2(\Omega)}^2$, and we denote this functional by $\btheta^*_\eta$. To distinguish the parameter $(\eta,\btheta^*_\eta)$ and the truth that generates the data, we denote $\eta_0$ to be the true physical system producing physical data $(y_i)_{i=1}^n$, and $\btheta_0^* = \btheta^*_{\eta_0}$. 

We now introduce the Bayesian projected calibration. The unknown physical process $\eta$ is assigned a mean-zero Gaussian process prior $\Pi=\mathrm{GP}(0,\tau^2\Psi)$, where $\Psi:\Omega\times\Omega\to\mathbb{R}_+$ is a positive definite covariance function, and $\tau>0$ is a scaling factor. 
We will discuss later in this section regarding the choice of an appropriate covariance function $\Psi$. 
Let $\calD_n=(\bx_i,y_i)_{i=1}^n$ be the physical data and $\Pi(\cdot\mid\calD_n)$ denote the posterior distribution given  $\calD_n$. It is straightforward to show that the posterior distribution of $\eta$ is also a Gaussian process with mean function $\widetilde{\eta}$ and covariance function $\widetilde{\Psi}$, where
\begin{align}
\label{eqn:GP_posterior_mean}
\widetilde{\eta}(\bx) &= \tau^2\boldsymbol{\Psi}(\bx_{1:n},\bx)\transpose{}(\tau^2\boldsymbol{\Psi}(\bx_{1:n},\bx_{1:n})+\sigma^2\eye_n)^{-1}\by,\\
\label{eqn:GP_posterior_covariance}
\widetilde{\Psi}(\bx,\bx') &= \tau^2\Psi(\bx,\bx') - \tau^2\boldsymbol{\Psi}(\bx_{1:n},\bx)\transpose{}(\tau^2\boldsymbol{\Psi}(\bx_{1:n},\bx_{1:n})+\sigma^2\eye_n)^{-1}\tau^2\boldsymbol{\Psi}(\bx_{1:n},\bx').
\end{align}
Here $\boldsymbol{\Psi}(\bx_{1:n},\bx) = \left[\Psi(\bx_1,\bx),\cdots,\Psi(\bx_n,\bx)\right]\transpose{}\in\mathbb{R}^{n}$
and $\by = [y_1,\cdots,y_n]\transpose{}\in\mathbb{R}^{n}$. 
The predictive mean  $\widetilde{\eta}(\bx)$ given   $\calD_n$ coincides with the kernel ridge regression estimate $\widehat{\eta}$ for some suitably chosen $\tau$ (see, for example, \citealp{rasmussen2006gaussian}). 
The Gaussian process prior $\mathrm{GP}(0,\tau^2\Psi)$ on $\eta$ naturally induces a prior distribution on $\btheta^*_\eta$ through the constrained manifold $\calM$ in \eqref{eqn:manifold_calibration}. Therefore, after collecting $\calD_n$ from \eqref{eqn:physical_system}, one can obtain the posterior distribution of $\eta$, and hence, that of $\btheta_\eta^*$ given $\calD_n$ and the computer model $y^s$. We refer to the procedure of computing the posterior distribution of $\btheta_\eta^*$ as the Bayesian projected calibration. 
It can be regarded as a Bayesian version of the $L_2$-projected calibration method, because they estimate the ``true'' value of $\btheta$ over the constrained manifold $\calM$ from the Bayesian and the frequentist perspective, respectively. Furthermore, in Section \ref{sec:asymptotic_properties} we will show that the posterior of $\btheta_\eta^*$ under the Bayesian projected calibration is asymptotically centered at the $L_2$-projected calibration estimate $\widehat{\btheta}_{L_2}$.

Choosing an appropriate covariance function $\Psi$ for the Gaussian process prior is of fundamental importance in computer model calibration. One of the most popular choices is the class of Mat\'ern covariance functions. Formally, given $\alpha > p/2$, the Mat\'ern covariance function with roughness parameter $\alpha$ and range parameter $\psi$ is given by
\begin{align}\label{eqn:matern_covariance_function}
\Psi_\alpha(\bx,\bx'\mid\psi)=\frac{1}{\Gamma(\alpha)2^{\alpha-1}}\left(\frac{\sqrt{2\alpha}\|\bx-\bx'\|}{\psi}\right)^\alpha K_\alpha\left(\frac{\sqrt{2\alpha}\|\bx-\bx'\|}{\psi}\right),
\end{align}
where $K_\alpha$ is the modified Bessel function of the second kind. 
Throughout this work we use Mat\'ern covariance functions for the sake of theoretical analysis. As will be seen in Section \ref{sec:asymptotic_properties}, when the smoothness parameter $\alpha$ matches with the smoothness level of the underlying true physical system, the resulting convergence rate is minimax-optimal. In practice the Mat\'ern covariance functions with roughness parameter $\alpha = 3/2$ and $\alpha = 5/2$ are also popular due to their closed-form expression, but users can select other covariance functions if desired. 

\begin{remark}[Expensive computer model]
In the literature of computer experiments, it is common that the computer model $y^s$ is not directly available or time-consuming to run, in which case $y^s$ can be only computed at given design points. Thus finding an emulator $\hat{y}^s$ for $y^s$ using  data  from the computer outputs at given design points is needed. One first collects a set of data $(\bx^s_j,\btheta_j^s, y_j^s)_{j=1}^m$ from $m$ runs of the computer model, where $y_j^s=y^s(\bx_j^s,\btheta_j^s)$ is the output at the design point $\bx_j^s$ with $\btheta = \btheta_j^s$,  then estimate the emulator $\widehat{y}^s$ using the data $(\bx^s_j,\btheta_j^s, y_j^s)_{j=1}^m$.  
There are varieties of methods for constructing emulators for computer experiments, including Gaussian process models \citep{santner2013design}, radial basis function interpolation \citep{wendland2004scattered}, polynomial chaos approximation \citep{xiu2010numerical}, or the smoothing spline ANOVA \citep{wahba1990spline}. To perform computer model calibration when the computer model $y^s$ is not directly available or time-consuming to run, the calibration parameter $\btheta_\eta^*$ can be estimated by replacing $y^s$ with the corresponding emulator $\widehat{y}^s$.
\end{remark}


\section{Theoretical Properties} 
\label{sec:asymptotic_properties}
In this section we provide the large sample justification of the proposed Bayesian projected calibration. In particular, asymptotic characterization of the posterior distribution $\Pi(\btheta_\eta^*\in\cdot\mid\calD_n)$ 
is offered. The posterior of $\btheta_\eta^*$ has similar behavior as the $L_2$-projected calibration estimator $\widehat{\btheta}_{L_2}$: $\Pi(\btheta_\eta^*\in\cdot\mid\calD_n)$ is not only $\sqrt{n}$-consistent, but also asymptotically normal with efficient covariance matrix \emph{a posteriori}. The asymptotic normality of the Bayesian posterior is also referred to as Bernstein-von Mises (BvM) limit (see Chapter 10 in \citealp{van2000asymptotic}). The development of semiparametric BvM theorem had not been established until \cite{bickel2012semiparametric}. For a unifying treatment of BvM limits for smooth functionals in semiparametric models, we refer to \cite{castillo2015}.

We first introduce some notations and definitions. Given an integer vector $\bk = [k_1,\cdots,k_p]\transpose{}$ and a function $f(x_1,\cdots,x_p):\Omega\to\mathbb{R}$, denote $D^\bk $ to be the mixed partial derivative operator defined by $D^\bk f=\partial^{|\bk|}f/\partial x_1^{k_1}\cdots\partial x_p^{k_p}$, where $|\bk|:=\sum_{j=1}k_j$. Let $\alpha>0$ be positive, and $\underline{\alpha}$ be the greatest integer strictly smaller than $\alpha$. The $\alpha$-H\"older norm of a function $f:\Omega\to\mathbb{R}$ is defined by
\[
\|f\|_{\mathfrak{C}_\alpha(\Omega)}:=\max_{\bk:|\bk|\leq\underline{\alpha}}\left\|D^\bk f\right\|_{L_\infty(\Omega)}+\max_{\bk:|\bk|=\underline{\alpha}}\sup_{\bx\neq\bx'}\frac{|D^\bk f(\bx)-D^\bk f(\bx')|}{\|\bx-\bx'\|^{\alpha-\underline{\alpha}}}.
\]
The $\alpha$-H\"older space of functions on $\Omega$, denoted by $\mathfrak{C}_\alpha(\Omega)$, is the set of functions with finite $\alpha$-H\"older norm. 
The $\alpha$-Sobolev space of functions, denoted by $\calH_\alpha(\Omega)$, is the set of functions $f:\Omega\to\mathbb{R}$ that can be extended to $\mathbb{R}^p$ such that the Fourier transformation $\widehat{f}(\bt)=(2\pi)^{-p}\int_{\mathbb{R}^p} \mathrm{e}^{i\bt\transpose{}\bx}f(\bx)\mathrm{d}\bx$ satisfies \citep{vaart2011information}
\[
\int_{\mathbb{R}^p} \left(1+\|\bt\|^2\right)^\alpha\left|\widehat{f}(\bt)\right|^2\mathrm{d}\bt<\infty.
\] 

To study the asymptotic behavior of the posterior of $\btheta_\eta^*$, we first explore convergence properties of the physical system $\eta$.  
For the ease of mathematical treatment, we assume that the design points $(\bx_i)_{i = 1}^n$ are independent  samples uniformly drawn from the unit hypercube $[0, 1]^p$, and the computer model $y^s$ is directly available to us. 
The true but unknown physical system $\eta_0$ is assumed to lie in the intersection of the $\alpha$-H\"older space $\mathfrak{C}_\alpha(\Omega)$ and $\alpha$-Sobolev space $\calH_\alpha(\Omega)$ for some $\alpha>p/2$. We assume that the prior $\Pi$ for $\eta$ is the mean-zero Gaussian process $\mathrm{GP}(0, \tau^2\Psi_\alpha(\cdot,\cdot\mid\psi))$ and without loss of generality, the scaling factor $\tau$ is fixed at $1$. We shall also assume that the range parameter $\psi$ is fixed, since fixing $\psi$ does not change the asymptotic analyses of both $\eta$ and $\btheta$. 
We use shorthand notation $\Psi_\alpha$ to denote $\Psi_\alpha(\cdot,\cdot\mid\psi)$. 
\begin{remark}[Design points]
In practice it is possible to encounter design points that are either randomly sampled or fixed \emph{a priori}, and the design space $\Omega$ may not be as regular as a hypercube. However, in this section we assume that   $\Omega$ is the unit hypercube $[0,1]^p$ and $(\bx_i)_{i=1}^n$ are independently and uniformly sampled from $\Omega$ for the ease of mathematical treatment. The theory developed here can   be easily extended to the case where $(\bx_i)_{i=1}^n$ are independently drawn from a distribution with a density that is bounded away from $0$ and $\infty$ and supported on a compact domain in $\mathbb{R}^p$. 
\end{remark}
\begin{remark}[More on expensive computer model]
The theoretical results in this section are still valid   when the computer model is not directly available to us
for the following reasons. Firstly, the amount of data from computer experiments is typically much larger than the sample size of the physical data. Furthermore, 
the approximation error between $y^s$ and $\widehat{y}^s$, when sufficiently small as the number $m$ of runs gets large, does not affect the stochastic analysis here. Therefore, we can assume that the approximation error between $\widehat{y}^s$ and $y^s$ is negligible when the computer model is not directly available to us. 
\end{remark}
We now present the convergence result for $\eta$ in Theorem \ref{thm:matern_process_contraction}. In particular, the first result \eqref{eqn:convergence_eta} directly follows Theorem 5 of \cite{vaart2011information}, and the proof of the second result  is given in the Supplementary Material. 

\begin{theorem}[Convergence of $\eta$]
\label{thm:matern_process_contraction}
Suppose $\eta$ is assigned the Gaussian process prior $\Pi = \mathrm{GP}(0,\Psi_\alpha)$, and $\eta_0\in\mathfrak{C}_\alpha(\Omega)\cap\calH_\alpha(\Omega)$, where $\alpha>p/2$. Then for any sequence $M_n\to\infty$,
\begin{align}\label{eqn:convergence_eta}
&\mathbb{E}_0\left[\Pi\left(\|\eta-\eta_0\|_{L_2(\Omega)}>M_nn^{-\alpha/(2\alpha+p)}\mid\calD_n\right)\right]\to 0,
\end{align}
and there exists some constant $M>0$ such that
$\Pi\left(\|\eta-\eta_0\|_{L_\infty(\Omega)}> M\mid\calD_n\right)\to 0$
in $\mathbb{P}_0$-probability. 
\end{theorem}
The resulting rate $n^{-\alpha/(2\alpha+p)}$ is proven to be optimal when the underlying true function $\eta_0$ is an $\alpha$-H\"older or $\alpha$-Sobolev function  \citep{stone1982optimal, van1996weak, ghosal2017fundamentals}.

We next discuss the property of $\btheta_\eta^*$ as a functional: $\eta\mapsto\btheta_\eta^*$. Under the regularity conditions A1-A4 to be stated next, $\btheta_\eta^*$ yields a first-order Taylor expansion with respect to $\eta$ locally around $\eta_0$.
The smoothness property
of $\btheta_\eta^*$ serves as the building block to derive the asymptotic normality of the posterior of $\btheta_\eta^*$.  
\begin{itemize}
	\item[A1] $\btheta^*_\eta$ is the unique solution to \eqref{eqn:L_2_projection} and is in the interior of $\Theta$ for $\eta$ in an $L_2$-neighborhood of $\eta_0$.
	\item[A2] $\sup_{\btheta\in\Theta}\|y^s(\cdot,\btheta)\|_{L_2(\Omega)}<\infty$.
	\item[A3] The Hessian matrix
	\[
	\bV_\eta=\int_\Omega\left\{\frac{\partial^2}{\partial\btheta\partial\btheta\transpose{}}[\eta(\bx)-y^s(\bx,\btheta)]^2\right\}\mathrm{d}\bx\mathrel{\bigg|}_{\btheta = \btheta_\eta^*}
	\]
	is strictly positive definite for all $\eta$ in an $L_2$-neighborhood of $\eta_0$. 
	\item[A4] For all $j,k=1,\ldots,q$, it holds that
	\[
	\sup_{\btheta\in\Theta}\left\|\frac{\partial y^s}{\partial\theta_j}(\cdot,\btheta)\right\|_{\mathbb{H}_{\Psi_{\alpha}}(\Omega)}<\infty,\quad\frac{\partial^2 y^s}{\partial\theta_j\partial\theta_k}(\cdot,\cdot)\in\mathfrak{C}_1(\Omega\times\Theta).
	\]
\end{itemize}
The proof of the following lemma is given in the Supplementary Material. 
\begin{lemma}[Taylor Expansion]
\label{lemma:taylor_expansion_functional}
Assume that $\eta_0\in\mathfrak{C}_\alpha(\Omega)\cap\calH_\alpha(\Omega)$ for some $\alpha > p/2$. 
Under conditions A1-A4, there exists some $\eps>0$ and some positive constants $L_{\eta_0}^{(1)}$ and $L_{\eta_0}^{(2)}$ depending on $\eta_0$ only, such that $\|\btheta_\eta^*-\btheta_0^*\|\leq L_{\eta_0}^{(1)}\|\eta-\eta_0\|_{L_2(\Omega)}$ and
 \begin{align}
 \left\|\btheta_\eta^*-\btheta_0^*-2\int_\Omega[\eta(\bx)-\eta_0(\bx)]\bV_0^{-1}\frac{\partial y^s}{\partial\btheta}(\bx,\btheta_0^*)\mathrm{d}\bx\right\|\leq L_{\eta_0}^{(2)}\|\eta-\eta_0\|_{L_2(\Omega)}^2
 \end{align}
 whenever $\|\eta-\eta_0\|_{L_2(\Omega)}<\eps$, where $\bV_0 = \bV_{\eta_0}$. 
 Furthermore, if A1 and A3 hold for all $\eta$ in an $L_2$-neighborhood $\calU$ of $\eta_0$, then the set
 $\calM(\calU) : = \{(\eta, \btheta_\eta^*):\eta \in \calU\}$ forms a Banach manifold, and if $\calU$ is the entire $L_2(\Omega)$ space, then $\calM$ defined by \eqref{eqn:manifold_calibration} is a Banach manifold. 
\end{lemma}
It follows immediately from the convergence results of   $\eta$ (Theorem \ref{thm:matern_process_contraction}) and the Taylor expansion property of $\btheta_\eta^*$ (Lemma \ref{lemma:taylor_expansion_functional}) that the posterior of $\btheta_\eta^*$ is consistent.
\begin{corollary}[Consistency of $\btheta_\eta^*$]
\label{corr:consistency}
Suppose $\eta$ is assigned the Gaussian process prior $\Pi = \mathrm{GP}(0,\Psi_\alpha)$, and  $\eta_0\in \mathfrak{C}_\alpha(\Omega)\cap\calH_\alpha(\Omega)$. Then the posterior of $\btheta_\eta^*$ is consistent, \emph{i.e.}, $\Pi(\|\btheta_\eta^*-\btheta_0^*\|>\eps\mid\calD_n)\to 0$ in $\mathbb{P}_0$-probability for any $\eps>0$.
\end{corollary} 

Now we are in a position to characterize the asymptotic behavior of the posterior distribution of $\btheta_\eta^*$,  
which is the main result of this paper and the proof is deferred to Appendix.  
Under the aforementioned regularity conditions A1-A4, the posterior distribution of $\sqrt{n}(\btheta_\eta^*-\widehat{\btheta}_{L_2})$ is asymptotically normal, where $\widehat{\btheta}_{L_2}$ is the frequentist $L_2$-projected calibration estimate of $\btheta$ proposed by \cite{tuo2015efficient}. 
We describe the $L_2$-projected calibration procedure in our context for completeness: 
\begin{align}
\widehat{\eta}&=\argmin_{f\in\mathbb{H}_{\Psi_{\nu}}(\Omega)}\frac{1}{n}\sum_{i=1}^n[y_i-f(\bx_i)]^2+\lambda_n\|f\|_{\mathbb{H}_{\Psi_{\nu}}(\Omega)}^2,\nonumber\\
\widehat{\btheta}_{L_2}&=\argmin_{\btheta\in\Theta}\left\|\widehat{\eta}(\cdot)-y^s(\cdot,\btheta)\right\|_{L_2(\Omega)}^2,\nonumber
\end{align}
where $\nu = \alpha-p/2$, and $\lambda_n\asymp n^{-2\alpha/(2\alpha+p)}$ is a sequence depending on the sample size of the physical data $\calD_n$. 

\begin{theorem}[Asymptotic Normality]
\label{thm:BvM_limit}
Suppose $\eta$ is assigned the Gaussian process prior $\Pi = \mathrm{GP}(0,\Psi_\alpha)$, and $\eta_0\in\mathfrak{C}_\alpha(\Omega)\cap\calH_\alpha(\Omega)$, where $\alpha>p/2$. Under conditions A1-A4, it holds that
\begin{align}
\sup_{A}\left|\Pi\left(\sqrt{n}(\btheta_\eta^*-\widehat{\btheta})\in A\mid\calD_n\right)-\mathrm{N}\left(\zero, 4\sigma^2\bV_0^{-1}\bW\bV_0^{-1}\right)(A)\right|=o_{\mathbb{P}_0}(1),\nonumber
\end{align}
provided that
\[
\bW=\int_\Omega\left[\frac{\partial y^s}{\partial\btheta}(\bx,\btheta_0^*)\frac{\partial y^s}{\partial\btheta\transpose{}}(\bx,\btheta_0^*)\right]\mathrm{d}\bx
\]
is strictly positive definite, where the supremum is taken over all measurable subsets in $\mathbb{R}^q$, and $\widehat\btheta$ is any estimator of $\btheta$ satisfying 
\[
\widehat{\btheta}-\btheta^*_0=2\bV_0^{-1}\left[\frac{1}{n}\sum_{i=1}^ne_i\frac{\partial y^s}{\partial \btheta}(\bx_i,\btheta_0^*)\right]+o_{\mathbb{P}_0}(n^{-1/2})\nonumber.
\]
In particular, $\widehat\btheta$ can be taken as the $L_2$-calibration estimate $\widehat\btheta_{L_2}$ if $\lambda_n \asymp n^{-2\alpha/(2\alpha + p)}$ and $\nu = \alpha - p/2$ are chosen in the kernel ridge regression \eqref{eqn:kernel_ridge_regression}. 
\end{theorem}

\cite{tuo2015efficient} proved that the $L_2$-projected calibration estimate $\widehat{\btheta}_{L_2}$ is asymptotically normal: $\sqrt{n}(\widehat{\btheta}_{L_2}-\btheta_0^*)\overset{\calL}{\to}\mathrm{N}(\zero, 4\sigma^2\bV_0^{-1}\bW\bV_0^{-1})$. Furthermore, the covariance matrix $4\sigma^2\bV_0^{-1}\bW\bV_0^{-1}$ achieves semiparametric efficiency in the sense that there does not exist a regular estimate with a smaller asymptotic covariance matrix (in spectrum). 
The posterior of $\btheta_\eta^*$ possesses a similar optimal behavior as the $L_2$-calibration. 
 Firstly, $\Pi(\btheta_\eta^*\in\cdot\mid\calD_n)$ is \emph{a posteriori} consistent, 
and the covariance matrix of the asymptotic posterior of $\sqrt{n}(\btheta_\eta^*-\widehat{\btheta}_{L_2})$ coincides with that of $\widehat{\btheta}_{L_2}$. Secondly, the 
coordinate-wise 
posterior median of $\btheta_\eta^*$, as a Bayes estimator resulting from the full posterior distribution, is asymptotically equivalent to $\widehat\btheta_{L_2}$ in the coordinate-wise sense, which is unveiled in 
the following corollary.
The proof  is given in the Supplementary Material. 
\begin{corollary}\label{corr:Bayes_estimator}
Suppose $\eta$ is assigned the Gaussian process prior $\Pi = \mathrm{GP}(0,\Psi_\alpha)$, and $\eta_0\in\mathfrak{C}_\alpha(\Omega)\cap\calH_\alpha(\Omega)$, where $\alpha>p/2$. Let $\widehat\btheta^*$ be the coordinate-wise posterior median of $\btheta_\eta^*$. Then under the conditions of Theorem \ref{thm:BvM_limit}, for each $k = 1,\ldots,q$, 
\[
\sqrt{n}\left[\widehat\btheta^* - \btheta_0^*\right]_k\overset{\calL}{\to}\mathrm{N}\left(0, 4\sigma^2\left[\bV_0^{-1}\bW\bV_0^{-1}\right]_{kk}\right),
\]
where $[\cdot]_k$ is the $k$th component of the argument vector and $[\cdot]_{kk}$ is the $(k,k)$th element of the argument matrix.
\end{corollary}

We finish this section with the following $\sqrt{n}$-consistency result of $\btheta_\eta^*$, which is a refinement of Corollary \ref{corr:consistency}. It follows immediately from theorem \ref{thm:BvM_limit} and the asymptotic normality of $\widehat{\btheta}_{L_2}$. 
\begin{corollary}[$\sqrt{n}$-Consistency of $\btheta_\eta^*$]
\label{corr:root_n_consistency}
Suppose $\eta$ is assigned the Gaussian process prior $\Pi = \mathrm{GP}(0,\Psi_\alpha)$, and $\eta_0\in\mathfrak{C}_\alpha(\Omega)\cap\calH_\alpha(\Omega)$, where $\alpha>p/2$. Under the conditions of Theorem \ref{thm:BvM_limit}, the posterior of $\btheta_\eta^*$ is $\sqrt{n}$-consistent, \emph{i.e.}, for any sequence $M_n\to\infty$, it holds that
$\Pi\left(\sqrt{n}\|\btheta_\eta^*-\btheta_0^*\|>M_n\mid\calD_n\right)\to 0$ in $\mathbb{P}_0$-probability.
\end{corollary}




\section{Computational Strategies} 
\label{sec:algorithmic_issue}

We discuss computational strategies to obtain the posterior distribution of $\btheta^*_\eta$ given   $\calD_n = (\bx_i,y_i)_{i = 1}^n$ in this section. By definition, to draw $T$ independent samples from $\Pi(\btheta_\eta^*\in\cdot\mid\calD_n)$, one needs to first draw $T$ independent sample paths $\eta^{(1)},\ldots,\eta^{(T)}$ from the posterior distribution of $\eta$ using formulas  \eqref{eqn:GP_posterior_mean} and \eqref{eqn:GP_posterior_covariance}, then compute the minimizer $\btheta_{\eta^{(t)}}^* = \argmin_\btheta\|y^s(\cdot,\btheta) - \eta^{(t)}(\cdot)\|_{L_2(\Omega)}^2$ for each $\eta^{(t)}$, $t = 1,\ldots,T$. 
Although drawing sample paths from the posterior distribution  of $\eta$ is straightforward (see Section \ref{sub:bayesian_projected_calibration}), it is 
non-trivial to compute the corresponding $\btheta_\eta^*$ due to the generally intractable integral $\|y^s(\cdot,\btheta) - \eta(\cdot)\|_{L_2(\Omega)}^2$. 
One naive strategy is to discretize the integral by the Monte Carlo method. Specifically, we draw $N$ independent samples $\bx_1^d,\ldots,\bx_N^d$ (the superscript $d$ here indicates that these points are drawn for discretizing the integral)
uniformly from $\Omega$, and then approximate $\btheta_\eta^*$ with respect to a sample path $\eta$ by minimizing the discretized integral, \emph{i.e.}, find
\begin{align}
\label{eqn:discretized_integral}
\btheta_\eta^*\approx\argmin_{\btheta\in\Theta}\frac{1}{N}\sum_{j = 1}^N[y^s(\bx_j^d, \btheta) - \eta(\bx_j)]^2.
\end{align}
This strategy becomes accurate as $N\to\infty$ by the law of large numbers, but is not recommended since finding the minimizer  of the discretized integral often requires iterative optimization procedures due to the lack of closed-form expression for $\btheta_\eta^*$ except in rare cases. 
 Assuming that at least $R$ iterations are needed to obtain $\btheta_{\eta^{(t)}}^*$ for each $\eta^{(t)}$, the total computational complexity  becomes $O(NTR)$. In particular, 
$N$ is typically made sufficiently large  to ensure the quality of the approximation in practice, 
especially when $\Omega$ is multi-dimensional. 
 In what follows we will borrow ideas from stochastic optimization to reduce the computational burden. 

\subsection{Stochastic Approximation for the Projected Calibration} 
\label{sub:stochastic_optimization_for_projected_calibration}
Stochastic approximation methods can be dated back to \cite{robbins1951stochastic}, and have been gaining enormous progress in the recent decade thanks to the rapid development of advanced machine learning techniques and the emerging big data problems.
They focus on minimizing the objective function $f(\btheta)$ that is the expected value of a function $g(\bw, \btheta)$ depending on a random variable $\bw$ with density $p(\bw)$, namely, $f(\btheta) = \mathbb{E}_\bw[g(\bw, \btheta)]$. Here $f(\btheta)$ cannot be observed directly, but can only be estimated via the noisy version $g(\bw, \cdot)$.  The key idea of stochastic approximation methods is to generate iterates of the form: $\btheta^{(t + 1)} = \btheta^{(t)} - \alpha_t\nabla_\btheta g(\bw_t, \btheta^{(t)})$, where $(\alpha_t)_{t\geq 1}$ is a sequence of suitable step sizes, and $(\bw_t)_{t\geq 1}$ are independent copies of $\bw\sim p(\bw)$. 
There is vast literature
on how to select the step sizes $(\alpha_t)_{t\geq 1}$ for convex and non-convex $f$,
 among which the \emph{AdaGrad} method \citep{duchi2011adaptive} is one of the most popular choices. Specifically, \cite{li2018convergence} proposed the following coordinate-wise step sizes:
\begin{align}\label{eqn:AdaGrad_stepsize}
\alpha_{tk} = {a_0}\left\{b_0 + \sum_{j = 1}^{t - 1}\left[\frac{\partial g(\bw_j, \btheta^{(j)})}{\partial\theta_k}\right]^2\right\}^{-(1/2 + \eps)},\quad k = 1,\ldots,q,
\end{align}
where $a_0, b_0 > 0$ and $\eps\in(0, 1/2]$ are some constants, and then update $\btheta^{(t + 1)}$ by
\begin{align}\label{eqn:AdaGrad}
\btheta^{(t + 1)} = \btheta^{(t)} - 
\mathrm{diag}(\alpha_{t1},\ldots,\alpha_{tq})\frac{\partial g}{\partial\btheta}(\bw_t,\btheta^{(t)}).
\end{align}
Convergence of the \emph{AdaGrad} approach for convex and non-convex $f$ was addressed in \cite{li2018convergence}. In what follows, we modify the \emph{AdaGrad} 
method for the projected calibration to reduce the aforementioned computational burden. 

Recall that in the projected calibration procedure, we are interested in computing $\btheta_\eta^* = \argmin_{\btheta\in\Theta}\|y^s(\cdot,\btheta) - \eta(\cdot)\|_{L_2(\Omega)}^2$ given a sample path $\eta$ drawn from the posterior distribution. Denote the integral $f_\eta(\btheta) = \|y^s(\cdot,\btheta) - \eta(\cdot)\|_{L_2(\Omega)}^2$. By introducing a uniform random variable $\bw\sim\mathrm{Unif}(\Omega)$, $f_\eta(\btheta)$ can be expressed as the expected value of a function of $\bw$: 
$f_\eta(\btheta)= \mathbb{E}_\bw\{[y^s(\bw, \btheta) - \eta(\bw)]^2\}$. 
Since the parameter space $\Theta$ for the calibration parameter is compact, we modify the \emph{AdaGrad} to avoid the search outside the boundary of $\Theta$.
Specifically, whenever the updated $\btheta^{(t + 1)}$ stays outside the parameter space, a step-halving procedure is applied until it falls back to $\Theta$. We formally organize the modified \emph{AdaGrad} 
for the projected calibration in Algorithm \ref{alg:AdaGrad_calibration}.


\begin{algorithm}[h] 
  \renewcommand{\algorithmicrequire}{\textbf{Input:}}
  \renewcommand{\algorithmicensure}{\textbf{Output:} }
  \caption{Modified \emph{AdaGrad} for the Projected Calibration} 
  \label{alg:AdaGrad_calibration} 
  \begin{algorithmic}[1] 
    \State{\textbf{Input: }
      Computer model $y^s(\cdot,\cdot)$ and its derivative $\nabla_\btheta y^s(\cdot, \cdot)$; Sample path $\eta(\cdot)$; }
    \State{\textbf{Initialize: } 
        Initialize $\btheta^{(1)}\sim\mathrm{Unif}(\Theta)$; 
        Set $N$ to be number of samples from $\Omega$;
    }
    \State{\textbf{For }$t = 1:(N - 1)$
        \State{\indent\textbf{Draw} $\bw_t \sim\mathrm{Unif}(\Omega)$;}
        \State{\indent\textbf{For }$k = 1:q$
            \State{\indent\indent\textbf{Compute} the step size $\alpha_{tk}$ using formula \eqref{eqn:AdaGrad_stepsize};}
            \State{\indent\textbf{End For}}
            }
            \State{\indent \textbf{Compute}
            \[
            \btheta^{(t + 1)} = \btheta^{(t)} - 2[y^s(\bw_t, \btheta^{(t)}) - \eta(\bw_t)]\mathrm{diag}(\alpha_{t1},\ldots,\alpha_{tq})\frac{\partial y^s}{\partial\btheta}(\bw_t,\btheta^{(t)});
            \]
            \State{\indent \textbf{If }$\btheta^{(t + 1)}\notin \Theta\backslash\partial\Theta$, then set $\alpha_{tk} \leftarrow \alpha_{tk} / 2$ for $k  = 1,\ldots,q$ and go to line 10;}
            }
    \State{\textbf{End For}}
    }
    \State{\textbf{Output: } The last iterate $\btheta^{(N)}$. }
  \end{algorithmic}
\end{algorithm}


Compared to optimizing the discretized integral in \eqref{eqn:discretized_integral}, the computational complexity of  sampling $\btheta_{\eta^{(t)}}^*$, $t = 1,\ldots,T$, is reduced to  $O(NT)$ using Algorithm \ref{alg:AdaGrad_calibration}.  
The convergence of Algorithm \ref{alg:AdaGrad_calibration} to a stationary point can be guaranteed by the following theorem, the proof of which is provided in the Supplementary Material. 
\begin{theorem}\label{thm:convergence_AdaGrad}
Assume that the sample path $\eta$ is continuous over $\Omega$. Then under conditions A2 and A4, the output $\btheta^{(N)}$ of Algorithm \ref{alg:AdaGrad_calibration} converges to a stationary point of $f_\eta(\btheta)$ as $N\to\infty$ a.s. with respect to the distribution of $(\bw_t)_{t\geq 0}$. 
\end{theorem} 

Although Theorem \ref{thm:convergence_AdaGrad} guarantees that Algorithm \ref{alg:AdaGrad_calibration} converges to a local minimizer, it is challenging to 
provide a theory for finding the global minimizer, since the objective function $f_\eta(\cdot)$ of is non-convex. However, this can be addressed by trying multiple starting points in practice.


\subsection{Approximate Computation of the Projected Calibration} 
\label{sub:approximate_computation_of_projected_calibration}

Although Algorithm \ref{alg:AdaGrad_calibration} adopts stochastic approximation techniques, the resulting samples of $\btheta_{\eta^{(t)}}^*$'s are drawn exactly from the posterior distribution of $\btheta_\eta^*$ for any sample size $n$ as $N\to\infty$ (recall that $N$ can be made arbitrarily large).
In this section, we 
seek an approximate computational method that can further reduce the computational cost of the projected calibration for a relatively large sample size $n$.

The major computational bottleneck in finding $\btheta_\eta^*$ 
is that there is no closed-form formula for $\btheta_\eta^*$ using $y^s$ and $\eta$. It is, however, feasible to approximate $\btheta_\eta^*$ in certain ways. Recall that in Lemma \ref{lemma:taylor_expansion_functional} we show that $\btheta_\eta^*$ can be linearly approximated by Taylor's expansion locally around $\eta_0$:
\[
\btheta_\eta^* = \btheta_0^* + 2\int_\Omega[\eta(\bx) - \eta_0(\bx)]\bV_0^{-1}\frac{\partial y^s}{\partial\btheta}(\bx, \btheta_0^*)\mathrm{d}\bx + O(\|\eta - \eta_0\|_{L_2(\Omega)}^2).
\]
Since $\eta_0$ is unknown, a kernel ridge regression estimator 
$\widehat\eta$ (details in Section \ref{sec:asymptotic_properties}) 
can replace $\eta_0$ to estimate $\btheta_0^*$ and $\bV_0$, leading to the following approximation:
\begin{align}
\label{eqn:approximate_PC}
\btheta_\eta^* \approx \widetilde\btheta_\eta: = \widehat\btheta_{L_2} + 2\int_\Omega\left[\eta(\bx) - \widehat\eta(\bx)\right]\bV_{\widehat\eta}^{-1}\frac{\partial y^s}{\partial\btheta}(\bx, \widehat\btheta_{L_2})\mathrm{d}\bx.
\end{align}
Alternatively, the above approximation can  be treated as Taylor's expansion of $\btheta^*_\eta$ locally around $\widehat\eta$. The following theorem, the proof of which is  in the Supplementary Material, proves that $\widetilde\btheta_\eta$ is asymptotically equivalent to $\btheta_\eta^*$ for large sample size $n$.  
\begin{theorem}\label{thm:approximate_PC_convergence}
Assume the conditions in Theorem \ref{thm:BvM_limit} hold, and $\widetilde\btheta_\eta$ is computed using formula \eqref{eqn:approximate_PC}. Then it holds that
\begin{align*}
\sup_{A}\left|\Pi\left(\sqrt{n}(\widetilde\btheta_\eta-\widehat{\btheta}_{L_2})\in A\mid\calD_n\right)-\mathrm{N}\left(\zero, 4\sigma^2\bV_0^{-1}\bW\bV_0^{-1}\right)(A)\right|=o_{\mathbb{P}_0}(1),\nonumber
\end{align*}
where $\bW$ is given in Theorem \ref{thm:BvM_limit}.
\end{theorem}

Thanks to the closed-form expression \eqref{eqn:approximate_PC},  computing the posterior distribution of $\widetilde\btheta_\eta$ is much more convenient than computing that of $\btheta_\eta^*$. 
By discretizing the involved  integral using $N$ independent uniform samples $\bx_1^d,\ldots,\bx_N^d$ from $\Omega$, we obtain
\[
\widetilde\btheta_\eta  \approx\widetilde\btheta_\eta^{(N)}:= \widehat\btheta_{L_2} - \frac{2}{N}\sum_{j = 1}^N\widehat\eta(\bx_j^d)\bV_{\widehat\eta}^{-1}\frac{\partial y^s}{\partial\btheta}(\bx_j^d, \widehat\btheta_{L_2}) + \frac{2}{N}\sum_{j = 1}^N\eta(\bx_j^d)\bV_{\widehat\eta}^{-1}\frac{\partial y^s}{\partial\btheta}(\bx_j^d,\widehat\btheta_{L_2}).
\]
Note that the posterior distribution of $\eta$ is $\mathrm{GP}(\widetilde\eta,\widetilde\Psi)$ by formulas \eqref{eqn:GP_posterior_mean} and \eqref{eqn:GP_posterior_covariance}, it follows that a posteriori, $\widetilde\btheta_\eta^{(N)}$ follows a normal distribution with mean
\begin{align}\label{eqn:approximate_PC_mean}
\widehat\btheta_{L_2} + \frac{2}{N}\sum_{j = 1}^N\left[\widetilde\eta(\bx_j^d) - \widehat\eta(\bx_j^d)\right]\bV_{\widehat\eta}^{-1}\frac{\partial y^s}{\partial\btheta}(\bx_j^d, \widehat\btheta_{L_2})
\end{align}
and covariance matrix
\begin{align}\label{eqn:approximate_PC_covariance}
\frac{4}{N^2}\sum_{j = 1}^N\sum_{\ell = 1}^N\bV_{\widehat\eta}^{-1}\frac{\partial y^s}{\partial\btheta}(\bx_j^d, \widehat\btheta_{L_2})\widetilde\Psi(\bx_j^d, \bx_\ell^d)\bV_{\widehat\eta}^{-1}\frac{\partial y^s}{\partial\btheta\transpose}(\bx_j^d, \widehat\btheta_{L_2}).
\end{align}
The detailed algorithm of computing the approximate projected calibration is shown in Algorithm \ref{alg:approximate_PC}, which further reduces the overall computational  complexity from $O(NT)$ to $O(N)$.
In numerical studies we find that Algorithm \ref{alg:approximate_PC} provides good approximation to the exact posterior when $n \geq 30$. 
\begin{algorithm}[h] 
  \renewcommand{\algorithmicrequire}{\textbf{Input:}}
  \renewcommand{\algorithmicensure}{\textbf{Output:} }
  \caption{Approximate computation for the Projected Calibration} 
  \label{alg:approximate_PC} 
  \begin{algorithmic}[1] 
    \State{\textbf{Input:} Computer model $y^s(\cdot,\cdot)$ and its derivative $\nabla_\btheta y^s(\cdot, \cdot)$; Physical data $(\bx_i, y_i)_{i = 1}^n$; }
    \State{\textbf{Compute} the kernel ridge regression estimate $\widehat\eta$; }
    \State{\textbf{Call} Algorithm \ref{alg:AdaGrad_calibration} with input $y^s$, $\nabla_\btheta^s$, and $\widehat\eta$ and output $\widehat\btheta_{L_2}$; }
    \State{\textbf{Generate} $N$ independent samples $\bx_1^d,\ldots,\bx_N^d$ uniformly from $\Omega$;}
    \State{\textbf{Compute} the mean vector $\widehat\btheta_{\mathrm{APC}}$ using formula \eqref{eqn:approximate_PC_mean};}
    \State{\textbf{Compute} the covariance matrix $\widehat\bSigma_{\mathrm{APC}}$ using formula \eqref{eqn:approximate_PC_covariance};}
    \State{\textbf{Output: }$\widehat\btheta_{\mathrm{APC}}$ and $\widehat\bSigma_{\mathrm{APC}}$.}
  \end{algorithmic}
\end{algorithm}


\section{Numerical Examples} 
\label{sec:numerical_examples}
This section provides extensive numerical examples to evaluate the proposed Bayesian projected calibration. Subsection \ref{sub:simulation_studies} presents simulation studies via three synthetic examples. Two real-world data examples are included in Subsections \ref{sub:ion_channel_example} and \ref{sub:spot_welding_example}, respectively. 

\subsection{Simulation Studies} 
\label{sub:simulation_studies}

\begin{figure}[htbp!]
	\centerline{\includegraphics[width=.6\textwidth]{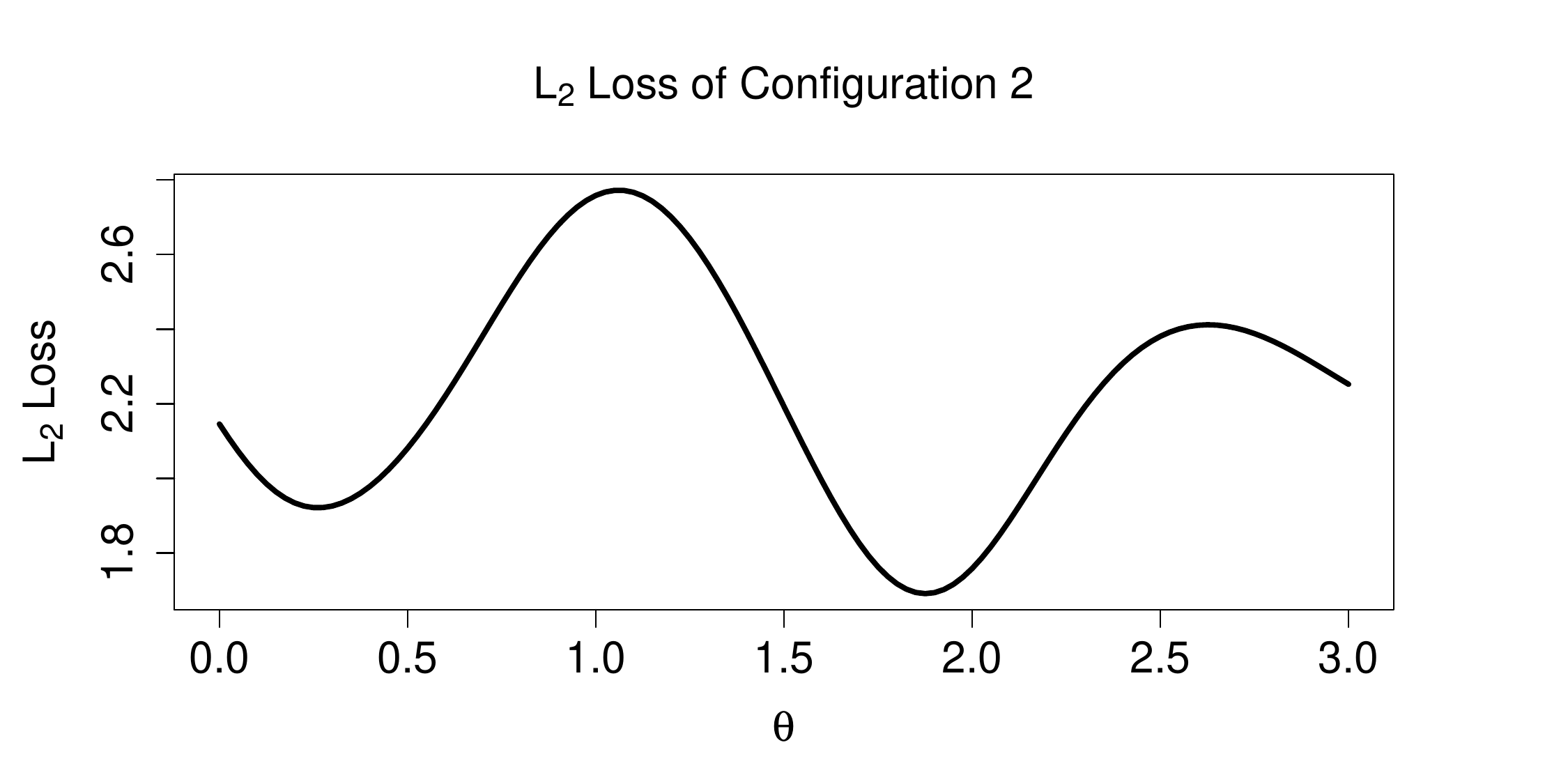}}
	\caption{The $L_2$-discrepancy $\|\eta_0(\cdot)-y^s(\cdot,\theta)\|_{L_2(\Omega)}$ between the computer model $y^s$ and the physical system $\eta_0$ as a function of $\theta$ for configuration 2. }
	\label{fig:L2_loss_example1}
\end{figure}

 We consider three configurations. 
\begin{itemize}
	\item \textbf{Configuration 1. }The computer model is 
	\[y^s(x,\btheta) = 7[\sin(2\pi\theta_1-\pi)]^2+2[(2\pi\theta_2-\pi)^2\sin(2\pi x-\pi)],\]
	and the physical system coincides with the computer model when $\btheta_0^* = [0.2, 0.3]\transpose{}$, \emph{i.e.}, $\eta_0(x) = y^s(x,\btheta_0^*)$. The design space $\Omega$ is $[0,1]$, and the parameter space $\Theta$ for $\btheta$ is $[0,0.25]\times[0,0.5]$.  We simulate $n = 50$ observations from the randomly perturbed physical system $y_i = \eta_0(x_i)+e_i$, where $(x_i)_{i=1}^n$ are uniformly sampled from $\Omega$, and the variance for the noises $(e_i)_{i=1}^n$ is set to $0.2^2$. 
	\item \textbf{Configuration 2. } We follow an example provided in \cite{gu2017improved}. The computer model is 
	$y^s(x,\theta) = \sin(5\theta x)+5x,$
	and the physical system is $\eta_0(x) = 5x\cos(15x/2)+5x$. The design space $\Omega$ is $[0,1]$, and the parameter space $\Theta$ for $\theta$ is $[0,3]$.  We simulate $n = 30$ observations from $y_i = \eta_0(x_i)+e_i$ with $\mathrm{var}(e_i)=0.2^2$, and $(x_i)_{i=1}^n$ are equidistant on $\Omega$. 
  The $L_2$-discrepancy $\|\eta_0(\cdot)-y^s(\cdot,\theta)\|_{L_2(\Omega)}$ between $y^s$ and  $\eta_0$ as a function of $\theta$ is depicted in Figure \ref{fig:L2_loss_example1}. 
  The minimizer of the $L_2$-discrepancy is at $\theta_0^* = 1.8771$. 
	\item \textbf{Configuration 3. } We use the pedagogical example in \cite{plumlee2017bayesian}. The physical system is $\eta_0(x) = 4x+x\sin (5x)$ and the computer model is $y^s(x,\theta) = \theta x$, where $x\in\Omega=[0,1]$ and $\theta\in\Theta=[2,4]$. We take $(x_i)_{i=1}^n=\{0, 0.05, 0.1, 0.15, 0.2, \cdots,0.8\}$, and the responses are given by $y_i=\eta_0(x_i)+e_i$ with $\mathrm{var}(e_i)=0.02^2$. The $L_2$-discrepancy as a function of $\theta$ is given by 
	\[
	\|\eta(\cdot)-y^s(\cdot,\theta)\|_{L_2(\Omega)}=\sqrt{0.33(4-\theta)^2-0.2898(4-\theta)+0.201714,}
	\]
	and is minimized at $\theta_0^* = 3.5609$. 
\end{itemize}

For the three configurations described above, we assign the Gaussian process prior $\mathrm{GP}(0,\tau^2\Psi_\alpha)$ on $\eta$, where $\Psi_\alpha$ is the Mat\'ern covariance function given by \eqref{eqn:matern_covariance_function} with $\alpha = 5/2$. Here the scaling factor $\tau$ is set to  1 in all three configurations for the ease of implementation.
To draw posterior samples of $\btheta_\eta^*$, we first draw posterior samples of $\eta$ using formulas \eqref{eqn:GP_posterior_mean} and \eqref{eqn:GP_posterior_covariance}, then compute $\btheta_\eta^*$ by $\btheta_\eta^*=\argmin_\btheta\|\eta(\cdot)-y^s(\cdot,\btheta)\|_{L_2(\Omega)}^2$ using Algorithm \ref{alg:AdaGrad_calibration}. 
For all three configurations, $1000$ samples of $\btheta_\eta^*$ are drawn from the posterior distribution for subsequent analysis, and the number of random samples $N$ for the modified \emph{AdaGrad} in Algorithm \ref{alg:AdaGrad_calibration} is set to $2000$. For configurations 1 and 2, We also draw $1000$ samples  from the approximate projected calibration using Algorithm \ref{alg:approximate_PC}.

For comparison, we implement the calibration method by \cite{kennedy2001bayesian} (abbreviated as KO) and the orthogonal Gaussian process method by \cite{plumlee2017bayesian} (abbreviated as OGP). 
For the KO calibration approach, we follow the suggestion of \cite{van2009adaptive} and let the range parameter $\psi$ follow an inverse-Gamma prior distribution $\psi\sim\pi(\psi)\propto \psi^{-a_\psi - 1}\exp(-b_\psi/\psi)$ for some $a_\psi, b_\psi > 0$. We set $a_\psi = b_\psi = 2$ in all numerical examples. 
For both methods in all three configurations, Markov chain Monte Carlo is applied to draw $1000$ posterior samples after discarding $1000$ burn-in samples. 

\begin{figure}[htbp]
	\centerline{\includegraphics[width=1\textwidth]{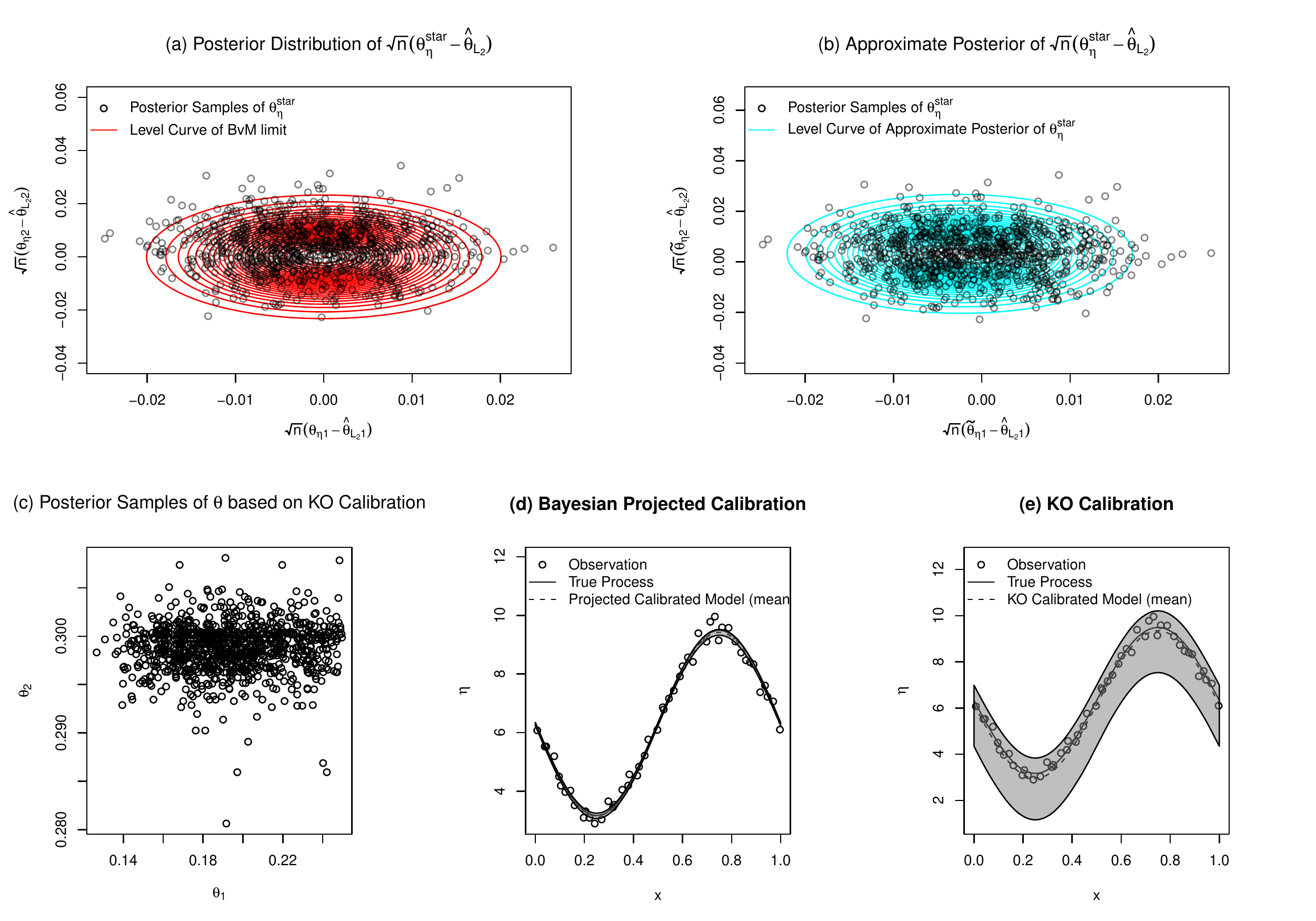}}
	\caption{Visualization of the posterior inference for configuration 1 in simulation studies. Panels (a) and (b) show the scatter plot of the posterior samples of $\sqrt{n}(\btheta_\eta^*-\widehat{\btheta}_{L_2})$ against the level curves of the corresponding BvM limit and the approximate projected calibration density from Algorithm \ref{alg:approximate_PC}, respectively. Panel (c) presents the scatter plot of the posterior samples of $\btheta$ using the KO approach. Panels (d) and (e) display the calibrated computer models (in dashed lines) using the Bayesian projected calibration and the KO approach, respectively, together with their corresponding point-wise $95\%$-credible intervals (in shaded area). }
	\label{fig:Projected_calibration_histogram_configuration1}
\end{figure}

\begin{figure}[htbp]
  \centerline{\includegraphics[width=1\textwidth]{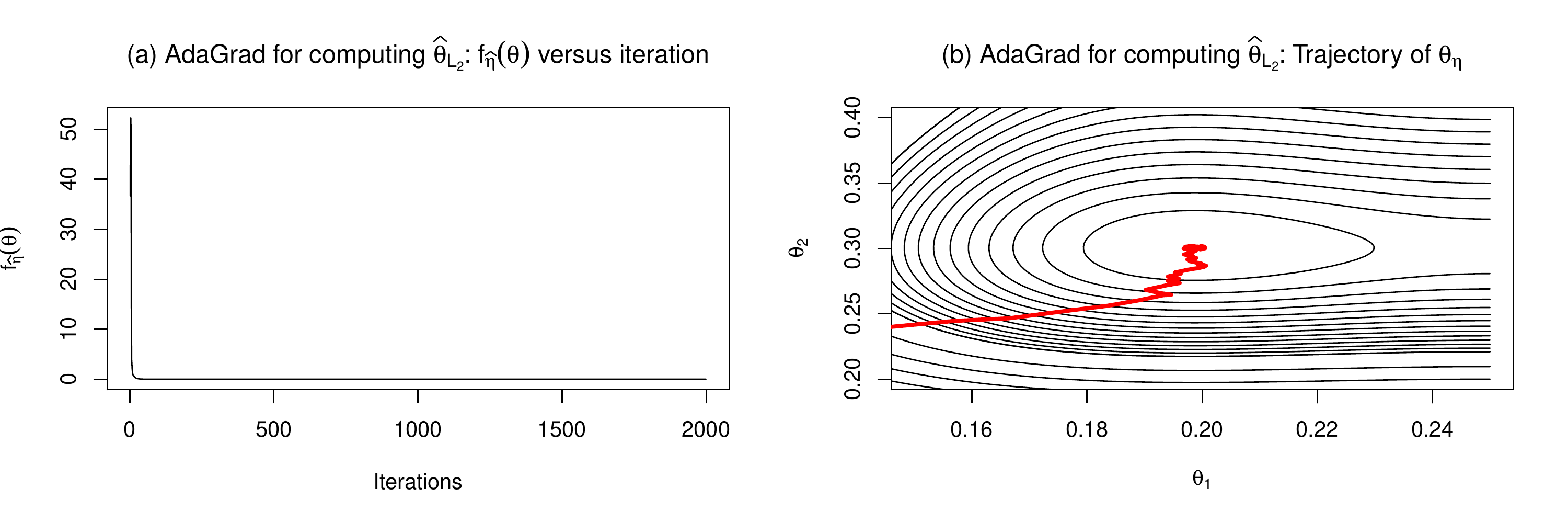}}
  \caption{Convergence behavior of the modified \emph{AdaGrad} for computing $\widehat\btheta_{L_2}$ for configuration 1. }
  \label{fig:Projected_calibration_convergence_configuration1}
\end{figure}

For configuration 1, the summary statistics of the posterior distribution of $\btheta^*_\eta$ are provided in Table \ref{table:summary_stat_configuration1}, together with those using Algorithm \ref{alg:approximate_PC}, the KO approach, and the OGP method. We can see that the Bayesian projected calibration, the approximate projected calibration using Algorithm \ref{alg:approximate_PC}, and the OGP method outperform the KO approach in terms of both the accuracy of the point estimates (posterior means) and the uncertainty quantification (length of credible intervals and standard deviations of the posterior samples).
Although the OGP provides comparable posterior inference to the Bayesian projected calibration, the computational runtime is significantly longer than other   methods.
The computational bottleneck of the OGP was also discussed in Section 6 of \cite{plumlee2017bayesian}. 
\begin{table}[htbp]
	\centering
	\caption{Summary Statistics of Posterior of $\btheta$ for Configuration 1 (the simulation truth is $\btheta_0^*=[0.2,0.3]\transpose{}$); Projected refers the Bayesian projected calibration, and Approximate refers to the approximate projected calibration using Algorithm \ref{alg:approximate_PC}. }
	\begin{tabular}{c|c c|c c|c c|c c}
		\hline\hline
		&\multicolumn{2}{c}{Projected}&\multicolumn{2}{c}{KO}&\multicolumn{2}{c}{OGP}
    &\multicolumn{2}{c}{Approximate}
    \\
		\hline
		$\btheta$& $\theta_1$ & $\theta_2$ & $\theta_1$ & $\theta_2$ & $\theta_1$ & $\theta_2$& $\theta_1$ & $\theta_2$\\
		\hline
		Mean&$0.1984$& $0.3009$& 
         $0.1934$ & $0.2988$ & 
         $0.2002$ & $0.2987$ &  
         $0.1986$ & $0.3004$\\
		Standard Deviation& $0.0011$ & $0.0013$ & 
                        $0.0269$ & $0.0025$ & 
                        $0.0005$ & $0.0006$ & 
                        $0.0011$ & $0.0013$\\
		$97.5\%$-Quantile &$0.2006$ & $0.3034$& 
                       $0.2439$ & $0.3182$ & 
                       $0.2013$ & $0.2999$ & 
                       $0.2007$ & $0.3029$\\
		$2.5\%$-Quantile &$0.1963$ & $0.2984$& 
                      $0.1445$ & $0.2938$ & 
                      $0.1992$ & $0.2975$ & 
                      $0.1965$ & $0.2979$\\
		\hline
		Runtime&\multicolumn{2}{c}{279s} & \multicolumn{2}{c}{0.834s} & \multicolumn{2}{c}{40562s}& \multicolumn{2}{c}{7.365s}\\
		\hline\hline
	\end{tabular}%
	\label{table:summary_stat_configuration1}
\end{table}%
Figure \ref{fig:Projected_calibration_histogram_configuration1}(a) presents the scatter plot of the posterior samples of $\sqrt{n}(\btheta_\eta^*-\widehat{\btheta}_{L_2})$. The level curves of the BvM limit shows that the asymptotic distribution of $\Pi(\sqrt{n}(\btheta_\eta^*-\widehat{\btheta}_{L_2})\mid\calD_n)$ developed in Section \ref{sec:asymptotic_properties} offers a decent approximation to the exact posterior. 
Figure \ref{fig:Projected_calibration_histogram_configuration1}(b) presents the scatter plot of the posterior samples of $\sqrt{n}(\btheta_\eta^*-\widehat{\btheta}_{L_2})$ against the level curves of the approximate projected calibration density, showing that the approximate projected calibration is satisfactory as well. We provide the trace plot of the loss function $f_{\widehat\eta}(\btheta) = \|\widehat\eta(\cdot) - y^s(\cdot,\btheta)\|_{L_2(\Omega)}^2$ and the trajectory of the calibration parameter $\btheta$ in Figure \ref{fig:Projected_calibration_convergence_configuration1} to demonstrate the convergence behavior of the modified \emph{AdaGrad} in Algorithm \ref{alg:approximate_PC}.  
Comparing Figures \ref{fig:Projected_calibration_histogram_configuration1}(a) and (c), the Bayesian projected calibration outperforms the KO method in terms of the
uncertainty quantification. We also investigate the performance of the calibrated computer model in Figures \ref{fig:Projected_calibration_histogram_configuration1}(d) and (e). The point-wise $95\%$-credible bands for the computer model also validate that the Bayesian projected calibration produces less uncertainty in contrast to the KO approach in calibrating the computer model $y^s$. 

\begin{figure}[htb]
	\centerline{\includegraphics[width=1\textwidth]{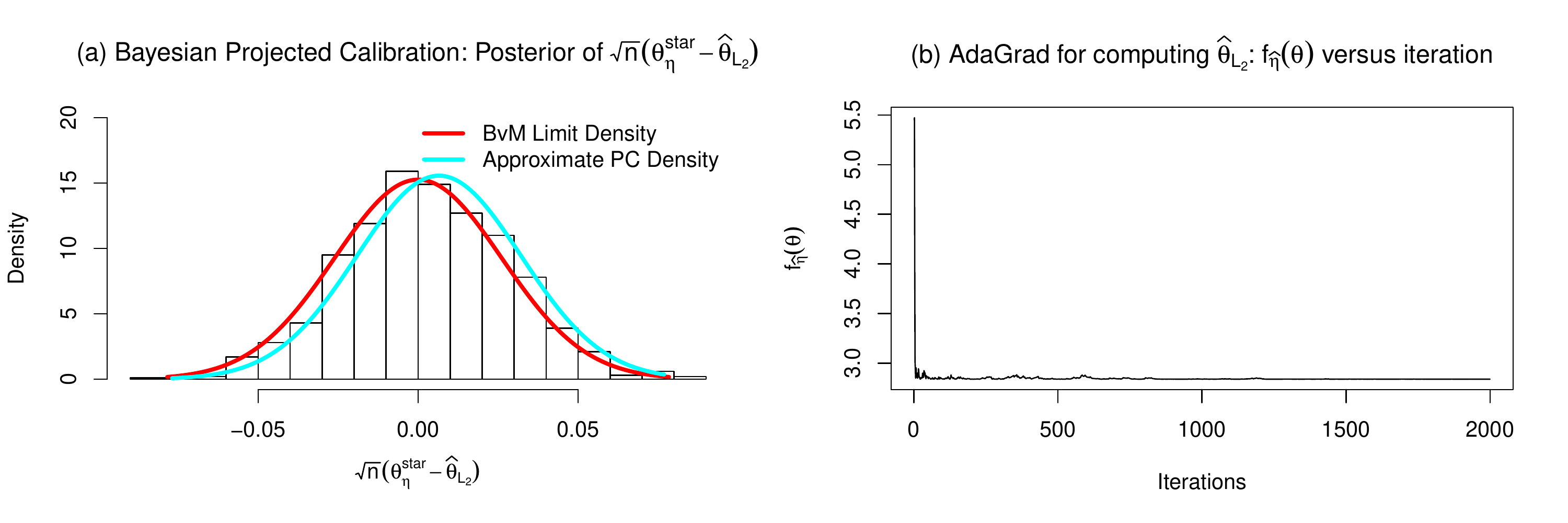}}
	\caption{Simulation study configuration 2: Panel (a) is the histogram of the posterior samples of $\sqrt{n}(\btheta_\eta^*-\widehat{\btheta}_{L_2})$, together with the theoretical BvM limit density (red solid line); Panel (b) presents the trace plot of the loss function $f_{\widehat\eta}(\btheta)$ versus the number of iterations. }
	\label{fig:Projected_calibration_theta_configuration2}
\end{figure}

Similarly, for configuration 2, the advantages of the Bayesian/approximate projected calibration in terms of the uncertainty quantification and the computational cost can be summarized from the statistics reported in Table \ref{table:summary_stat_configuration2}. 
It can be seen that the Bayesian/approximate projected calibration provide smaller uncertainty compared to the KO calibration. 
We also provide the histogram of the projected calibration and the density of the approximate projected calibration (blue curve) in Figure \ref{fig:Projected_calibration_theta_configuration2}(a), and it can be seen that the approximate projected calibration density provides a decent approximation to the exact posterior. Furthermore, the red curve in Figure \ref{fig:Projected_calibration_theta_configuration2}(a) shows that the asymptotic BvM limit approximates the exact posterior well even though the sample size is only $n = 30$.  The convergence of the modified \emph{AdaGrad} in Algorithm \ref{alg:approximate_PC} can be assessed via the trace plot of the loss function $f_{\widehat\eta}(\btheta)$ in Figure \ref{fig:Projected_calibration_theta_configuration2}(b). 
\begin{table}[htbp]
	\centering
	\caption{Summary Statistics of Posterior of $\theta$ for Configuration 2 (simulation truth is $\theta_0^* = 1.8771$); Projected refers the projected calibration, and Approximate refers to the approximate projected calibration using Algorithm \ref{alg:approximate_PC}. }
	\begin{tabular}{c|c|c|c|c}
		\hline\hline
		&{Projected}&{KO}&{OGP}&Approximate\\
		\hline
		Mean&$1.8816$& $1.8805$ & $1.8825$ & $1.8822$\\
		Standard Deviation& $0.0047$ & $0.0661$ & $0.0023$ & $0.0047$\\
		$97.5\%$-Quantile &$1.8907$ & $2.0089$ & $1.8766$ & $1.8915$\\
		$2.5\%$-Quantile &$1.8725$ & $1.7480$ & $1.8678$ & $1.8731$\\
		\hline
		Runtime & 237.289s & 1.034s & 31843s & 6.269s\\
		\hline\hline
	\end{tabular}%
	\label{table:summary_stat_configuration2}
\end{table}%

\begin{figure}[htbp]
  \centerline{\includegraphics[width=1.1\textwidth]{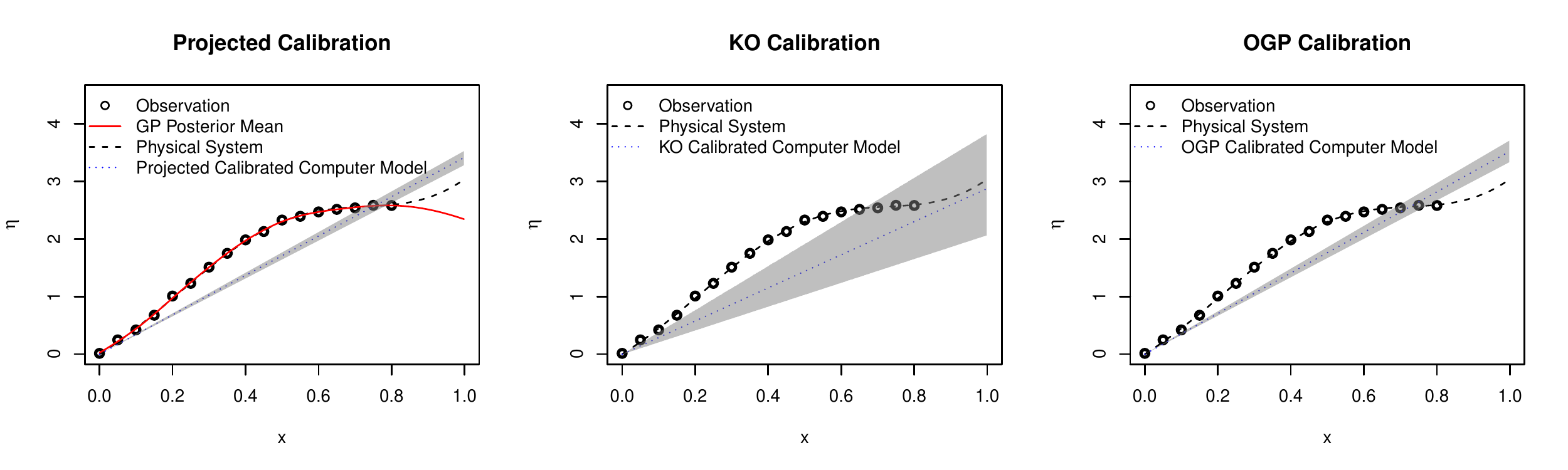}}
  \caption{Visualization of the posterior inference for configuration 3. The three panels show the calibrated computer models (in dotted lines) using the Bayesian projected calibration, the KO calibration approach, and the OGP calibration, respectively, together with their corresponding point-wise $95\%$-credible intervals (in shaded area). The dashed lines are the physical system. }
  \label{fig:Projected_KO_OGP_config3}
\end{figure}

The scenario for configuration 3 is slightly challenging due to the fact that no physical data are available on $(0.8,1]$, and the physical data are relatively sparse (see Figure \ref{fig:Projected_KO_OGP_config3}). 
In such a scenario, we do not recommend using Algorithm \ref{alg:approximate_PC} for approximate posterior inference. We provide the corresponding summary statistics for the Bayesian projected calibration, the KO approach, and the OGP method, in Table \ref{table:summary_stat_configuration3}. When the design points are not regularly spread over $\Omega$, 
the KO approach yields larger uncertainty when estimating $\btheta$ compared to the Bayesian projected calibration and the OGP method. Note that it is unfair to compare the point estimate of the KO approach with those of the other two competitors, since the ``true'' values of $\btheta$ are different. 
For the uncertainty quantification performance measured by the width of credible intervals and standard deviation, the OGP   and the Bayesian projected calibration are similar, and both outperform the KO approach. The calibrated computer models using the three approaches are visualized in Figure \ref{fig:Projected_KO_OGP_config3}.
\begin{table}[htbp]
	\centering
	\caption{Summary Statistics of Posterior of $\theta$ for Configuration 3 (simulation truth is $\theta_0^*=3.5609$)}
	\begin{tabular}{c|c|c|c}
		\hline\hline
		&{Projected Calibration}&{KO Calibration}&{OGP Calibration}\\
		\hline
		Mean&$3.4064$& $3.1109$ & $3.6001$\\
		Standard Deviation& $0.0614$ & $0.4760$ & $0.0911$\\
		$97.5\%$-Quantile &$3.5964$ & $3.9385$ & $3.7733$\\
		$2.5\%$-Quantile &$3.3624$ & $2.1467$ & $3.4167$\\
		\hline\hline
	\end{tabular}%
	\label{table:summary_stat_configuration3}
\end{table}%


\subsection{Ion Channel Example} 
\label{sub:ion_channel_example}
\begin{figure}[htb]
	\centerline{\includegraphics[width=1\textwidth]{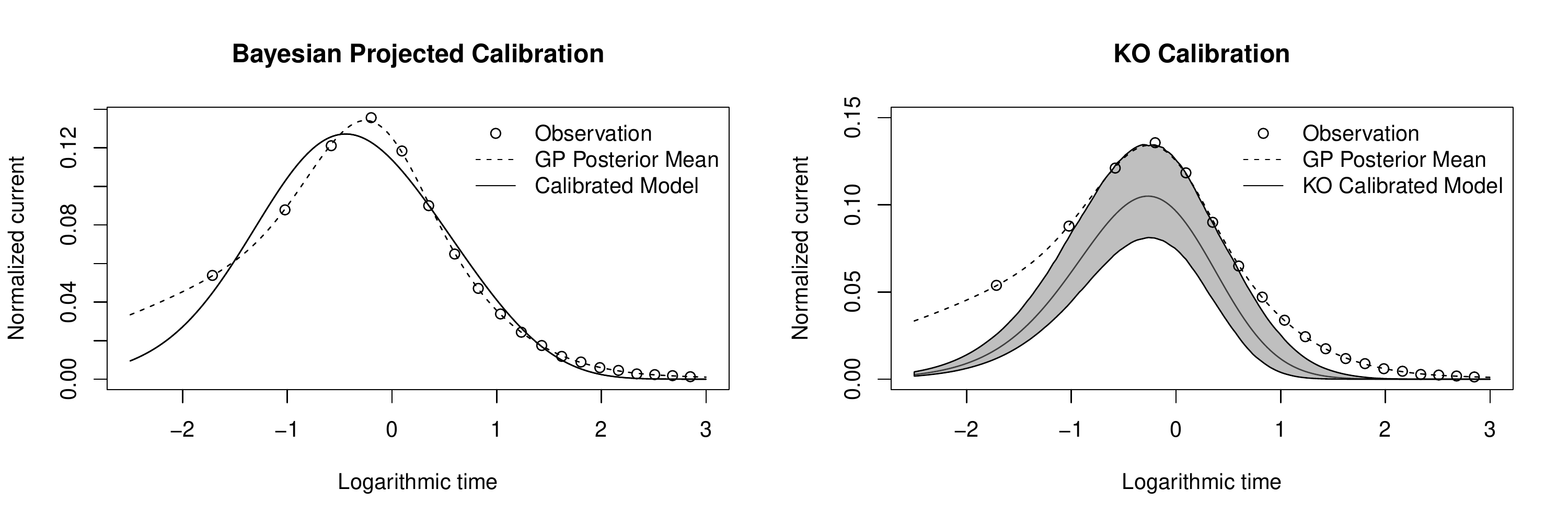}}
	\caption{Visualization of the computer model calibration for the ion channel example. The left and right panels present the calibrated computer models (dashed lines) using the proposed approach and the KO calibration approach, respectively. The shaded area is the point-wise $95\%$-credible intervals for the KO calibrated computer model. The physical data (circles) and the Gaussian process (GP) estimates of the physical system (dashed lines) are also displayed. }
	\label{fig:Projected_vs_KO_ion_channel}
\end{figure}

We apply the Bayesian projected calibration to the ion channel example used in \cite{plumlee2016calibrating}. This dataset involves measurements from experiments concerning ion channels of cardiac cells. Specifically, the output of the experiment is the current through the sodium channels in a cardiac cell membrane, and the input is the time. For detailed description of the experiment, we refer to \cite{plumlee2016calibrating}. Here we consider a subset of the original dataset, which consists of $19$ normalized current records needed to maintain the membrane potential fixed at $-35$mV, together with the logarithm of the corresponding time as the inputs. The same dataset was also studied in \cite{plumlee2017bayesian}. For the computer model, \cite{clancy1999linking} suggests the following Markov model for ion channels:
\[
y^s(x,\btheta) = \mathbf{e}_1\transpose{}\exp[\exp(x)A(\btheta)]\mathbf{e}_4,
\]
where $\mathbf{e}_i$ is the column vector with $1$ at the $i$th coordinate and $0$ for the rest components, the outer $\exp$ is the matrix exponential function, and the $A(\btheta)$ matrix is 
\[
A(\btheta) = \left[\begin{array}{cccc}
-\theta_2-\theta_3&\theta_1&0&0\\
\theta_2&-\theta_1-\theta_2&\theta_1&0\\
0&\theta_2&-\theta_1-\theta_2&\theta_1\\
0&0&\theta_2&-\theta_1
\end{array}\right].
\]
We implement Algorithm \ref{alg:AdaGrad_calibration} developed in Section \ref{sub:stochastic_optimization_for_projected_calibration} to collect $1000$ posterior samples of $\btheta$ under the Bayesian projected calibration approach.
We also collect $1000$ posterior samples of $\btheta$ under the KO   approach. The OGP approach, however, fails to provide adequate samples from the posterior distribution for subsequent inference within 20 hours. The roughness parameter $\alpha$ for the Mat\'ern covariance function is set to $5/2$, and we set $\tau = 0.02$, $\sigma = 0.001$, as suggested by \cite{plumlee2017bayesian}. Table \ref{table:summary_stat_ion_channel} presents the corresponding comparison of summary statistics. The calibrated computer models are also visualized in Figure \ref{fig:Projected_vs_KO_ion_channel}. Clearly, the Bayesian projected calibration provides better estimates to both the calibration parameter $\btheta$ and the computer model in terms of lower uncertainty (smaller standard deviation and thinner credible intervals). It can also be seen that the Bayesian projected calibrated computer model yields better approximation to the physical data than the KO approach. 
\begin{table}[htbp]
	\centering
	\caption{Summary Statistics of Posterior of $\btheta$ for the Ion Channel Example}
	\begin{tabular}{c|c|c|c|c|c|c}
		\hline\hline
		&\multicolumn{3}{c}{Projected Calibration}&\multicolumn{3}{c}{KO Calibration}\\
		\hline
		$\btheta$& $\theta_1$ & $\theta_2$ & $\theta_3$ & 
               $\theta_1$ & $\theta_2$ & $\theta_3$\\
		\hline
		Mean& $6.011166$ & $5.578567$ & $3.500813$ & 
          $3.4713447$ & $0.9325514$ & $6.7811932$ \\
		Standard Deviation& $0.000012$ & $0.000006$ & $0.000006$ & 
                        $0.2974497$ & $0.5369031$ & $1.1803662$\\
		$97.5\%$-Quantile &$6.011191$ & $5.578578$ & $3.500824$ & 
                       $4.154933$ & $2.034486$ & $9.148351$\\
		$2.5\%$-Quantile &$6.011143$ & $5.578556$ & $3.500802$ & 
                      $3.009278$ & $0.114780$ & $4.536802$\\
		\hline\hline
	\end{tabular}%
	\label{table:summary_stat_ion_channel}
\end{table}%


\subsection{Spot Welding Example} 
\label{sub:spot_welding_example}
Now we consider the spot welding example studied in 
\cite{bayarri2007framework} and \cite{chang2014model}. Three control variables in the experiment are the load, the current, and the gauge. The physical experiments are only conducted for gauge being $1$ and $2$. Since the computer model fails to produce enough meaningful outputs when the gauge is set to $1$, here we only consider the case where the gauge is $2$, \emph{i.e.}, the control variables are the load and the current only.
The physical data are provided in Table 4 of \cite{bayarri2007framework}. For each fixed design point, the mean of the $10$ replicates of the output is taken as the response. 

The computer model for the spot welding system, on the other hand, is not directly available to us. In short, the computer model consists of a time-consuming finite element method (FEM) for numerically solving a system of partial differential equations (PDEs). 
There are $21$ available runs for the computer code, as presented in Table 3 of \cite{bayarri2007framework}. 
Besides the three control variables (the load, the current, and the gauge) in the physical experiment, the computer model also involves another unknown parameter $\theta$ (denoted as $u$ in \citealp{bayarri2007framework}) that summarizes the material and surface. This parameter needs to be tuned with the physical data, and is exactly the calibration parameter in our context. 
As discussed in Section \ref{sec:problem_formulation}, an emulator is needed as a surrogate for the computer model. 
Here we apply the \verb|RobustGaSP| package \citep{gu2018robustgasp} to emulate the expensive FEM computer model. For theoretical background on the \verb|RobustGaSP| emulator, we refer to \cite{gu2017robust}.

\begin{figure}[htbp!]
	\centerline{\includegraphics[width=1\textwidth]{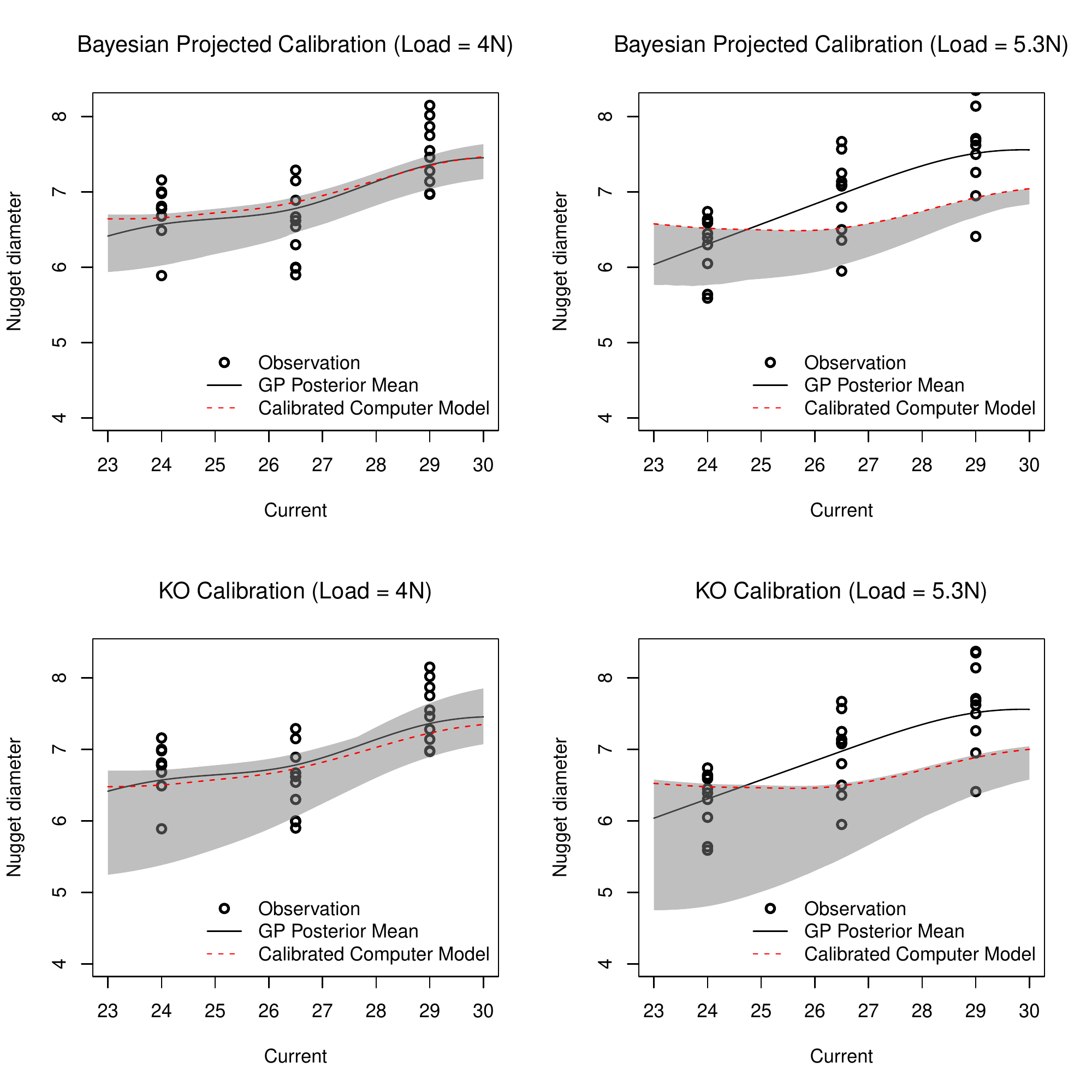}}
	\caption{Visualization of computer model calibration for the spot welding example. The left and right panels present the calibrated computer models (red dashed lines) as a function of the current with the load fixed at $4$N and $5.3$N, respectively. The shaded areas are the point-wise $95\%$-credible intervals for the corresponding calibrated computer models. The physical data (circles) and the Gaussian process (GP) estimates of the physical system (solid lines) are also displayed. }
	\label{fig:Calibration_computer_model_Spot_Welding}
\end{figure}

We follow similar approaches in Subsections \ref{sub:simulation_studies} and \ref{sub:ion_channel_example} to draw posterior samples under the Bayesian projected calibration and the KO approach. The only difference is that the non-available computer model $y^s$ is replaced by the predictive mean of the \verb|RobustGaSP| emulator based on the results of the $21$ runs on the FEM computer code. The summary statistics are presented in Table \ref{table:summary_stat_spot_welding}, indicating that the Bayesian projected calibration outperforms the KO approach in terms of the uncertainty quantification, \emph{i.e.}, smaller standard deviation and thinner credible interval. The computer models calibrated via the Bayesian projected calibration and the KO approach, together with their corresponding point-wise $95\%$-credible intervals, are depicted in Figure \ref{fig:Calibration_computer_model_Spot_Welding}. We can see that in terms of computer model calibration, both approaches behave similarly. The point-wise credible intervals, however, indicate that the Bayesian projected calibration method outperforms the KO approach regarding the uncertainty quantification. The imperfection of the computer model can also be seen from the discrepancy presented on the right two panels of Figure \ref{fig:Calibration_computer_model_Spot_Welding}.
\begin{table}[htbp!]
	\centering
	\caption{Summary Statistics of Posterior of $\theta$ for the Spot Welding Example}
	\begin{tabular}{c|c|c}
		\hline\hline
		&{Projected Calibration}&{KO Calibration}\\
		\hline
		Mean&$4.385933$& $4.126239$ \\
		Standard Deviation& $0.08455849$ & $1.440555$\\
		$97.5\%$-Quantile &$4.505187$ & $7.164378$\\
		$2.5\%$-Quantile &$4.183981$ & $1.604301$\\
		\hline\hline
	\end{tabular}%
	\label{table:summary_stat_spot_welding}
\end{table}%

\section{Conclusion and Discussion} 
\label{sec:conclusion_and_discussion}

We develop a novel Bayesian projected calibration method following the frequentist $L_2$-projected calibration method in \cite{tuo2015efficient}. The proposed approach is formulated in an identifiable way and naturally quantifies the uncertainty of the calibration parameters through the posterior distribution. Theoretical justification of the Bayesian projected calibration is provided: the marginal posterior distribution of the calibration parameter is not only $\sqrt{n}$-consistent, but also asymptotically normal with the efficient covariance matrix. 

To estimate the true value $\btheta_0^*$ of the calibration parameter (defined as the minimizer of the $L_2$-distance between the physical system and the computer model), the  
OGP calibration  \citep{plumlee2017bayesian} and the Bayesian projected calibration proposed in this work can be applied. Alternatively,  \cite{doi:10.1137/17M1159890} proposed  to directly apply a modified GP prior, referred to as the scaled Gaussian process, to the discrepancy function $\delta(\bx) =\eta(\bx) - y^s(\bx,\btheta)$ for computer model calibration. The scaled Gaussian process is defined by modifying the eigenvalues of the covariance function in some classical GPs (\emph{e.g.}, Mat\'ern processes or squared-exponential processes) such that the sample paths have smaller $L_2$-norms than the original GP. Its construction is slightly involved, but the resulting maximum \emph{a posteriori} estimator of $\btheta$ and $\delta$ can be expressed as the following doubly penalized kernel ridge regression problem \citep{gu2018theoretical}:
\begin{align*}
(\widehat\btheta, \widehat\delta) = \argmin_{\btheta\in\Theta, \delta\in\mathbb{H}(\Omega)}\frac{1}{n}\sum_{i = 1}^n[y_i - y^s(\bx_i,\btheta) - \delta(\bx_i)]^2 + \lambda_1\|\delta\|_{\mathbb{H}(\Omega)}^2 + \lambda_2\|\delta\|_{L_2(\Omega)}^2,
\end{align*}
where $\mathbb{H}(\Omega)$ is the RKHS associated with the original GP, and $\lambda_1,\lambda_2 > 0$ are tuning parameters. The motivation of the extra penalty term $\lambda_2\|\delta\|_{L_2(\Omega)}^2$ in comparison with \eqref{eqn:kernel_ridge_regression} exactly comes from the idea of minimizing the $L_2$-norm of $\delta$. When $\lambda_1$ and $\lambda_2$ are carefully selected, the resulting estimate $\widehat\btheta$ converges to $\btheta_0^*$, but the rate is slower than the $1/\sqrt{n}$ \citep{gu2018theoretical} in contrast to the Bayesian projected calibration and the $L_2$-calibration. Such a drawback may not be desired when efficient estimation of $\btheta$ is needed. 

In this work we follow the definition in \cite{tuo2015efficient} and \cite{wong2017frequentist} to define the true value $\btheta_0^*$  of the calibration parameter as the minimizer of the $L_2$-distance between the physical system $\eta(\cdot)$ and the computer model $y^s(\cdot,\btheta)$, in which we assume that $\btheta_0^*$ can be uniquely defined. 
When $\btheta_0^*$ is not  uniquely defined via the $L_2$-projection, we can define $\btheta_0^*$ using alternative loss functions.  
For example, if certain expert knowledge on  the mechanism of the computer model  or the calibration procedure results in a penalty function $\calP(\btheta, y^s)$,
one may define $\btheta_0^*$ as the minimizer of the following  penalized $L_2$-function 
\[
\|\eta_0(\cdot) - y^s(\cdot,\btheta)\|_{L_2(\Omega)} +  \calP(\btheta, y^s).
\]
Such a calibration procedure not only shrinks the discrepancy between the physical system and the computer model, but also integrates the experts' knowledge. 
The corresponding asymptotic theory and efficiency of the Bayes estimates can be developed following the same techniques adopted in this work, provided that $\calP$ is twice continuously differentiable.


The proposed Bayesian projected calibration can be extended to cases where the model discrepancy cannot be modeled by an additive stochastic process in the measurement equation. For instance, consider the following nonlinear state space model
	\[
	\eta''(x) = \theta_1\eta'(x) + \theta_2\eta(x) + \theta_3\eta^3(x) + q(x) + \delta(x),
	\]
	where $q(x)$ is some known process, $\delta$ is the model discrepancy, $\btheta = [\theta_1,\theta_2,\theta_3]\transpose$ is the calibration parameter, and noisy  measurements  $y_i$'s are taken from $y_i = \eta(\bx_i) + \eps_i$, where $\eps_i\sim\mathrm{N}(0,\sigma^2)$ independently. For simplicity let $q$ be deterministic.  Following the spirit of minimizing the $L_2$-norm of the discrepancy function, one may define $\btheta_\eta^*$ by
	\[
	\btheta_\eta^* = \argmin_{\btheta\in\Theta}\left\|\eta''(x) - \btheta\transpose\bv_\eta(\cdot) - q(\cdot)\right\|_{L_2(\Omega)}^2,
	\]
	where $\bv_\eta(x) = [\eta'(x),\eta(x),\eta^3(x)]\transpose$. Then we can model $\eta$, $\eta'$, and $\eta''$ jointly by assigning a GP prior on $\eta$ with sufficient smoothness (see, for example, Section 9.4 of \citealp{rasmussen2006gaussian}). Simple algebra leads to the following closed-form formula for $\btheta_\eta^*$:
	\[
	\btheta_\eta^* = \left\{\int_\Omega\bv_\eta(x)\bv_\eta(x)\transpose\mathrm{d}x\right\}^{-1}\int_\Omega\bv_\eta(x)[\eta''(x) - q(x)]\mathrm{d}x.
	\]
	It is easy to compute $\btheta_\eta^*$ once $\eta$ is appropriately modeled through a GP prior with sufficient smoothness, but the theoretical properties would require a separate exploration.

The estimation methods in this work can be viewed as the following two-step procedure: First estimate the physical system through Gaussian process models; Then estimate the calibration parameter using the $L_2$-projection criterion. On the other hand, it is possible to jointly estimate the calibration parameter and the discrepancy between the physical system and the computer model. For example, \cite{plumlee2017bayesian} proposed an orthogonal Gaussian process (OGP) model to tackle this joint estimation issue. 
The theoretical development for OGP, nevertheless, is only restricted to a point estimate \citep{tuo2017adjustments}: the \emph{maximum a posteriori} (MAP) estimate of $\btheta$ is asymptotically normal and semiparametric efficient. It will be non-trivial to apply the techniques developed here to the OGP calibration approach, and asymptotic characterization of the corresponding full posterior distribution will be an interesting topic. 

Similar to the OGP calibration method, the Bayesian projected calibration also involves intractable integrals, and we propose stochastic approximation methods to reduce computational complexity in Section \ref{sec:algorithmic_issue}. For a moderately large sample size, one can apply Algorithm \ref{alg:approximate_PC} to efficiently approximate  the projected calibration. However, for sparse data, one has to rely on Algorithm \ref{alg:AdaGrad_calibration} to perform the exact posterior inference. 
It is therefore desired that the computational barrier of Algorithm \ref{alg:AdaGrad_calibration} can be tackled via more efficient algorithms. 



\appendix
\bigskip
\begin{center}
{\large\bf APPENDIX}
\end{center}


\section{Auxiliary Results} 
\label{sec:auxiliary_results}
In this section we list some auxiliary results that are used to prove theorem \ref{thm:BvM_limit}. The proofs of the lemmas stated in this section are deferred to the supplementary material. We first introduce some notions and definitions that are widely used in  empirical processes studies. 
Suppose $\calF$ is a function space equipped with metric $d$. Given two functions $l$, $u\in\calF$, a bracket $[l,u]$ is a set of functions $f$ such that $l\leq f\leq u$ everywhere, and the size of the bracket is defined to be $d(l,u)$. 
The $\eps$-bracketing number of $\calF$ with respect to the metric $d$, denoted by $\calN_{[\bdot]}(\eps,\calF,d)$, is the minimum number of brackets of size $\eps$ that are needed to cover $\calF$. The bracketing integral $J_{[\bdot]}(\eps,\calF,d)$ is defined to be the integral of the logarithmic bracketing number as follows:
\[
J_{[\bdot]}(\eps,\calF,d)=\int_0^\eps\sqrt{\log\calN_{[\bdot]}(\delta,\calF,d)}\mathrm{d}\delta.
\]
Suppose $\calX$ is the space where random variables take values. Given a sequence $(\bx_i)_{i=1}^n$ of independent and identically distributed random variables, the empirical measure and the empirical process of a function $f:\calX\to\mathbb{R}$, denoted by $\mathbb{P}_nf$ and $\mathbb{G}_nf$, are defined by
\[
\mathbb{P}_nf=\frac{1}{n}\sum_{i=1}^nf(\bx_i),\quad\mathbb{G}_nf=\frac{1}{\sqrt{n}}\sum_{i=1}^n[f(\bx_i)-\mathbb{E}f(\bx_i)],
\]
respectively. For two variables $a$ and $b$, we use $a\lesssim b$ and $a\gtrsim b$ to denote the inequalities up to a universal multiplicative constant, and $a\asymp b$ if $a\lesssim b$ and $a\gtrsim b$. 

In the empirical processes theory, maximum inequalities are widely adopted to study the asymptotic behavior of nonparametric estimates. Here we cite one of them that is used in the proof of Theorem \ref{thm:BvM_limit} (see Lemma 19.36 in \citealp{van2000asymptotic}).
\begin{theorem}\label{thm:maximum_inequality}
Let $(\bx_i)_{i=1}^n$ be independent and identically distributed according to a distribution $\mathbb{P}_\bx$ over $\calX$, and let $\calF$ be a class of measurable functions $f:\calY\to\mathbb{R}$. If $\|f\|_{L_2(\mathbb{P}_\bx)}^2<\delta^2$ and $\|f\|_\infty\leq M$ for all $f\in\calF$, where $\delta$ and $M$ does not depend on $\calF$, then
\begin{align}
\mathbb{E}\left[\sup_{f\in\calF}\left|\mathbb{G}_nf\right|\right]
\lesssim J_{[\bdot]}\left(\delta,\calF,\|\cdot\|_{L_2(\mathbb{P}_\bx)}\right)\left[1+\frac{M}{\delta^2\sqrt{n}}J_{[\bdot]}\left(\delta,\calF,\|\cdot\|_{L_2(\mathbb{P}_\bx)}\right)\right]\nonumber.
\end{align}
\end{theorem}

The following lemma is the modification of a standard probabilistic theorem for Gaussian processes. For related literature, we refer to \cite{van2008rates} and \cite{ghosal2017fundamentals}. 
\begin{lemma}\label{lemma:sieve_construction}
Suppose $\eta$ is imposed the Mat\'ern Gaussian process with roughness parameter $\alpha$, and $\eta_0\in\mathfrak{C}_\alpha(\Omega)\cap\calH_\alpha(\Omega)$, where $\alpha>p/2$. Let $\eps_n = n^{-\alpha/(2\alpha+p)}$. Then there exists a measurable set $\calB_n$ in $\mathfrak{C}(\Omega)$ such that for sufficiently large $n$, the following hold:
\begin{align}
\Pi(\calB_n\mid\calD_n)&=1-o_{\mathbb{P}_0}(1),\nonumber\\
J_{[\bdot]}(\eps_n\log n, \calB_n, \|\cdot\|_{L_2(\Omega)})&\lesssim (\log n)^{2\alpha/(2\alpha+p)}\sqrt{n}\eps_n^2.\nonumber
\end{align}
\end{lemma}
Now denote
\[
\ell_n(\eta)=\sum_{i=1}^n\log p_\eta(y_i,\bx_i)=\sum_{i=1}^n\log \phi_\sigma(y_i-\eta(\bx_i))
\]
to be the log-likelihood function of $\eta$ given  $(\bx_i,y_i)_{i=1}^n$. Define the event 
\begin{align}
\calA_n=\left\{\|\eta-\eta_0\|_{L_2(\Omega)}\leq M_n\eps_n\right\}\cap\left\{\|\eta-\eta_0\|_{L_\infty(\Omega)}\leq M\right\}\cap\calB_n\nonumber,
\end{align}
where $M_n=\log n$, $M$ is given by Theorem \ref{thm:matern_process_contraction}, and $\calB_n$ is given by Lemma \ref{lemma:sieve_construction}. Then by Theorem \ref{thm:matern_process_contraction} and Lemma \ref{lemma:sieve_construction} we know that $\Pi(\calA_n\mid\calD_n)=1-o_{\mathbb{P}_0}(1)$. 
\begin{lemma}\label{lemma:integral_ratio_stability}
Suppose the conditions of Theorem \ref{thm:BvM_limit} hold. For each vector $\bt\in\mathbb{R}^q$ and each $\eta\in\calF$ define
\[
\eta_\bt(\bx)=\eta(\bx)-\frac{2\sigma^2}{\sqrt{n}}\bt\transpose{}\bV_0^{-1}\frac{\partial y^s}{\partial\btheta}(x,\btheta_0^*).
\]
Given a realization $\eta$ of the Mat\'ern Gaussian process $\mathrm{GP}(0,\Psi_\alpha)$, define the following isometry associated with $\eta$:
\begin{align}
U:\mathbb{H}_0=\left\{\sum_{k=1}^K a_k\Psi(\cdot,\bt_k):\bt_k\in\Omega,a_k\in\mathbb{R},K\in\mathbb{N}_+\right\}\to L_2(\mathbb{P}_0),\ 
\sum_{k=1}^Ka_k\Psi(\cdot,\bt_k)\mapsto \sum_{k=1}^Ka_k\eta(\bt_k)\nonumber,
\end{align}
and extend $U$ from $\mathbb{H}_0$ to $\overline{\mathbb{H}}_0=\mathbb{H}_{\Psi_\alpha}(\Omega)$ continuously. Define the event
\[
\calC_n=\left\{|U(g)|\leq L\sqrt{n}\eps_n\|g\|_{\mathbb{H}_{\Psi_\alpha}(\Omega)}\right\}.
\]
Then there exists a sufficiently large $L$ such that $\Pi(\calC_n^c\mid\calD_n)=o_{\mathbb{P}_0}(1)$, and the following holds:
\begin{align}
\int_{\calA_n\cap\calC_n} \exp\left[\ell_n(\eta_\bt)-\ell_n(\eta_0)\right]\Pi(\mathrm{d}\eta)=\left[1+o_{\mathbb{P}_0}(1)\right]\left\{\int \exp\left[\ell_n(\eta)-\ell_n(\eta_0)\right]\Pi(\mathrm{d}\eta)\right\}\nonumber.
\end{align}
\end{lemma}

The asymptotic normality result of the $L_2$-projected calibration estimate $\widehat{\btheta}_{L_2}$ from \cite{tuo2015efficient} is also useful to study the asymptotic behavior of $\Pi(\sqrt{n}(\btheta_\eta^*-\widehat{\btheta})\in\cdot\mid\calD_n)$ in the case where $\widehat\btheta$ is taken to be $\widehat\btheta_{L_2}$.
\begin{theorem}
\label{thm:asymptotic_normality_L2_calibration}
Under the conditions of Theorem \ref{thm:BvM_limit}, it holds that
\begin{align}
\widehat{\btheta}_{L_2}-\btheta^*_0=2\bV_0^{-1}\left[\frac{1}{n}\sum_{i=1}^ne_i\frac{\partial y^s}{\partial \btheta}(\bx_i,\btheta_0^*)\right]+o_{\mathbb{P}_0}(n^{-1/2})\nonumber.
\end{align}
\end{theorem}


\section{Proof of Theorem \ref{thm:BvM_limit}} 
\label{sec:proof_of_theorem_thm:bvm_limit}
Theorem \ref{thm:matern_process_contraction} and Lemma \ref{lemma:sieve_construction} imply that $\Pi(\calA_n\cap\calC_n\mid\calD_n)=1-o_{\mathbb{P}_0}(1)$. Let $\Theta_n = \left\{\btheta_\eta^*:\eta\in\calA_n\cap\calC_n\right\}$. It follows directly that $\Pi(\btheta_\eta^*\in\Theta_n\mid\calD_n) = 1-o_{\mathbb{P}_0}(1)$. 
Denote 
\[\Pi(\btheta_\eta^*\in\bdot\mid\calD_n,\Theta_n) = \frac{\Pi(\btheta_\eta^*\in\bdot\cap\Theta_n\mid\calD_n)}{\Pi(\btheta_\eta^*\in\Theta_n\mid\calD_n)}.\]
Following the argument in \cite{castillo2015}, it suffices to show that
\[
\sup_A\left|\Pi\left(\sqrt{n}(\btheta_\eta^*-\widehat{\btheta})\in A\mid\calD_n,\Theta_n\right)-\mathrm{N}\left(\zero,4\sigma^2\bV_0^{-1}\bW\bV_0^{-1}\right)\right|\overset{\mathbb{P}_0}{\to} 0.
\]
We prove the result by the method of moment generating function, namely, showing that for any fixed vector $\bt\in\mathbb{R}^q$, it holds that
\begin{align}
\int_{\calA_n\cap\calC_n} \exp\left[\bt\transpose{}\sqrt{n}(\btheta_\eta^*-\widehat{\btheta})\right]\Pi(\mathrm{d}\eta\mid\calD_n)\to \exp\left[\frac{1}{2}\bt\transpose{}\left(4\sigma^2\bV_0^{-1}\bW\bV_0^{-1}\right)\bt\right]\nonumber
\end{align}
in $\mathbb{P}_0$-probability. The rest part of the proof is completed by Lemma 1 and Lemma 2 in \cite{castillo2015supplement}. 

Let $\eps_n = n^{-\alpha/(2\alpha+p)}$. 
The left-hand side of the preceding display can be re-written as
\begin{align}
\left\{\int\exp\left[\ell_n(\eta)-\ell_n(\eta_0)\right]\Pi(\mathrm{d}\eta)\right\}^{-1}\left\{\int_{\calA_n\cap\calC_n}\exp\left[\bt\transpose{}\sqrt{n}(\btheta_\eta^*-\widehat{\btheta})+\ell_n(\eta)-\ell_n(\eta_0)\right]\Pi(\mathrm{d}\eta)\right\}.\nonumber
\end{align}
For the vector $\bt\in\mathbb{R}^q$, define 
\[
\eta_\bt(\bx)=\eta(\bx)-\frac{2\sigma^2}{\sqrt{n}}\bt\transpose{}\bV_0^{-1}\frac{\partial y^s}{\partial\btheta}(\bx,\btheta_0^*),
\]
and for each $\eta$, define the remainder
\[
R_n(\eta,\eta_0)=\frac{n}{2}\|\eta-\eta_0\|_{L_2(\Omega)}^2-\frac{n}{2}\mathbb{P}_n(\eta-\eta_0)^2.
\]
Then simple algebra shows
\begin{align}
&\left[\ell_n(\eta_\bt)-\ell_n(\eta_0)\right]-\left[\ell_n(\eta)-\ell_n(\eta_0)\right]\nonumber\\
&\quad=-\frac{n}{2\sigma^2}\left[\|\eta_\bt-\eta_0\|_{L_2(\Omega)}^2-\|\eta-\eta_0\|_{L_2(\Omega)}^2\right]
-\frac{2}{\sqrt{n}}\sum_{i=1}^ne_i\bt\transpose{}\bV_0^{-1}\frac{\partial y^s}{\partial\btheta}(\bx_i,\btheta_0^*)\nonumber\\
&\quad\quad+\frac{1}{\sigma^2}\left[R_n(\eta_\bt,\eta_0)-R_n(\eta,\eta_0)\right]\nonumber\\
&\quad=2\sqrt{n}\int_\Omega[\eta(\bx)-\eta_0(\bx)]\bt\transpose{}\bV_0^{-1}\frac{\partial y^s}{\partial\btheta}(\bx,\btheta_0^*)\mathrm{d}\bx-\frac{1}{2}\bt\transpose{}\left(4\sigma^2\bV_0^{-1}\bW\bV_0^{-1}\right)\bt\nonumber\\
&\quad\quad-\frac{2}{\sqrt{n}}\sum_{i=1}^ne_i\bt\transpose{}\bV_0^{-1}\frac{\partial y^s}{\partial\btheta}(\bx_i,\btheta_0^*)+\frac{1}{\sigma^2}[R_n(\eta_\bt,\eta_0)-R_n(\eta,\eta_0)]\nonumber.
\end{align}
Denote the remainder of the Taylor expansion of $\btheta_\eta^*$ at $\btheta_0^*$ by
\[
\br(\eta,\eta_0)=\btheta_\eta^*-\btheta_0^*-2\int_\Omega[\eta(\bx)-\eta_0(\bx)]\bV_0^{-1}\frac{\partial y^s}{\partial\btheta}(\bx,\btheta_0^*)\mathrm{d}\bx.
\]
Then by assumption we have
\begin{align}
&\bt\transpose{}\sqrt{n}(\btheta_\eta^*-\widehat{\btheta})+\ell_n(\eta)-\ell_n(\eta_0)\nonumber\\
&\quad = \bt\transpose{}\sqrt{n}\left(\btheta_\eta^*-\btheta_0^*\right)-\frac{2}{\sqrt{n}}\sum_{i=1}^ne_i\bt\transpose{}\bV_0^{-1}\frac{\partial y^s}{\partial\btheta}(\bx_i,\btheta_0^*)+o_{\mathbb{P}_0}(1)+\ell_n(\eta)-\ell_n(\eta_0)\nonumber\\
&\quad = \bt\transpose{}\sqrt{n}\left(\btheta_\eta^*-\btheta_0^*\right)+o_{\mathbb{P}_0}(1)-
2\sqrt{n}\int_\Omega[\eta(\bx)-\eta_0(\bx)]\bt\transpose{}\bV_0^{-1}\frac{\partial y^s}{\partial\btheta}(\bx,\btheta_0^*)\mathrm{d}\bx\nonumber\\
&\quad\quad+\frac{1}{2}\bt\transpose{}\left(4\sigma^2\bV_0^{-1}\bW\bV_0^{-1}\right)\bt-\frac{1}{\sigma^2}\left[R_n(\eta_\bt,\eta_0)-R_n(\eta,\eta_0)\right]
+\ell_n(\eta_\bt)-\ell_n(\eta_0)\nonumber\\
&\quad = \frac{1}{2}\bt\transpose{}\left(4\sigma^2\bV_0^{-1}\bW\bV_0^{-1}\right)\bt+\sqrt{n}\bt\transpose{}\br(\eta,\eta_0)+\frac{1}{\sigma^2}\left[R_n(\eta,\eta_0)-R_n(\eta_\bt,\eta_0)\right]\nonumber\\
&\quad\quad+\ell_n(\eta_\bt)-\ell_n(\eta_0)+o_{\mathbb{P}_0}(1)\nonumber.
\end{align}
Now set $M_n =\log n$. By Lemma \ref{lemma:taylor_expansion_functional} we see that
\[
\sup_{\eta\in\calA_n\cap\calC_n}\left|\sqrt{n}\bt\transpose{}\br(\eta,\eta_0)\right|\leq L_{\eta_0}^{(2)}\|t\|\sqrt{n}M_n^2n^{-2\alpha/(2\alpha+p)}\lesssim M_n^2 n^{(p/2-\alpha)/(2\alpha+p)}=o(1).
\]
In addition, simple algebra and the law of large numbers imply that
\begin{align}
&R_n(\eta,\eta_0)-R_n(\eta_\bt,\eta_0)\nonumber\\
&\quad=\frac{2\sigma^4}{n}\sum_{i=1}^n\left[\bt\transpose{}\bV_0^{-1}\frac{\partial y^s}{\partial\btheta}(\bx_i,\btheta_0^*)\right]^2-2\sigma^4\bt\transpose{}\bV_0^{-1}\bW\bV_0^{-1}\bt-2\sigma^2\mathbb{G}_n\left[(\eta-\eta_0)(\cdot)\bt\transpose{}\bV_0^{-1}\frac{\partial y^s}{\partial\btheta}(\cdot,\btheta_0^*)\right]\nonumber\\
&\quad=-2\sigma^2\mathbb{G}_n\left[(\eta-\eta_0)(\cdot)\bt\transpose{}\bV_0^{-1}\frac{\partial y^s}{\partial\btheta}(\cdot,\btheta_0^*)\right]+o_{\mathbb{P}_0}(1)\nonumber.
\end{align}
We now claim that 
\[\sup_{\eta\in\calA_n}|R_n(\eta,\eta_0)-R_n(\eta_\bt,\eta_0)|=o_{\mathbb{P}_0}(1).\] Since $\|\eta-\eta_0\|_{L_2(\Omega)}\leq M_n\eps_n$, $\|\eta-\eta_0\|_{L_\infty(\Omega)}\leq M$ over $\calA_n$, and by Lemma \ref{lemma:sieve_construction} it holds that
\[
J_{[\bdot]}(M_n\eps_n,\calA_n,\|\cdot\|_{L_\infty(\Omega)})\lesssim M_n^{2\alpha/(2\alpha+p)}\sqrt{n}\eps_n^2=(\log n)^{2\alpha/(2\alpha+p)}\sqrt{n}\eps_n^2. 
\]
Following the maximal inequality for empirical process (Theorem \ref{thm:maximum_inequality}), we have 
\begin{align}
&\mathbb{E}_0\left\{\sup_{\eta\in\calA_n}\left|\mathbb{G}_n\left[(\eta-\eta_0)(\cdot) \bt\transpose{}\bV_0^{-1}\frac{\partial y^s}{\partial\btheta}(\cdot,\btheta_0^*)\right]\right|\right\}\nonumber\\
&\quad\lesssim J_{[\bdot]}(M_n\eps_n,\calA_n,\|\cdot\|_{L_2(\Omega)})\left[1+M\frac{J_{[\bdot]}(M_n\eps_n,\calA_n,\|\cdot\|_{L_2(\Omega)})}{M_n^2\eps_n^2\sqrt{n}}\right]\nonumber\\
&\quad\lesssim M_n^{2\alpha/(2\alpha+p)}\sqrt{n}\eps_n^2\left[1+\frac{M_n^{2\alpha/(2\alpha+p)}\sqrt{n}\eps_n^2}{M_n^2\sqrt{n}\eps_n^2}\right]\nonumber\\
&\quad\lesssim M_n\sqrt{n}\eps_n^2=o(1)\nonumber,
\end{align}
and hence, it holds that $\sup_{\eta\in\calA_n\cap\calC_n}|R_n(\eta,\eta_0)-R_n(\eta_\bt,\eta_0)|=o_{\mathbb{P}_0}(1)$. Therefore by applying Lemma \ref{lemma:integral_ratio_stability} we obtain
\begin{align}
&\int_{\calA_n\cap\calC_n}\exp\left[\bt\transpose{}\sqrt{n}(\btheta_\eta^*-\widehat{\btheta})+\ell_n(\eta)-\ell_n(\eta_0)\right]\Pi(\mathrm{d}\eta)\nonumber\\
&\quad=\exp\left[\frac{1}{2}\bt\transpose{}\left(4\sigma^2\bV_0^{-1}\bW\bV_0^{-1}\right)\bt+o_{\mathbb{P}_0}(1)\right]
\int_{\calA_n}\exp\left[\ell_n(\eta_\bt)-\ell_n(\eta_0)\right]\Pi(\mathrm{d}\eta)\nonumber\\
&\quad=\exp\left[\frac{1}{2}\bt\transpose{}\left(4\sigma^2\bV_0^{-1}\bW\bV_0^{-1}\right)\bt+o_{\mathbb{P}_0}(1)\right][1+o_{\mathbb{P}_0}(1)]
\int\exp\left[\ell_n(\eta)-\ell_n(\eta_0)\right]\Pi(\mathrm{d}\eta)\nonumber\\
&\quad=\left\{\exp\left[\frac{1}{2}\bt\transpose{}\left(4\sigma^2\bV_0^{-1}\bW\bV_0^{-1}\right)\bt\right]+o_{\mathbb{P}_0}(1)\right\}\int\exp[\ell_n(\eta)-\ell_n(\eta_0)]\Pi(\mathrm{d}\eta)\nonumber.
\end{align}
The proof is thus completed.

\section*{Supplementary Material}
The supplementary material contains the proofs of Lemma \ref{lemma:taylor_expansion_functional}, Corollary \ref{corr:Bayes_estimator} in Section \ref{sec:asymptotic_properties}, Theorem \ref{thm:convergence_AdaGrad}, Theorem \ref{thm:approximate_PC_convergence} in Section \ref{sec:algorithmic_issue}, Lemma \ref{lemma:sieve_construction}, Lemma \ref{lemma:integral_ratio_stability} in Appendix, and additional numerical results.

\bibliographystyle{apalike}
\bibliography{reference}

\clearpage
\begin{center}
  \begin{Large}
    \textbf{Supplementary Material for ``Bayesian Projected Calibration of Computer Models''}
  \end{Large}
\end{center}
\appendix
\counterwithin{lemma}{section}
\counterwithin{theorem}{section}

\section{Proof of Theorem \ref{thm:matern_process_contraction}} 
\label{sec:proof_of_theorem_thm:matern_process_contraction}

We first present a classic result regarding convergence rate of the Mat\'ern Gaussian process regression from \cite{vaart2011information}. 
\begin{theorem}
Suppose $\eta$ is imposed the Mat\'ern Gaussian process with roughness parameter $\alpha$, and $\eta_0\in\mathfrak{C}_\alpha(\Omega)\cap\calH_\alpha(\Omega)$, where $\alpha>p/2$. Then there exists some constant $C>0$, such that
\begin{align}
\mathbb{E}_0\left\{\int_\Omega\left[\|\eta-\eta_0\|_{L_2(\Omega)}^2\right]\Pi(\mathrm{d}\eta\mid\calD_n)\right\}\leq Cn^{-2\alpha/(2\alpha+p)}.
\end{align}
\end{theorem}
The first assertion follows immediately from the Markov's inequality:
\begin{align}
&\mathbb{E}_0\left[\Pi\left(\|\eta-\eta_0\|_{L_2(\Omega)}>M_nn^{-\alpha/(2\alpha+p)}\mid\calD_n\right)\right]
\nonumber\\ &\quad
\leq\frac{1}{M_n^2 n^{-2\alpha/(2\alpha+p)}}\mathbb{E}_0\left\{\int_\Omega\left[\|\eta-\eta_0\|_{L_2(\Omega)}^2\right]\Pi(\mathrm{d}\eta\mid\calD_n)\right\}
\nonumber\\ &\quad
\leq \frac{C}{M_n^2}\to 0\nonumber.
\end{align}
The posterior distribution of $\eta$ can be expressed by
\[
\Pi(\eta\in\calU\mid\calD_n) = \left[\int_\calU \prod_{i=1}^n\frac{p_\eta(y_i,\bx_i)}{p_0(y_i,\bx_i)}\Pi(\mathrm{d}\eta)\right]\left[\int\prod_{i=1}^n\frac{p_\eta(y_i,\bx_i)}{p_0(y_i,\bx_i)}\Pi(\mathrm{d}\eta)\right]^{-1},
\]
where $p_0(y_i,\bx_i)=\psi_\sigma(y_i-\eta_0(\bx_i))$ is the density of the true distribution. To prove the second assertion, we need the following result from \cite{xie2017theoretical} to bound the denominator of the preceding display. 

\begin{lemma}\label{lemma:evidence_LB}
Assume the conditions of Theorem \ref{thm:matern_process_contraction} hold. For any $D>0$, define the event
\[
\calH_n=\left\{\int\prod_{i=1}^n\frac{p_\eta(y_i,\bx_i)}{p_0(y_i,\bx_i)}\Pi(\mathrm{d}\eta)\geq\Pi(\|\eta-\eta_0\|_{L_\infty(\Omega)}< \eps_n)\exp\left[-\left(D+\frac{1}{\sigma^2}\right)n\eps_n^2\right]\right\}.
\]
Suppose $(\eps_n)_{n=1}^\infty$ is a sequence such that $n\eps_n^2\to \infty$ and $\eps_n\to 0$. Then $\mathbb{P}_0(\calH_n^c)\to 0$. 
\end{lemma}

Since $\alpha>p/2$, there exists some positive $\beta$ such that $\beta\in(\max\{\underline{\alpha},p/2\},\alpha)$. Define $\eps_n = n^{-\beta/(2\beta+p)}$. Since the Mat\'ern Gaussian process assigns prior probability one to the space $\mathfrak{C}_\beta(\Omega)$ (see, for example, section 3.1 in \citealp{vaart2011information}), then the Gaussian process prior on $\eta$ can be regarded as a mean-zero Gaussian random element in the Banach space $\mathfrak{C}_\beta(\Omega)$ equipped with the $\beta$-H\"older norm $\|\cdot\|_{\mathfrak{C}_\beta(\Omega)}$. Therefore by the Borell's inequality (see, for example, \citealp{ghosal2017fundamentals}) it holds that
\begin{align}
\label{eqn:Borell_inequality}
\Pi\left(\|\eta\|_{\mathfrak{C}_\beta(\Omega)}>4x\left[\int\|\eta\|_{\mathfrak{C}_\beta(\Omega)}^2\Pi(\mathrm{d}\eta)\right]^{1/2}\right)\leq 2\mathrm{e}^{-2x^2}.
\end{align}
for any positive $x$. 

By Lemma 15 in \cite{vaart2011information} there exists a constant $\tilde{M}>0$ such that $\|f\|_{L_\infty(\Omega)}\leq\tilde{M} \|f\|_{\mathfrak{C}_\beta(\Omega)}^{p/(2\beta+p)}\|f\|_{L_2(\Omega)}^{2\beta/(2\beta+p)}$ for any function $f\in\mathfrak{C}_\beta(\Omega)$. Let $s>0$ be a constant determined later. Then
\begin{align}
&\left\{\|\eta-\eta_0\|_{L_2(\Omega)}\leq M_n n^{-{\alpha}/{(2\alpha+p)}}\right\}\cap\left\{\|\eta\|_{\mathfrak{C}_\beta(\Omega)}\leq 4s\sqrt{n}\eps_n\left[\int \|\eta\|_{\mathfrak{C}_\beta(\Omega)}^2\Pi(\mathrm{d}\eta)\right]^{1/2}\right\}\nonumber\\
&\quad\subset
\left\{\|\eta-\eta_0\|_{L_\infty(\Omega)}\leq \tilde{M}\|\eta-\eta_0\|_{\mathfrak{C}_\beta(\Omega)}^{p/(2\beta+p)}M_n^{{2\beta}/{(2\beta+p)}}n^{-{(2\alpha\beta)}/{[(2\alpha+p)(2\beta+p)]}}
\right\}\nonumber\\
&\qquad\quad\cap\left\{
\|\eta\|_{\mathfrak{C}_\beta(\Omega)}\leq 4s\sqrt{n}\eps_n\left[\int \|\eta\|_{\mathfrak{C}_\beta(\Omega)}^2\Pi(\mathrm{d}\eta)\right]^{1/2}\right\}\nonumber\\
&\quad\subset
\left\{\|\eta-\eta_0\|_{L_\infty(\Omega)}\leq\tilde{M}\left(\|\eta\|_{\mathfrak{C}_\beta(\Omega)}+\|\eta_0\|_{\mathfrak{C}_\beta(\Omega)}\right)^{{p}/{(2\beta+p)}} M_n^{{2\beta}/{(2\beta+p)}}n^{-{2\alpha\beta}/{[(2\alpha+p)(2\beta+p)]}}
\right\}\nonumber\\
&\qquad\quad\cap\left\{
\|\eta\|_{\mathfrak{C}_\beta(\Omega)}\leq 4s\sqrt{n}\eps_n\left[\int \|\eta\|_{\mathfrak{C}_\beta(\Omega)}^2\Pi(\mathrm{d}\eta)\right]^{1/2}\right\}\nonumber\\
&\quad\subset\left\{\|\eta-\eta_0\|_{L_\infty(\Omega)}\leq M_1 M_n^{{2\beta}/{(2\beta+p)}} n^{{-2\alpha\beta}/{[(2\alpha+p)(2\beta+p)]}}n^{{p^2}/{[2(2\beta+p)^2]}}\right\}\nonumber
\end{align}
for some constant $M_1>0$ depending on $\eta_0$ only when $n$ is sufficiently large. Note that $-\alpha/(2\alpha+p)<-\beta/(2\beta+p)$, then taking $M_n = \log n$ yields
\begin{align}
&\left\{\|\eta-\eta_0\|_{L_\infty(\Omega)}\leq M_1 M_n^{{2\beta}/{(2\beta+p)}} n^{{-2\alpha\beta}/{[(2\alpha+p)(2\beta+p)]}}n^{{p^2}/{[2(2\beta+p)^2]}}\right\}\nonumber\\
&\quad\subset
\left\{\|\eta-\eta_0\|_{L_\infty(\Omega)}\leq M_1 (\log n)^{2\beta/(2\beta+p)}n^{{-(2\beta^2-p^2/2)}/{(2\beta+p)^2}}\right\}\nonumber\\
&\quad\subset\left\{\|\eta-\eta_0\|_{L_\infty(\Omega)}\leq M\right\}\nonumber
\end{align}
for some constant $M>0$, where $\beta>p/2$ is applied. Since by the first assertion $\Pi(\|\eta-\eta_0\|_{L_2(\Omega)}\leq M_n n^{-\alpha/(2\alpha+p)}\mid\calD_n)=1-o_{\mathbb{P}_0}(1)$, it suffices to show that $\mathbb{E}_0\left[\Pi(\calU_n\mid\calD_n)\right]\to0$, where $\calU_n$ is the event
\[
\calU_n = \left\{\|\eta\|_{\mathfrak{C}_\beta(\Omega)}> 4s\sqrt{n}\eps_n\left[\int\|\eta\|_{\mathfrak{C}_\beta(\Omega)}^2\Pi(\mathrm{d}\eta)\right]^{1/2}\right\}.
\]
The following argument is quite similar to that of Lemma 1 in \cite{ghosal2007convergence} and is included here for completeness. 
Let $\calH_n$ be defined as in Lemma \ref{lemma:evidence_LB} with the constant $D$ be determined later. Then $\mathbb{P}_0(\calH_n^c)\to0$, and we directly compute by Fubini's theorem
\begin{align}
\mathbb{E}_0\left[\Pi\left(\calU_n\mid\calD_n\right)\right]
\end{align}
By Lemma 3 and Lemma 4 in \cite{vaart2011information} we have
\[
\Pi(\|\eta-\eta_0\|_{L_\infty(\Omega)}\leq \eps_n)\geq\exp\left(-C\eps_n^{-p/\alpha}\right)\geq\exp\left(-Cn^{p\beta/[\alpha(2\beta+p)]}\right)
\]
for some constant $C>0$. Now take $D=1/(2\sigma^2)$, $s = 1/\sigma$, and we conclude
\begin{align}
\mathbb{E}_0\{\Pi(\calU_n\mid\calD_n)\}
&\leq \exp\left(\frac{3}{2\sigma^2}n\eps_n^2+Cn^{p\beta/[\alpha(2\beta+p)]}\right)\Pi(\calU_n)+o(1)\nonumber\\
&\leq 2\exp\left(\frac{3}{2\sigma^2}n\eps_n^2+Cn^{(p\beta/[\alpha(2\beta+p)]}-\frac{2}{\sigma^2}n\eps_n^2\right)+o(1)\to 0\nonumber,
\end{align}
where the last inequality is due to \eqref{eqn:Borell_inequality} and the fact $\beta<\alpha$.


\section{Proof of Lemma \ref{lemma:taylor_expansion_functional}} 
 \label{sec:proof_of_lemma_lemma:taylor_expansion_functional}
 We first prove the first assertion, \emph{i.e.}, the Taylor's expansion of $\btheta_\eta^*$ with respect to $\eta$. 
 Recall that
 $\btheta_\eta^*=\argmin_{\btheta\in\Theta}\|\eta(\cdot)-y^s(\cdot,\btheta)\|^2_{L_2(\Omega)}$. Since by condition A4 it is permitted to interchange the differentiation with respect to $\btheta$ and the integral, it follows that
 \begin{align}
 \zero=\left.\frac{\partial}{\partial\btheta}\|\eta(\cdot)-y^s(\cdot,\btheta)\|^2_{L_2(\Omega)}\right|_{\btheta=\btheta_\eta^*}=-2\int_\Omega\left[\eta(\bx)-y^s(\bx,\btheta_\eta^*)\right]\frac{\partial y^s}{\partial\btheta}(\bx,\btheta_\eta^*)\mathrm{d}\bx.\nonumber
 \end{align}
 Now define the function $\bF:\calF\times \Theta\to\mathbb{R}^q$ by 
 \begin{align}
 \bF(\eta,\btheta)=-2\int_{\Omega}[\eta(\bx)-y^s(\bx,\btheta)]\frac{\partial y^s}{\partial\btheta}(\bx,\btheta)\mathrm{d}\bx.\nonumber
 \end{align}
 It follows immediately that $\bF(\eta,\btheta_\eta^*)=\zero$. The partial derivative of $F$ with respect to $\btheta$ is given by
 \[
 \bF_\btheta(\eta,\btheta):=\frac{\partial}{\partial\btheta}\bF(\eta,\btheta)=\int_\Omega\left\{\frac{\partial^2}{\partial\btheta\partial\btheta\transpose{}}[\eta(\bx)-y^s(\bx,\btheta)]^2\right\}\mathrm{d}\bx,
 \]
 and the partial Fr\'echet derivative of $\bF$ with respect to $\eta$ is a function $\bF_\eta:\calF\to\mathbb{R}^q$ given by 
 \[
 [\bF_\eta(\eta,\btheta)](h)=-2\int_{\Omega}h(\bx)\frac{\partial y^s}{\partial\btheta}(\bx,\btheta)\mathrm{d}\bx,
 \]
 since $F$ is linear with respect to $\eta$. Therefore by the implicit function theorem on Banach space, there exists some $\eps>0$ such that over $\{\eta\in\calF:\|\eta-\eta_0\|_{L_2(\Omega)}<\eps\}$, the functional $\btheta_\eta^*:\eta\mapsto \argmin_{\btheta\in\Theta}\|\eta(\cdot)-y^s(\cdot,\btheta)\|_{L_2(\Omega)}^2$ is continuous, the Fr\'echet derivative $\dot{\btheta}_\eta^*:\calF\to\mathbb{R}^q$ for $\btheta^*_\eta$ exists, and can be computed by
 \begin{align}
 \dot{\btheta}_\eta^*(h)=-\left[\bF_\theta(\eta,\btheta_\eta^*)\right]^{-1}\left[\bF_\eta(\eta,\btheta_\eta^*)\right](h)
 =2\bV_\eta^{-1}\int_\Omega h(\bx)\frac{\partial y^s}{\partial\btheta}(x,\btheta_\eta^*)\mathrm{d}\bx.\nonumber
 \end{align}
 Therefore we obtain by the fundamental theorem of calculus and the mean-value theorem that
 \begin{align}
 \btheta_\eta^*-\btheta_0^*&=\int_0^1\frac{\mathrm{d}}{\mathrm{d}u}\btheta_{\eta[u]}^*\mathrm{d}u\nonumber\\
 &=\int_0^1 \dot\btheta_{\eta[u]}^*\left(\frac{\mathrm{d}}{\mathrm{d}u}\eta[u]\right)\mathrm{d}u\nonumber\\
 &=2\int_0^1 \bV_{\eta[u]}^{-1}\int_\Omega[\eta(\bx)-\eta_0(\bx)]\frac{\partial y^s}{\partial\btheta}(\bx,\btheta_{\eta[u]}^*)\mathrm{d}\bx\mathrm{d}u\nonumber\\
 &=2\int_\Omega [\eta(\bx)-\eta_0(\bx)] \bV_{\eta[u']}^{-1}\frac{\partial y^s}{\partial\btheta}(\bx,\btheta_{\eta[u']}^*)\mathrm{d}\bx\nonumber,
 \end{align}
 where $\eta[u]=\eta_0+(\eta-\eta_0)u$ for $0\leq u\leq 1$ and $u'\in[0,1]$. By condition A3, we know that the smallest eigenvalue $\lambda_{\min}(\bV_\eta)$ of $\bV_\eta$ is strictly positive in an $L_2$-neighborhood of $\eta_0$, and we can without loss of generality require that 
 $\inf_{\|\eta-\eta_0\|_{L_2(\Omega)}\leq\eps}\lambda_{\min}(\bV_\eta)>0.$
 Hence we proceed by condition A4 and Jensen's inequality that
 \begin{align}
 \|\btheta_\eta^*-\btheta_0^*\|&
 \leq 2\sup_{\|\eta-\eta_0\|_{L_2(\Omega)}\leq \eps}\left\|\bV_\eta^{-1}\right\|\sup_{(\bx,\btheta)\in\Omega\times\Theta}\left\|\frac{\partial y^s}{\partial\btheta}(\bx,\btheta)\right\|\int_\Omega|\eta(\bx)-\eta_0(\bx)|\mathrm{d}\bx\nonumber\\
 &\leq 2\left[\inf_{\|\eta-\eta_0\|_{L_2(\Omega)}\leq \eps}\lambda_{\min}(\bV_\eta)\right]^{-1}\sup_{(\bx,\btheta)\in\Omega\times\Theta}\left\|\frac{\partial y^s}{\partial\btheta}(\bx,\btheta)\right\|\left\{\int_\Omega[\eta(\bx)-\eta_0(\bx)]^2\mathrm{d}\bx\right\}^{1/2}\nonumber\\
 &= L_{\eta_0}^{(1)}\|\eta-\eta_0\|_{L_2(\Omega)}\nonumber
 \end{align}
 for some constant $L_{\eta_0}^{(1)}>0$ depending on $\eta_0$ only.

 We now analyze the property of $\bV_\eta$ as a functional $\{\eta\in\calF:\|\eta-\eta_0\|_{L_2(\Omega)}<\eps\}\to \in\mathbb{R}^{q\times q}$, $\eta\mapsto \bV_\eta$. For a matrix $\bA\in\mathbb{R}^{q\times q}$, denote $[\bA]_{ij}$ to be the $(i,j)$-th element of $\bA$. Directly compute
 \begin{align}
 [\bV_\eta]_{jk}-[\bV_0]_{jk}&=2\int_\Omega\left[\frac{\partial y^s}{\partial \theta_j}(\bx,\btheta_\eta^*)\frac{\partial y^s}{\partial \theta_k}(\bx,\btheta_\eta^*)-\frac{\partial y^s}{\partial \theta_j}(\bx,\btheta_0^*)\frac{\partial y^s}{\partial \theta_k}(\bx,\btheta_0^*)\right]\mathrm{d}\bx\nonumber\\
 &\quad-2\int_\Omega\left\{[\eta(\bx)-y^s(\bx,\btheta_\eta^*)]\left[\frac{\partial^2y^s}{\partial\theta_j\partial\theta_k}(\bx,\btheta_\eta^*)-\frac{\partial^2y^s}{\partial\theta_j\partial\theta_k}(\bx,\btheta_0^*)\right]\right\}\mathrm{d}\bx\nonumber\\
 &\quad-2\int_\Omega\left\{[\eta(\bx)-\eta_0(\bx)+y^s(\bx,\btheta_0^*)-y^s(\bx,\btheta_\eta^*)]\frac{\partial^2y^s}{\partial\theta_j\partial\theta_k}(\bx,\btheta_0^*)\right\}\mathrm{d}\bx\nonumber\\
 &:=2V_1-2V_2-2V_3.\nonumber
 \end{align}
 For $V_1$, by condition A4 we know that $\partial y^s/\partial\btheta$ is Lipschitz continuous on $\Omega\times\Theta$, and therefore
 \begin{align}
 |V_1|
 &\leq\int_\Omega\left|\frac{\partial y^s}{\partial \theta_j}(\bx,\btheta_\eta^*)\right|\left|\frac{\partial y^s}{\partial \theta_k}(\bx,\btheta_\eta^*)-\frac{\partial y^s}{\partial \theta_k}(\bx,\btheta_0^*)\right|\mathrm{d}\bx\nonumber\\
 &\quad+\int_\Omega\left|\frac{\partial y^s}{\partial \theta_j}(\bx,\btheta_\eta^*)-\frac{\partial y^s}{\partial \theta_j}(\bx,\btheta_0^*)\right|\left|\frac{\partial y^s}{\partial \theta_k}(\bx,\btheta_0^*)\right|\mathrm{d}\bx\nonumber\\
 &\leq\sup_{(\bx,\btheta)\in\Omega\times\Theta}\left\|\frac{\partial y^s}{\partial \btheta}(\bx,\btheta)\right\|\left[\left\|\frac{\partial y^s}{\partial \theta_k}(\cdot,\btheta_\eta^*)-\frac{\partial y^s}{\partial \theta_k}(\cdot,\btheta_0^*)\right\|_{L_1(\Omega)}+\left\|\frac{\partial y^s}{\partial \theta_j}(\cdot,\btheta_\eta^*)-\frac{\partial y^s}{\partial \theta_j}(\cdot,\btheta_0^*)\right\|_{L_1(\Omega)}\right]\nonumber\\
 &\leq 2\sup_{(\bx,\btheta)\in\Omega\times\Theta}\left\|\frac{\partial y^s}{\partial \btheta}(\bx,\btheta)\right\|\sup_{(\bx,\btheta)\in\Omega\times\Theta}\left\|\frac{\partial^2 y^s}{\partial \btheta\partial\btheta\transpose{}}(\bx,\btheta)\right\|\|\btheta^*_\eta-\btheta_0^*\|\nonumber\\
 &\lesssim \|\eta-\eta_0\|_{L_2(\Omega)}\nonumber.
 \end{align}
 Condition A4 also implies that $\partial^2y^s/(\partial\theta_j\partial\theta_k)$ is Lipschitz continuous on $\Omega\times\Theta$. Hence
 \begin{align}
 |V_2|&\lesssim \int_\Omega[|\eta(\bx)-\eta_0(\bx)|+|\eta_0(\bx)-y^s(\bx,\btheta_\eta^*)]\|\btheta_\eta^*-\btheta_0^*\|\mathrm{d}\bx\nonumber\\
 &\leq L_{\eta_0}^{(1)}\|\eta-\eta_0\|_{L_2(\Omega)}\left\{2\int_\Omega[\eta(\bx)-\eta_0(\bx)]^2\mathrm{d}\bx+2\int_\Omega[\eta_0(\bx)-y^s(\bx,\btheta_\eta^*)]^2\mathrm{d}\bx\right\}^{1/2}\nonumber\\
 &\leq L_{\eta_0}^{(1)}\|\eta-\eta_0\|_{L_2(\Omega)}\left(2\eps^2+4\|\eta_0\|_{L_2(\Omega)}^2+4\sup_{\btheta\in \Theta}\|y^s(\cdot,\btheta)\|_{L_2(\Omega)}^2\right)^{1/2}\nonumber\\
 &\lesssim L_{\eta_0}^{(1)}\|\eta-\eta_0\|_{L_2(\Omega)}\nonumber.
 \end{align}
 Now we consider $V_3$:
 \begin{align}
 |V_3|&\leq \sup_{(\bx,\btheta)\in\Omega\times\Theta}\left|\frac{\partial^2y^s}{\partial\theta_j\partial\theta_k}\right|\int_\Omega[|\eta(\bx)-\eta_0(\bx)|+|y^s(\bx,\btheta_0)-y^s(\bx,\btheta_\eta^*)|]\mathrm{d}\bx\nonumber\\
 &\leq \sup_{(\bx,\btheta)\in\Omega\times\Theta}\left|\frac{\partial^2y^s}{\partial\theta_j\partial\theta_k}\right|\left[\|\eta-\eta_0\|_{L_2(\Omega)}+\sup_{(\bx,\btheta)\in\Omega\times\Theta}\left\|\frac{\partial y^s}{\partial\btheta}(\bx,\btheta)\right\|\|\btheta_\eta^*-\btheta_0^*\|\mathrm{d}x\right]\nonumber\\
 &\leq \sup_{(\bx,\btheta)\in\Omega\times\Theta}\left|\frac{\partial^2y^s}{\partial\theta_j\partial\theta_k}\right|\left[1+\sup_{(\bx,\btheta)\in\Omega\times\Theta}\left\|\frac{\partial y^s}{\partial\btheta}(\bx,\btheta)\right\|L_{\eta_0}^{(1)}\right]\|\eta-\eta_0\|_{L_2(\Omega)}\nonumber.
 \end{align}
 We conclude that $|[\bV_\eta]_{jk}-[\bV_0]_{jk}|\leq C_{\eta_0} \|\eta-\eta_0\|_{L_2(\Omega)}$ for all $j,k=1,\ldots,q$ for some constant $C_{\eta_0}>0$ depending on $\eta_0$ only. By the fact that
 \[
 \sum_{j=1}^q|\lambda_j(\bA)-\lambda_j(\bB)|\leq \|\bA-\bB\|_F^2
 \]
 holds for any positive definite matrices $\bA,\bB\in\mathbb{R}^{q\times q}$ (see, for example, \citealp{hoffman2003variation}), we obtain
 \begin{align}
 |\lambda_{\min}(\bV_\eta)-\lambda_{\min}(\bV_0)|\leq \|\bV_\eta-\bV_0\|_F^2=\sum_{j=1}^q\sum_{k=1}^q|[\bV_\eta]_{jk}-[\bV_0]_{jk}|^2\leq q^2C_{\eta_0}^2\|\eta-\eta_0\|_{L_2(\Omega)}^2\nonumber.
 \end{align}
 We may also assume without loss of generality that $\eps$ is sufficiently small such that $|\lambda_{\min}(\bV_\eta)-\lambda_{\min}(\bV_0)|\leq \lambda_{\min}(\bV_0)/2$ whenever $\|\eta-\eta_0\|_{L_2(\Omega)}\leq\eps$, in which case it holds that $\|\bV_\eta^{-1}\|\geq 2\|\bV_0^{-1}\|$. Hence
 \begin{align}
 \left\|\bV_\eta^{-1}-\bV_0^{-1}\right\|&=\left\|\bV_0^{-1}(\bV_0-\bV_\eta)\bV_\eta^{-1}\right\|\nonumber\\
 &\leq \left\|\bV_0^{-1}\|\|\bV_0-\bV_\eta\|\|\bV_\eta^{-1}\right\|\nonumber\\
 &\leq 2\left\|\bV_0^{-1}\right\|^2\left\|\bV_\eta-\bV_0\right\|_F\nonumber\\
 &\leq 2qC_{\eta_0}\left\|\bV_0^{-1}\right\|\left\|\eta-\eta_0\right\|_{L_2(\Omega)}\nonumber
 \end{align}
 whenever $\|\eta-\eta_0\|_{L_2(\Omega)}<\eps$. Hence
 \begin{align}
 \br(\eta,\eta_0)&=\btheta_\eta^*-\btheta_0^*-2\int_\Omega [\eta(\bx)-\eta_0(\bx)]\bV_0^{-1}\frac{\partial y^s}{\partial\btheta}(\bx,\btheta_0^*)\mathrm{d}\bx\nonumber\\
 &=2\int_\Omega [\eta(\bx)-\eta_0(\bx)]\left[\bV_{\eta[u']}^{-1}\frac{\partial y^s}{\partial\btheta}(\bx,\btheta_{\eta[u']}^*) - \bV_0^{-1}\frac{\partial y^s}{\partial\btheta}(\bx,\btheta_0^*)\right]\mathrm{d}\bx\nonumber\\
 &=2\int_\Omega [\eta(\bx)-\eta_0(\bx)]\left[(\bV_{\eta[u']}^{-1} - \bV_0^{-1})\frac{\partial y^s}{\partial\btheta}(\bx,\btheta_{\eta[u']}^*)\right]\mathrm{d}\bx\nonumber\\
 &\quad + 2\int_\Omega [\eta(\bx)-\eta_0(\bx)]\bV_0^{-1}\left[\frac{\partial y^s}{\partial\btheta}(\bx,\btheta_{\eta[u']}^*)-\frac{\partial y^s}{\partial\btheta}(\bx,\btheta_0^*)\right]\mathrm{d}\bx\nonumber,
 \end{align}
 and hence,
 \begin{align}
 \|\br(\eta,\eta_0)\|
 &\leq 2\int_\Omega|\eta(\bx)-\eta_0(\bx)|\left[\left\|\bV_{\eta[u']}^{-1} - \bV_0^{-1}\right\|\sup_{(\bx,\btheta)\in\Omega\times\Theta}\left\|\frac{\partial y^s}{\partial\btheta}(\bx,\btheta)\right\|\right]\mathrm{d}\bx\nonumber\\
 &\quad + 2\int_\Omega|\eta(\bx)-\eta_0(\bx)|\left[\|\bV_0^{-1}\|\left\|\frac{\partial y^s}{\partial\btheta}(\bx,\btheta_{\eta[u']}^*)-\frac{\partial y^s}{\partial\btheta}(\bx,\btheta_0^*)\right\|\right]\mathrm{d}\bx\nonumber\\
 &\lesssim \|\eta-\eta_0\|_{L_2(\Omega)}q^2C_{\eta_0}^2\|\bV_0^{-1}\|\|\eta-\eta_0\|_{L_2(\Omega)}+\|\bV_0^{-1}\|\|\eta-\eta_0\|_{L_2(\Omega)}^2,\nonumber
 \end{align}
 implying that $\|\br(\eta,\eta_0)\|\leq L_{\eta_0}^{(2)}\|\eta-\eta_0\|_{L_2(\Omega)}^2$ for some constant $L_{\eta_0}^{(2)}$ depending on $\eta_0$ only. This completes the proof of the first assertion. 

 To prove the second assertion, note that if A1 and A3 hold for all $\eta$ in an $L_2$-neighborhood $\calU$ of $\eta_0$, then for any $\eta_1\in \calU$, A1 and A3 hold for all $\eta$ in an $L_2$-neighborhood of $\eta_1$ inside $\calU$. Therefore, the first assertion can be applied to $\eta_1$: For all $\eta_1\in\calU$, $\btheta_\eta^*$ is a continuous functional of $\eta$ at $\eta = \eta_1$. Namely, $\btheta_\eta^*$ is a continuous functional of $\eta\in\calU$. Therefore, $\calM(\calU) = \{(\eta,\btheta_\eta^*):\eta\in\calU\}$ becomes the graph of this continuous functional. It follows directly that the maps $T_1:\calM(\calU)\to\calU:(\eta, \btheta_\eta^*) \mapsto \eta$ and $T_2:\calU\to\calM(\calU):\eta\mapsto (\eta, \btheta_\eta^*)$ are continuous and invertible to each other. Therefore, the transition map $T_2\circ T_1^{-1}$ is the identity on $\calU$, showing that $\calM(\calU)$ is a Banach manifold. 
 

\section{Proof of Lemma \ref{lemma:sieve_construction}} 
\label{sec:proof_of_lemma_lemma:sieve_construction}
Before proceeding, we introduce the notion of \emph{covering number} for a metric space $(\mathfrak{X},d)$. The $\epsilon$-covering number of $(\mathfrak{X},d)$ for $\eps>0$, is the smallest number of $\eps$-balls (with respect to the metric $d$) that are needed to cover $\mathfrak{X}$.

Since $\eta$ is imposed the Mat\'ern Gaussian process with roughness parameter $\alpha$, it follows that the concentration function
\[
\varphi_{\eta_0}(\eps)=\inf_{\eta_1\in\mathbb{H}_{\Psi_{\alpha}}(\Omega):\|\eta_1-\eta_0\|_{L_\infty(\Omega)}\leq\eps}\frac{1}{2}\|\eta_1\|_{\mathbb{H}_{\Psi_\alpha}(\Omega)}^2-\log\Pi(\|\eta\|_{L_\infty(\Omega)}<\eps)
\]
satisfies $\varphi_{\eta_0}(\eps)\leq C\eps^{-p/\alpha}$ for some constant $C>0$ for sufficiently small $\eps>0$. Then by Theorem 2.1 in \cite{van2008rates}, it holds that
\begin{align}
\label{eqn:prior_concentration_condition}
\Pi(\|\eta-\eta_0\|_{L_\infty(\Omega)}<\eps_n)&\geq\exp(-C^2n\eps_n^2),
\end{align}
where $\eps_n = n^{-\alpha/(2\alpha+p)}$. 
Pick $\beta>0$ such that $\beta\in(\max\{\underline{\alpha},p/2\},\alpha)$. Then we know that the Mat\'ern Gaussian process $\mathrm{GP}(0,\Psi_\alpha)$ assigns prior probability one to $\mathfrak{C}_\beta(\Omega)$. 
Now set $\calB_n = \eps_n\mathfrak{C}_\beta^1(\Omega)+m_n\mathbb{H}_{\Psi_\alpha}^1(\Omega)$, where 
\[\mathfrak{C}_\beta^1(\Omega)=\left\{f\in\mathfrak{C}_\beta(\Omega):\|f\|_{\mathfrak{C}_\beta(\Omega)}\leq 1\right\},\quad\mathbb{H}_{\Psi_\alpha}^1(\Omega)=\left\{f\in\mathbb{H}_{\Psi_\alpha}(\Omega):\|f\|_{\mathbb{H}_{\Psi_\alpha}(\Omega)}\leq 1\right\},\]
$m_n$ is some sequence determined later, and $\mathbb{H}_{\Psi_\alpha}(\Omega)$ is the reproducing kernel Hilbert space (abbreviated as RKHS) associated with the Mat\'ern covariance function $\Psi_\alpha$. Denote $\Phi$ to be the cumulative distribution function of the standard normal distribution and set $m_n = -2\Phi^{-1}(\exp[-(2C+1/\sigma^2)n\eps_n^2])$. Since $\eta\sim\mathrm{GP}(0,\Psi_\alpha)$ can be viewed as a Gaussian random element in the Banach space $\mathfrak{C}_\beta(\Omega)$ with the norm $\|\cdot\|_{\mathfrak{C}_\beta(\Omega)}$, then by the Borell's inequality \citep{van2008rates} we have
\begin{align}
\Pi(\calB_n)&\geq 
\Phi\left(\Phi^{-1}\left(\exp\left(-Cn\eps_n^2\right)\right)+m_n\right)\nonumber\\
&= 
\Phi\left(\Phi^{-1}\left(\exp\left(-Cn\eps_n^2\right)\right)-2\Phi^{-1}\left(\exp\left[-\left(2C+\frac{1}{\sigma^2}\right)n\eps_n^2\right]\right)\right)
\nonumber\\&
\geq\Phi\left(-\Phi^{-1}\left(\exp\left[-\left(2C+\frac{1}{\sigma^2}\right)n\eps_n^2\right]\right)\right)
\nonumber\\
&=1-\exp\left[-\left(2C+\frac{1}{\sigma^2}\right)n\eps_n^2\right]\nonumber.
\end{align}
Hence
\begin{align}
\label{eqn:prior_mass_condition}
\Pi(\eta\in\calB_n^c)&\leq\exp\left[-\left(2C+\frac{1}{\sigma^2}\right)n\eps_n^2\right].
\end{align}
Now we prove the first inequality using \eqref{eqn:prior_concentration_condition} and \eqref{eqn:prior_mass_condition}. Let $\calH_n$ be defined as in Lemma \ref{lemma:evidence_LB}. Denote $M_n = \log n$. Then
\begin{align}
\mathbb{E}_0[\Pi(\calB_n^c\mid\calD_n)]&\leq \mathbb{E}_0[\mathbbm{1}(\calH_n)\Pi(\calB_n^c\mid\calD_n)]+\mathbb{P}_0(\calH_n^c)\nonumber\\
&=\mathbb{E}_0\left\{\mathbbm{1}(\calH_n)\left[\int\prod_{i=1}^n\frac{p_\eta(y_i,\bx_i)}{p_0(y_i,\bx_i)}\Pi(\mathrm{d}\eta)\right]^{-1}\left[\int_{\calB_n^c}\prod_{i=1}^n\frac{p_\eta(y_i,\bx_i)}{p_0(y_i,\bx_i)}\Pi(\mathrm{d}\eta)\right]\right\}+o(1)\nonumber\\
&\leq \frac{\exp[(D+\sigma^{-2})n\eps_n^2]}{\Pi(\|\eta-\eta_0\|_{L_\infty(\Omega)}<\eps_n)}\Pi(\calB_n^c)+o(1)\nonumber\\
&\leq \exp\left[\left(D+\frac{1}{\sigma^2}\right)n\eps_n^2+Cn\eps_n^2-\left(2C+\frac{1}{\sigma^2}\right)n\eps_n^2\right]+o(1)\nonumber\\
&\leq \exp\left[\left(D-C\right)n\eps_n^2\right]+o(1)\nonumber.
\end{align}
Hence taking $D=C/2$ yields $\mathbb{E}_0[\Pi(\calB_n^c\mid\calD_n)]\to 0$. 

Finally we prove the second inequality involving the bracketing integral. Since $\mathbb{H}_{\Psi_\alpha}(\Omega)$ is the RKHS of the Mat\'ern covariance function with roughness parameter $\alpha$, then $\mathbb{H}_{\Psi_\alpha}(\Omega)$ coincides with the Sobolev space $\calH_{\alpha+p/2}(\Omega)$ (see, for example, Corollary 1 of \citealp{tuo2016theoretical}). The logarithm of the covering number of $\rho\mathbb{H}^1_{\Psi_\alpha}(\Omega)$ is bounded by \citep{edmunds2008function}
\[
\log\calN\left(\eps, \rho\mathbb{H}^1_{\Psi_\alpha}(\Omega), \|\cdot\|_{L_\infty(\Omega)}\right)\lesssim \left(\frac{\rho}{\eps}\right)^{2p/(2\alpha+p)}
\]
for sufficiently small $\eps>0$. The metric entropy for the $\alpha$-H\"older space $\eps_n\mathfrak{C}_\alpha^1(\Omega)$ is also known in the literature (see, for example, \citealp{van1996weak}):
\[
\log\calN\left(\eps, \eps_n\mathfrak{C}_\beta^1(\Omega), \|\cdot\|_{L_\infty(\Omega)}\right)\lesssim\left(\frac{\eps_n}{\eps}\right)^{p/\beta}.
\]
Hence for sufficiently small $\eps>0$,
\[
\log\calN(\eps, \calB_n, \|\cdot\|_{L_\infty(\Omega)})
\lesssim \left(\frac{m_n}{\eps}\right)^{2p/(2\alpha+p)}+\left(\frac{\eps_n}{\eps}\right)^{p/\beta},
\]
and it follows by simple algebra that
\begin{align}
J_{[\bdot]}(M_n\eps_n,\calB_n,\|\cdot\|_{L_2(\Omega)})&\lesssim \int_0^{M_n\eps_n}\sqrt{\log\calN\left(\eps,\calB_n,\|\cdot\|_{L_\infty(\Omega)}\right)}\mathrm{d}\eps\nonumber\\
&\lesssim m_n^{p/(2\alpha+p)}\left(M_n\eps_n\right)^{2\alpha/(2\alpha+p)}+\eps_n^{p/2\beta}(M_n\eps_n)^{(2\beta-p)/(2\beta)}\nonumber\\
&\asymp M_n^{2\alpha/(2\alpha+p)}\sqrt{n}\eps_n^2+M_n^{(2\beta-p)/(2\beta)}\eps_n\nonumber\\
&\lesssim M_n^{2\alpha/(2\alpha+p)}\sqrt{n}\eps_n^2\nonumber
\end{align}
for sufficiently large $n$.


\section{Proof of Lemma \ref{lemma:integral_ratio_stability}} 
\label{sec:proof_of_lemma_lemma:integral_ratio_stability}

Before proceeding, we establish the following fact: if $(\calW_n)_{n=1}^\infty$ is a sequence of event such that $\Pi(\calW_n\mid\calD_n)=o_{\mathbb{P}_0}(1)$, then 
\begin{align}
\int_{\calW_n}\exp(\ell_n(\eta)-\ell_n(\eta_0))\Pi(\mathrm{d}\eta)
&=\Pi(\calW_n\mid\calD_n)\int\exp(\ell_n(\eta)-\ell_n(\eta_0))\Pi(\mathrm{d}\eta)\nonumber\\
&=o_{\mathbb{P}_0}(D_n),
\end{align}
where
\[
D_n:=\int\exp(\ell_n(\eta)-\ell_n(\eta_0))\Pi(\mathrm{d}\eta).
\]

Recall that the RKHS $\mathbb{H}_{\Psi_\alpha}(\Omega)$ of the Mat\'ern Gaussian process with roughness parameter $\alpha>p/2$ coincides with the Sobolev space $\calH_{\alpha+p/2}(\Omega)$ \citep{wendland2004scattered,tuo2016theoretical}, and the RKHS norm $\|\cdot\|_{\mathbb{H}_{\Psi_\alpha}(\Omega)}$ is equivalent to the Sobolev norm $\|\cdot\|_{\calH_{\alpha+p/2}(\Omega)}$. 
Recall the definition of the isometry $U$. Then under the prior distribution $\Pi$, for any $h\in\mathbb{H}_{\Psi_\alpha}(\Omega)$, $U(h)\sim\mathrm{N}\left(0,\|h\|_{\mathbb{H}_{\Psi_\alpha}(\Omega)}^2\right)$. Hence by Lemma 17 in \cite{castillo2012semiparametric}, for any measurable function $T:\mathfrak{C}(\Omega)\to\mathbb{R}$, any $g,h\in\mathbb{H}_{\Psi_\alpha}(\Omega)$, and any $\rho>0$,
\begin{align}
\label{eqn:change_of_measure}
&\mathbb{E}_\Pi\left[\mathbbm{1}\{|U(g)|\leq\rho\}T(\eta-h)\right]\nonumber\\
&\quad=\mathbb{E}_\Pi\left\{\mathbbm{1}\left[\left|U(g)+\langle g,h\rangle_{\mathbb{H}_{\Psi_\alpha}(\Omega)}\right|\leq\rho\right]T(\eta)\exp\left[U(-h)-\frac{1}{2}\|h\|_{\mathbb{H}_{\Psi_\alpha}(\Omega)}^2\right]\right\}.
\end{align}
Let $\eps_n = n^{-\alpha/(2\alpha+p)}$. Denote $\calA_{1n}=\{\|\eta-\eta_0\|_{L_2(\Omega)}\leq M_n\eps_n\}$, $\calA_{2n}=\{\|\eta-\eta_0\|_{L_\infty(\Omega)}\leq M\}$, and take 
\begin{align}
g(\bx)&=2\sigma^2\bt\transpose{}\bV_0^{-1}\frac{\partial y^s}{\partial\btheta}(\bx,\btheta_0^*),\quad h(\bx)=\frac{2\sigma^2}{\sqrt{n}}\bt\transpose{}\bV_0^{-1}\frac{\partial y^s}{\partial\btheta}(\bx,\btheta_0^*).
\nonumber
\end{align}
Since $U(g/\|g\|_{\mathbb{H}_{\Psi_\alpha}(\Omega)})$ follows the standard normal distribution under the prior, it follows that for sufficiently large $L$, 
\[
\Pi(\calC_n^c)=\Pi\left\{\left|U\left(\frac{g}{\|g\|_{\mathbb{H}_{\Psi_\alpha}(\Omega)}}\right)\right|>L\sqrt{n}\eps_n\right\}\leq 2\exp\left(-\frac{L}{2}n\eps_n^2\right)\nonumber.
\]
Then by the proof of Lemma \ref{lemma:sieve_construction}, we know that $\Pi(\calC_n^c\mid\calD_n)=o_{\mathbb{P}_0}(1)$ by taking a sufficiently large $L$. This completes the proof of the first assertion.

Now we focus on proving the second assertion. Observe that
\begin{align}
\left|\langle g,h\rangle_{\mathbb{H}_{\Psi_\alpha}(\Omega)}\right|&=\frac{4\sigma^4}{\sqrt{n}}\left\|\bt\transpose{}\bV_0^{-1}\frac{\partial y^s}{\partial\btheta}(\cdot,\btheta_0^*)\right\|^2_{\mathbb{H}_{\Psi_\alpha}(\Omega)}\nonumber\\
&\leq\frac{4\sigma^4}{\sqrt{n}}\|\bV_0^{-1}\bt\|^2\sum_{j=1}^q\sup_{\btheta\in\Theta}\left\|\frac{\partial y^s}{\partial\theta_j}(\cdot,\btheta)\right\|^2_{\mathbb{H}_{\Psi_\alpha}(\Omega)}=o(\sqrt{n}\eps_n),\nonumber
\end{align}
which implies that for sufficiently large $n$,
\begin{align}
\label{eqn:change_of_measure_event_bounds}
\left\{\left|U(g)\right|\leq (L/2)\sqrt{n}\eps_n\|g\|_{\mathbb{H}_{\Psi_\alpha}(\Omega)}\right\}
&\subset
\left\{\left|U(g)+\langle g, h\rangle_{\mathbb{H}_{\Psi_\alpha}(\Omega)}\right|\leq L\sqrt{n}\eps_n\|g\|_{\mathbb{H}_{\Psi_\alpha}(\Omega)}\right\}\nonumber\\
&\subset\left\{\left|U(g)\right|\leq 2L\sqrt{n}\eps_n\|g\|_{\mathbb{H}_{\Psi_\alpha}(\Omega)}\right\}.
\end{align}
On the other hand, 
\[\|h\|_{L_2(\Omega)}\leq\frac{2q\sigma^2}{\sqrt{n}}\|\bV_0^{-1}\bt\|\max_{j=1,\cdots,q}\sup_{\btheta\in\Theta}\left\|\frac{\partial y^s}{\partial\theta_j}(\bdot,\btheta)\right\|_{L_2(\Omega)}=o(\eps_n),\]
implying that
\begin{align}
\label{eqn:calA_1n_upper_bound}
\calA_{1n}& = \left\{\|\eta_\bt-\eta_0+h\|_{L_2(\Omega)}\leq M_n\eps_n\right\}\nonumber\\
&\subset
\left\{
\|\eta_\bt-\eta_0\|_{L_2(\Omega)}\leq M_n\eps_n+\|h\|_{L_2(\Omega)}
\right\}\nonumber\\
&\subset\left\{
\|\eta_\bt-\eta_0\|_{L_2(\Omega)}\leq 2M_n\eps_n
\right\}: = \calA_{1n}^u(\bt)
\end{align}
for sufficiently large $n$, where the fact $n^{-1/2}\leq \eps_n$ is applied. Similarly, for sufficiently large $n$ it holds that 
\begin{align}
\label{eqn:calA_1n_lower_bound}
\calA_{1n}\supset \{\|\eta_\bt-\eta_0\|_{L_2(\Omega)}\leq M_n\eps_n/2\}:=\calA_{1n}^l(\bt).
\end{align}
Similarly, by taking $\calA_{2n}^l(\bt) = \{\|\eta_\bt - \eta_0\|_{L_\infty(\Omega)}\leq M/2\}$ one can also show that $\calA_{2n}^l(\bt)\subset\calA_{2n}$. 

\noindent
We break the rest of the proof  into two components. 

\noindent
\textbf{Step 1: We provide an upper bound for $\int_{\calA_n\cap\calC_n}\exp(\ell_n(\eta_\bt)-\ell_n(\eta_0))\Pi(\mathrm{d}\eta)$.}

\noindent 
 Write
\begin{align}
&\int_{\calA_n\cap\calC_n}\exp(\ell_n(\eta_\bt)-\ell_n(\eta_0))\Pi(\mathrm{d}\eta)\nonumber\\
&\quad\leq\int\mathbbm{1}\left\{|U(g)|\leq L\sqrt{n}\eps_n\|g\|_{\mathbb{H}_{\Psi_\alpha}(\Omega)}
\right\}\mathbbm{1}(\calA_{1n}^u(\bt))\exp(\ell_n(\eta_\bt)-\ell_n(\eta_0))\Pi(\mathrm{d}\eta)\nonumber.
\end{align}
We obtain the upper bound of the right-hand side of the last display using the change of measure formulas \eqref{eqn:change_of_measure}, \eqref{eqn:change_of_measure_event_bounds}, and \eqref{eqn:calA_1n_upper_bound}:
\begin{align}
&\int\mathbbm{1}\left\{|U(g)|\leq L\sqrt{n}\eps_n\|g\|_{\mathbb{H}_{\Psi_\alpha}(\Omega)}
\right\}\mathbbm{1}(\calA_{1n}^u(\bt))\exp(\ell_n(\eta_\bt)-\ell_n(\eta_0))\Pi(\mathrm{d}\eta)\nonumber\\
&\quad\leq\int\mathbbm{1}\left\{\left|U(g)+\langle g,h\rangle_{\mathbb{H}_{\Psi_\alpha}(\Omega)}\right|\leq L\sqrt{n}\eps_n\|g\|_{\mathbb{H}_{\Psi_\alpha}(\Omega)}
\right\}\mathbbm{1}(\|\eta-\eta_0\|_{L_2(\Omega)}\leq 2M_n\eps_n)\nonumber\\
&\qquad\quad\times\exp(\ell_n(\eta)-\ell_n(\eta_0))\exp\left[U(-h)-\frac{2\sigma^4}{n}\left\|\bt\transpose{}\bV_0^{-1}\frac{\partial y^s}{\partial\btheta}(\cdot,\btheta_0^*)\right\|^2_{\mathbb{H}_{\Psi_\alpha}(\Omega)}\right]\Pi(\mathrm{d}\eta)\nonumber\\
&\quad\leq \int\mathbbm{1}\left\{|U(g)|\leq 2L\sqrt{n}\eps_n\|g\|_{\mathbb{H}_{\Psi_\alpha}(\Omega)}
\right\}\mathbbm{1}(\|\eta-\eta_0\|_{L_2(\Omega)}\leq 2M_n\eps_n)\nonumber\\
&\qquad\quad\times\exp(\ell_n(\eta)-\ell_n(\eta_0))
\exp\left[U\left(-\frac{g}{\sqrt{n}}\right)
\right]
\Pi(\mathrm{d}\eta)\nonumber\\
&\quad\leq \int_{\{\|\eta-\eta_0\|_{L_2(\Omega)}\leq 2M_n\eps_n\}}\exp(\ell_n(\eta)-\ell_n(\eta_0))\exp\left(2L\eps_n\|g\|_{\mathbb{H}_{\Psi_\alpha}(\Omega)}\right)\Pi(\mathrm{d}\eta)\nonumber\\
&\quad\leq [1+o(1)]\int\exp(\ell_n(\eta)-\ell_n(\eta_0))\Pi(\mathrm{d}\eta)\nonumber.
\end{align}
Therefore we conclude that
\begin{align}
\label{eqn:integral_upper_bound}
\int_{\calA_n}\exp(\ell_n(\eta_\bt)-\ell_n(\eta_0))\Pi(\mathrm{d}\eta)
&\leq 
[1+o_{\mathbb{P}_0}(1)]\int\exp(\ell_n(\eta)-\ell_n(\eta_0))\Pi(\mathrm{d}\eta).
\end{align}

\noindent\textbf{Step 2: We provide a lower bound for $\int_{\calA_n\cap\calC_n}\exp(\ell_n(\eta_\bt)-\ell_n(\eta_0))\Pi(\mathrm{d}\eta)$.} 

\noindent Recall the construction of $\calB_n$ in the proof of Lemma \ref{lemma:sieve_construction}: $\calB_n = \eps_n\mathfrak{C}_\beta^1(\Omega) + m_n\mathbb{H}_{\Psi_\alpha}(\Omega)$, where 
\[
\mathfrak{C}_\beta^1(\Omega)=\left\{f\in\mathfrak{C}_\beta(\Omega):\|f\|_{\mathfrak{C}_\beta(\Omega)}\leq 1\right\},\quad\mathbb{H}_{\Psi_\alpha}^1(\Omega)=\left\{f\in\mathbb{H}_{\Psi_\alpha}(\Omega):\|f\|_{\mathbb{H}_{\Psi_\alpha}(\Omega)}\leq 1\right\},
\]
and $m_n = -2\Phi^{-1}(\exp[-(2C+1/\sigma^2)n\eps_n^2])$. 
Now take $\widetilde\calB_n = \eps_n\mathfrak{C}_\beta^1(\Omega) + (3m_n/4)\mathbb{H}_{\Psi_\alpha}(\Omega)$.
Then again by the Borell's inequality \citep{van2008rates} we have
\begin{align}
\Pi(\widetilde\calB_n)&\geq 
\Phi\left(\Phi^{-1}\left(\exp\left(-Cn\eps_n^2\right)\right)+\frac{3m_n}{4}\right)\nonumber\\
&= 
\Phi\left(\Phi^{-1}\left(\exp\left(-Cn\eps_n^2\right)\right)-\frac{3}{2}\Phi^{-1}\left(\exp\left[-\left(2C+\frac{1}{\sigma^2}\right)n\eps_n^2\right]\right)\right)
\nonumber\\&
\geq\Phi\left(-\frac{1}{2}\Phi^{-1}\left(\exp\left[-\left(2C+\frac{1}{\sigma^2}\right)n\eps_n^2\right]\right)\right)
\nonumber.
\end{align}
Using the facts that $\Phi^{-1}(u)\leq (-1/2)\sqrt{\log(1/u)}$ for $u\in(0,1/2)$, $1-\Phi(x)\leq (1/2)\mathrm{e}^{-x^2/2}$ for sufficiently large $x$ (see, for example, Lemma K.6 in \citealp{ghosal2017fundamentals}), and $n\eps_n^2\to\infty$, we further lower bound the last display as follows:
\begin{align*}
\Pi(\widetilde\calB_n)&\geq \Phi\left(-\frac{1}{2}\Phi^{-1}\left(\exp\left[-\left(2C+\frac{1}{\sigma^2}\right)n\eps_n^2\right]\right)\right)\\
&\geq \Phi\left(\frac{1}{4}\sqrt{\left(2C + \frac{1}{\sigma^2}\right)n\eps_n^2}\right)\geq 1 - \frac{1}{2}\exp\left[-\frac{1}{32}\left(2C + \frac{1}{2\sigma^2}\right)n\eps_n^2\right].
\end{align*}
Then we conclude that $\Pi(\widetilde\calB_n\mid\calD_n)= o_{\mathbb{P}_0}(1)$ by following an argument that is similar to that for proving $\Pi(\calB_n\mid\calD_n)= o_{\mathbb{P}_0}(1)$. Furthermore, for any $\eta\in\widetilde\calB_n$, there exists $\eta_1\in\mathfrak{C}_\beta^1(\Omega)$ and $\eta_2\in\mathbb{H}_{\Psi_\alpha}(\Omega)$ such that $\eta = \eps_n\eta_1 + (3m_n/4)\eta_2$. Consequently, if $\eta_\bt\in\widetilde\calB_n$, then
\[
\eta = \eta_\bt + h = \eps_n(\eta_\bt)_1 + (3m_n/4)(\eta_\bt)_2 + h = \eps_n(\eta_\bt)_1 + m_n\left(\frac{3(\eta_\bt)_2}{4} + \frac{h}{m_n}\right).
\]
Then we directly conclude that $\eta\in\calB_n$, namely, $\mathbbm{1}(\eta_\bt\in\widetilde\calB_n)\leq \mathbbm{1}(\eta\in\calB_n)$, by noting that
\[
\left\|\frac{3(\eta_\bt)_2}{4} + \frac{h}{m_n}\right\|_{\Psi_\alpha(\Omega)}
\leq \frac{3}{4}\left\|\eta_\bt\right\|_{\Psi_\alpha(\Omega)} + \frac{1}{m_n}\left\|h\right\|_{\Psi_\alpha(\Omega)}\leq 1.
\]

\noindent
Now we turn to the computation of the desired lower bound. Write
\begin{align}
&\int_{\calA_n\cap\calC_n}\exp(\ell_n(\eta_\bt)-\ell_n(\eta_0))\Pi(\mathrm{d}\eta)\nonumber\\
&\quad\geq\int\mathbbm{1}\left\{|U(g)|\leq L\sqrt{n}\eps_n\|g\|_{\mathbb{H}_\Psi(\Omega)}\right\}\mathbbm{1}(\calA_{1n}^l(\bt))\mathbbm{1}(\calA_{2n}^l(\bt))\mathbbm{1}(\eta_\bt\in\widetilde\calB_n)\exp(\ell_n(\eta_\bt)-\ell_n(\eta_0))\Pi(\mathrm{d}\eta)\nonumber.
\end{align}
We lower bound the preceeding display using \eqref{eqn:change_of_measure}, \eqref{eqn:change_of_measure_event_bounds}, and \eqref{eqn:calA_1n_lower_bound}:
\begin{align}
&\int\mathbbm{1}\left\{|U(g)|\leq L\sqrt{n}\eps_n\|g\|_{\mathbb{H}_{\Psi_\alpha}(\Omega)}\right\}\mathbbm{1}(\calA_{1n}^l(\bt))\mathbbm{1}(\calA_{2n}^l(\bt))\mathbbm{1}(\eta_\bt\in\widetilde\calB_n)\exp(\ell_n(\eta_\bt)-\ell_n(\eta_0))\Pi(\mathrm{d}\eta)\nonumber\\
&\quad=\int\mathbbm{1}\left\{\left|U(g)+\langle g,h\rangle_{\mathbb{H}_{\Psi_\alpha}(\Omega)}\right|\leq L\sqrt{n}\eps_n\|g\|_{\mathbb{H}_{\Psi_\alpha}(\Omega)}\right\}\mathbbm{1}\left\{\|\eta-\eta_0\|_{L_2(\Omega)}\leq M_n\eps_n/2\right\}\nonumber\\
&\qquad\quad\times \exp(\ell_n(\eta)-\ell_n(\eta_0))\exp\left[U\left(-\frac{g}{\sqrt{n}}\right)-\frac{2\sigma^2}{n}\left\|\bt\transpose{}\bV_0^{-1}\frac{\partial y^s}{\partial\btheta}(\cdot,\btheta_0^*)\right\|_{\mathbb{H}_{\Psi_\alpha}(\Omega)}^2\right]\Pi(\mathrm{d}\eta)\nonumber\\
&\quad\geq\int\mathbbm{1}\left\{\left|U(g)\right|\leq (L/2)\sqrt{n}\eps_n\|g\|_{\mathbb{H}_{\Psi_\alpha}(\Omega)}\right\}\mathbbm{1}\left\{\|\eta-\eta_0\|_{L_2(\Omega)}\leq M_n\eps_n/2\right\}\nonumber\\
&\qquad\quad\times \mathbbm{1}\left\{\|\eta-\eta_0\|_{L_\infty(\Omega)}\leq M/2\right\}\mathbbm{1}\left(\eta\in\widetilde\calB_n\right)\exp(\ell_n(\eta)-\ell_n(\eta_0))\nonumber\\
&\qquad\quad\times \exp\left(-\frac{1}{\sqrt{n}}\left|U\left(g\right)\right|\right)[1-o(1)]\Pi(\mathrm{d}\eta)\nonumber\\
&\quad\geq[1-o(1)]\int\mathbbm{1}\left\{\left|U(g)\right|\leq (L/2)\sqrt{n}\eps_n\|g\|_{\mathbb{H}_{\Psi_\alpha}(\Omega)}\right\}\mathbbm{1}\left\{\|\eta-\eta_0\|_{L_2(\Omega)}\leq M_n\eps_n/2\right\}
\nonumber\\&\qquad\quad\quad\quad\quad\quad\times 
\mathbbm{1}\left\{\|\eta-\eta_0\|_{L_\infty(\Omega)}\leq M/2\right\}\mathbbm{1}(\widetilde\calB_n)
\exp(\ell_n(\eta)-\ell_n(\eta_0))\Pi(\mathrm{d}\eta)\nonumber.
\end{align}
Since $\Pi(\|\eta-\eta_0\|_{L_2(\Omega)}> M_n\eps_n/2\mid\calD_n)=o_{\mathbb{P}_0}(1)$, $\Pi(\widetilde\calB_n^c)=o_{\mathbb{P}_0}(1)$, and for sufficiently large $L$ and $M$, $\Pi(|U(g)|>(L/2)\sqrt{n}\eps_n\|g\|_{\mathbb{H}_\Psi(\Omega)}\mid\calD_n)=o_{\mathbb{P}_0}(1)$, $\Pi(\|\eta-\eta_0\|_{L_\infty(\Omega)}>M/2\mid\calD_n) = o_{\mathbb{P}_0}(1)$, the last display can be further computed
\begin{align}
&\int\mathbbm{1}\left\{\left|U(g)\right|\leq (L/2)\sqrt{n}\eps_n\|g\|_{\mathbb{H}_{\Psi_\alpha}(\Omega)}\right\}\mathbbm{1}\left\{\|\eta-\eta_0\|_{L_2(\Omega)}\leq M_n\eps_n/2\right\}\nonumber\\
&\quad\times\left\{\|\eta-\eta_0\|_{L_\infty(\Omega)}\leq M/2\right\}\mathbbm{1}(\widetilde\calB_n)
\exp(\ell_n(\eta)-\ell_n(\eta_0))\Pi(\mathrm{d}\eta)\nonumber\\
&\quad \geq \int\exp(\ell_n(\eta)-\ell_n(\eta_0))\Pi(\mathrm{d}\eta)
-\int_{\left\{\left|U(g)\right|> (L/2)\sqrt{n}\eps_n\|g\|_{\mathbb{H}_{\Psi_\alpha}(\Omega)}\right\}}\exp(\ell_n(\eta)-\ell_n(\eta_0))\Pi(\mathrm{d}\eta)\nonumber\\
&\quad\quad - \int_{\left\{\|\eta-\eta_0\|_{L_2(\Omega)}> M_n\eps_n/2\right\}}\exp(\ell_n(\eta)-\ell_n(\eta_0))\Pi(\mathrm{d}\eta)
- \int_{\widetilde\calB_n}\exp(\ell_n(\eta)-\ell_n(\eta_0))\Pi(\mathrm{d}\eta)
\nonumber\\
&\quad\quad - \int_{\left\{\|\eta-\eta_0\|_{L_\infty(\Omega)}> M/2\right\}}\exp(\ell_n(\eta)-\ell_n(\eta_0))\Pi(\mathrm{d}\eta)\nonumber\\
&\quad = \int\exp(\ell_n(\eta)-\ell_n(\eta_0))\Pi(\mathrm{d}\eta)-o_{\mathbb{P}_0}(D_n).\nonumber
\end{align}
Hence we conclude that
\begin{align}
\label{eqn:integral_lower_bound}
\int_{\calA_n}\exp(\ell_n(\eta)-\ell_n(\eta_0))\Pi(\mathrm{d}\eta)&\geq[1-o(1)]\int\exp(\ell_n(\eta)-\ell_n(\eta_0))\Pi(\mathrm{d}\eta)-o_{\mathbb{P}_0}(D_n)\nonumber\\
&=[1-o_{\mathbb{P}_0}(1)]\int\exp(\ell_n(\eta)-\ell_n(\eta_0))\Pi(\mathrm{d}\eta).
\end{align}
The proof is completed by combining \eqref{eqn:integral_upper_bound} and \eqref{eqn:integral_lower_bound}.


\section{Proof of Corollary \ref{corr:Bayes_estimator}} 
\label{sec:proof_of_corollary_corr:bayes_estimator2}
The proof is similar to that of Corollary of \cite{yang2015semiparametric} and is included here for completeness. For each $k = 1,\ldots,q$, let the event $A = \mathbb{R}\times\ldots\times A_s\times\ldots\times\mathbb{R}$ in 
Theorem \ref{thm:BvM_limit}, where the $s$th component is $A_s$ and the rest are $\mathbb{R}$. Then it follows directly from Theorem \ref{thm:BvM_limit} that
\[
\sup_{A_s\subset\mathbb{R}}\left|
\Pi\left([\btheta_\eta^*]_k\in A_s\mathrel{\big|}\calD_n\right)
- \mathrm{N}\left([\widehat\btheta_{L_2}]_k, \frac{4\sigma^2}{n}[\bV_0^{-1}\bW\bV_0^{-1}]_{kk}\right)(A_s)
\right| = o_{\mathbb{P}_0}(1),
\]
where $[\cdot]_k$ is the $k$th component of the argument vector and $[\cdot]_{kk}$ is the $(k,k)$th element of the argument matrix. Now set $A_s = (-\infty, [\widehat\btheta^*]_k]$. It follows that
\[
\left|
\Phi\left(\sqrt{\frac{n}{4\sigma^2[\bV_0^{-1}\bW\bV_0^{-1}]_{kk}}}\left([\widehat\btheta^*]_k - [\widehat\btheta_{L_2}]_k\right)\right) - \frac{1}{2}
\right| = o_{\mathbb{P}_0}(1),
\]
where $\Phi$ is the cumulative distribution function (CDF) of the standard normal distribution. By the continuity of $\Phi^{-1}$, we have 
$[\widehat\btheta^*]_k - [\widehat\btheta_{L_2}]_k = o_{\mathbb{P}_0}(1/\sqrt{n})$. Invoking the asymptotic normality of $\widehat\btheta_{L_2}$ completes the proof. 


\section{Proof of Theorem \ref{thm:convergence_AdaGrad}} 
\label{sec:proof_convergence_adagrad}
Before presenting the proof, we need several auxiliary Lemmas from \cite{NIPS2013_5129} and \cite{li2018convergence}.

\begin{lemma}[\cite{NIPS2013_5129}, Lemma A.5]
\label{lemma:Lemma_3}
Let $(a_t)_{t\geq1},(b_t)_{t\geq1}$ be two non-negative real sequences such that $b_t$'s are bounded, $\sum_{t = 1}^\infty a_tb_t$ converges and $\sum_{t = 1}^\infty a_t$ diverges, and $|b_{t + 1} - b_t| \lesssim a_t$. Then $b_t\to 0$ as $t\to\infty$. 
\end{lemma}

\begin{lemma}[Lemma 4, \cite{li2018convergence}]
\label{lemma:Lemma_4}
Let $(a_t)_{t = 1}^N$ be a non-negative real sequences such that $a_0 > 0$, and $\beta > 1$. Then $\sum_{t = 1}^Na_t/(a_0 + \sum_{j = 1}^ta_t)^\beta\leq (\beta - 1)^{-1}a_0^{1 - \beta}$. 
\end{lemma}

\begin{lemma}[Lemma 5, \cite{li2018convergence}]
\label{lemma:Lemma_5}
Assume conditions A2 and A4 hold, and the sample path $\eta$ is squared-integrable. Then the iterates of Algorithm \ref{alg:AdaGrad_calibration} satisfy the following inequality
\begin{align*}
&\mathbb{E}_\bw\left[\sum_{t = 1}^N\left\langle \frac{\partial f_\eta(\btheta^{(t)})}{\partial\btheta}, 
\sum_{k = 1}^q\alpha_{tk}\frac{\partial f_\eta(\btheta^{(t)})}{\partial\theta_k}
\right\rangle\right]\\
&\quad\leq f_\eta(\btheta^{(1)}) - f_\eta(\btheta_\eta^*) + \frac{1}{2}\sup_{\btheta\in\Theta}\left\|\int_\Omega\frac{\partial}{\partial\btheta}[y^s(\bx,\btheta^{(t)}) - \eta(\bx)]^2\mathrm{d}\bx\right\|_{L_2(\Omega)}\\
&\quad\quad\times\mathbb{E}_\bw\left\{\sum_{t = 1}^N\sum_{k = 1}^q\alpha_{tk}^2\left[\frac{\partial }{\partial\theta_k}(y^s(\bw_t, \btheta^{(t)}) - \eta(\bw_t))^2\right]^2\right\}
\end{align*}
\end{lemma}

The proof is based on a modification of the Theorem 2 in \cite{li2018convergence}, which is provided here for completeness. 
Observe that by Lemma \ref{lemma:Lemma_4}, conditions A2 and A4, and the fact that $\eta$ is continuous over $\Omega$, we have,
\begin{align*}
&\sum_{t = 1}^\infty\sum_{k = 1}^q\alpha_{tk}^2\left[\frac{\partial }{\partial\theta_k}(y^s(\bw_t, \btheta^{(t)}) - \eta(\bw_t))^2\right]^2\\
&\quad = \sum_{t = 1}^\infty\sum_{k = 1}^q\alpha_{(t + 1)k}^2\left[\frac{\partial }{\partial\theta_k}(y^s(\bw_t, \btheta^{(t)}) - \eta(\bw_t))^2\right]^2\\
&\quad\quad + \sum_{t = 1}^\infty\sum_{k = 1}^q(\alpha_{tk}^2 - \alpha_{(t + 1)k}^2)\left[\frac{\partial }{\partial\theta_k}(y^s(\bw_t, \btheta^{(t)}) - \eta(\bw_t))^2\right]^2\\
&\quad \leq \frac{a_0^2}{2\eps b_0^{2\eps}} + \sup_{(\bw,\btheta)\in\Omega\times\Theta}\max_{1\leq k\leq q}\left|\frac{\partial }{\partial\theta_k}\left[y^s(\bw_t, \btheta^{(t)}) - \eta(\bw_t)\right]^2\right|^2\sum_{t = 1}^\infty\sum_{k = 1}^q(\alpha_{tk}^2 - \alpha_{(t + 1)k}^2)\\
&\quad\leq \frac{a_0^2}{2\eps b_0^{2\eps}} + \sup_{(\bw,\btheta)\in\Omega\times\Theta}\max_{1\leq k\leq q}\left|\frac{\partial }{\partial\theta_k}\left[y^s(\bw_t, \btheta^{(t)}) - \eta(\bw_t)\right]^2\right|^2\alpha_{1k}^2 < \infty.
\end{align*}
Therefore, for any $m\in\mathbb{N}_+$, we obtain by Cauchy-Schwarz inequality that
\begin{align*}
\|\btheta^{(N + m)} - \btheta^{(N)}\|^2
& = \left\|\sum_{t = N}^{N + m - 1}(\btheta^{(t + 1)} - \btheta^{(t)})\right\|^2
\leq m\sum_{t = N}^{N + m - 1}\|\btheta^{(t + 1)} - \btheta^{(t)}\|^2
\\
& \leq m\sum_{t = N}^{N + m - 1}\left\|2[y^s(\bw_t, \btheta^{(t)}) - \eta(\bw_t)]\mathrm{diag}(\alpha_{t1},\ldots,\alpha_{tq})\frac{\partial y^s}{\partial\btheta}(\bw_t,\btheta^{(t)})\right\|^2
\\
& \leq m\sum_{t = N}^{N + m - 1}\sum_{k = 1}^q\alpha_{tk}^2\left|\frac{\partial}{\partial\theta_k}[y^s(\bw_t, \btheta^{(t)}) - \eta(\bw_t)]^2\right|^2,
\end{align*}
and the previous infinite sum being finite implies that $\lim_{N\to\infty}\|\btheta^{(N + m)} - \btheta^{(N)}\| = 0$ a.s., \emph{i.e.}, $(\btheta^{(N)})_N$ forms a Cauchy sequence, and thus must converges to some point $\btheta^*\in\Theta$ a.s.. Note that $\btheta^*$ is still a random variable. 

Next we show that $\btheta^*$ is a stationary point of $f_\eta$. We obtain, by Lemma \ref{lemma:Lemma_5} and taking $N\to\infty$ that
\begin{align*}
\mathbb{E}_\bw\left[\sum_{t = 1}^\infty\sum_{k = 1}^q\alpha_{tk}\left(\frac{\partial f_\eta(\btheta^{(t)})}{\partial\theta_k}\right)^2\right]
&\leq f_\eta(\btheta^{(1)}) - f_\eta(\btheta_\eta^*) + \frac{1}{2}\sup_{\btheta\in\Theta}\left\|\int_\Omega\frac{\partial}{\partial\btheta}[y^s(\bx,\btheta^{(t)}) - \eta(\bx)]^2\mathrm{d}\bx\right\|_{L_2(\Omega)}\\
&\quad\times\mathbb{E}_\bw\left\{\sum_{t = 1}^\infty\sum_{k = 1}^q\alpha_{tk}^2\left[\frac{\partial }{\partial\theta_k}(y^s(\bw_t, \btheta^{(t)}) - \eta(\bw_t))^2\right]^2\right\} < \infty.
\end{align*}
Therefore, $\sum_{t = 1}^\infty\alpha_{tk}[\partial f_\eta(\btheta^{(t)})/\partial\theta_k]^2 < \infty$ a.s., for all $k = 1,\ldots, q$. In addition, observe that
\begin{align*}
&\sup_{\bw_t,\btheta^{(t)}}\left\|2[y^s(\bw_t, \btheta^{(t)}) - \eta(\bw_t)]\mathrm{diag}(\alpha_{t1},\ldots,\alpha_{tq})\frac{\partial y^s}{\partial\btheta}(\bw_t,\btheta^{(t)})\right\|\\
&\quad\leq \max_{t,k}\alpha_{tk}\sup_{(\bx,\btheta)\in\Omega\times\Theta}\left\|\frac{\partial y^s}{\partial\btheta}(\bx,\btheta)\right\|\sup_{(\bx,\btheta)\in\Omega\times\Theta}\left|2[y^s(\bx, \btheta) - \eta(\bx)]\right| < \infty.
\end{align*}
Since by the construction of Algorithm \ref{alg:AdaGrad_calibration}, $\btheta^{(t)}\in\Theta\backslash\partial\Theta$, we see that there exists an integer $m^*$, such that for all $t\in\mathbb{N}_+$, the number of times that line 11 of Algorithm \ref{alg:AdaGrad_calibration} is called is no greater than $m^*$. This implies that
\[
 \frac{a_0}{2^{m^*}}\left\{b_0 + \sum_{j = 1}^{t - 1}\left[\frac{\partial g(\bw_j, \btheta^{(j)})}{\partial\theta_k}\right]^2\right\}^{-(1/2 + \eps)}\leq
 \alpha_{tk}\leq 
 a_0\left\{b_0 + \sum_{j = 1}^{t - 1}\left[\frac{\partial g(\bw_j, \btheta^{(j)})}{\partial\theta_k}\right]^2\right\}^{-(1/2 + \eps)}
\]
for all $t\in\mathbb{N}_+$ and all $k = 1,\ldots,q$, where 
$g(\bx,\btheta) = [y^s(\bx, \btheta) - \eta(\bw_t)]^2$. This further implies that
\[
\sum_{t = 1}^\infty\alpha_{tk}\geq\frac{a_0}{2^{m^*}}\sum_{t = 1}^\infty\left\{b_0 + (t - 1)\sup_{(\bx,\btheta)\in\Omega\times\Theta}\left[\frac{\partial g(\bw_j, \btheta^{(j)})}{\partial\theta_k}\right]^2\right\}^{-(1/2 + \eps)} = \infty.
\]
Since condition A4 implies that almost surely,
\begin{align*}
\left|
\frac{\partial f_\eta(\btheta^{(t + 1)})}{\partial\theta_k} - \frac{\partial f_\eta(\btheta^{(t)})}{\partial\theta_k}
\right|&\leq|\theta_{k}^{(t + 1)} - \theta_k^{(t)}| \int_\Omega\sup_{(\bx,\btheta)\in\Omega\times\Theta}\frac{\partial^2}{\partial\theta_k^2}[y^s(\bx,\btheta) - \eta(\bx)]^2\mathrm{d}\bx\\
&\leq \alpha_{tk}\sup_{(\bx,\btheta)\in\Omega\times\Theta}\left|\frac{\partial g}{\partial\theta_k}g(\bx,\btheta)\right|\int_\Omega\sup_{(\bx,\btheta)\in\Omega\times\Theta}\frac{\partial^2}{\partial\theta_k^2}[y^s(\bx,\btheta) - \eta(\bx)]^2\mathrm{d}\bx\\
&\lesssim \alpha_{tk},
\end{align*}
then by the facts that $\sum_{t = 1}^\infty\alpha_{tk}[\partial f_\eta(\btheta^{(t)})/\partial\theta_k]^2 < \infty$ a.s., and $\sum_{t = 1}^\infty\alpha_{tk} = \infty$, we invoke Lemma \ref{lemma:Lemma_3} to conclude that $\lim_{N\to\infty}\partial f_\eta(\btheta^{(N)})/\partial\theta_k = 0$ a.s., for all $k = 1,\ldots,q$. The continuity of $\nabla f_\eta(\btheta)$ and the almost sure convergence of $\btheta^{(N)}\to\btheta^*$ directly yield that $\nabla f_\eta(\btheta^*) = \zero$ a.s., completing the proof. 


\section{Proof of Theorem \ref{thm:approximate_PC_convergence}} 
\label{sec:proof_of_theorem_4}
The idea of the proof is based on the proof of Theorem \ref{thm:BvM_limit} and a fine control between $\btheta_\eta^*$ and $\widetilde\btheta_\eta$. 
By the proof of Lemma \ref{lemma:taylor_expansion_functional}, there exists some $\eps>0$ such that over $\{\eta\in\calF:\|\eta-\eta_0\|_{L_2(\Omega)}<\eps\}$, the functional $\btheta_\eta^*:\eta\mapsto \argmin_{\btheta\in\Theta}\|\eta(\cdot)-y^s(\cdot,\btheta)\|_{L_2(\Omega)}^2$ is continuous, the Fr\'echet derivative $\dot{\btheta}_\eta^*:\calF\to\mathbb{R}^q$ for $\btheta^*_\eta$ exists, and can be computed by
 \begin{align}
 \dot{\btheta}_\eta^*(h)=-\left[\bF_\theta(\eta,\btheta_\eta^*)\right]^{-1}\left[\bF_\eta(\eta,\btheta_\eta^*)\right](h)
 =2\bV_\eta^{-1}\int_\Omega h(\bx)\frac{\partial y^s}{\partial\btheta}(x,\btheta_\eta^*)\mathrm{d}\bx.\nonumber
 \end{align}
By Proposition 1 in \cite{tuo2015efficient}, $\|\widehat\eta - \eta_0\|_{L_2(\Omega)} = O_{\mathbb{P}_0}(n^{-\alpha/(2\alpha + p)})$, since the RKHS $\mathbb{H}_{\Psi_\nu}$ coincides with $\calH_\alpha(\Omega)$ for $\nu = \alpha - p/2$. Therefore, with probability tending to one, $\|\widehat\eta - \eta_0\|_{L_2(\Omega)} < \eps$. We now assume this event occurs and denote it by $\calE_n$. Then similar to the proof of Lemma \ref{lemma:taylor_expansion_functional}, for any $\eta$ in the $L_2(\Omega)$-neighborhood of $\eta_0$ with radius $\eps$, we apply the fundamental theorem of calculus and mean-value theorem to obtain
\begin{align}
 \btheta_\eta^*-\widehat\btheta_{L_2}=\int_0^1\frac{\mathrm{d}}{\mathrm{d}u}\btheta_{\eta[u]}^*\mathrm{d}u\nonumber=2\int_\Omega [\eta(\bx)-\widehat\eta(\bx)] \bV_{\eta[u']}^{-1}\frac{\partial y^s}{\partial\btheta}(\bx,\btheta_{\eta[u']}^*)\mathrm{d}\bx\nonumber,
 \end{align}
 where $\eta[u] = \widehat\eta + (\eta - \widehat\eta)u$ for $0\leq u\leq 1$ and $u'\in[0,1]$. Then following the same argument in the proof of Lemma \ref{lemma:taylor_expansion_functional}, we have, $\|\btheta_\eta^*-\widehat\btheta_{L_2}\|\leq  L_{\eta_0}^{(1)}\|\eta-\widehat\eta\|_{L_2(\Omega)}$ for some constant $L_{\eta_0}^{(1)}>0$ depending on $\eta_0$ only.  Furthermore, $\|\bV_\eta^{-1} - \bV_0^{-1}\|\leq 2qC_{\eta_0}\|\bV_0^{-1}\|\|\eta - \eta_0\|_{L_2(\Omega)}$ for some constant $C_{\eta_0} > 0$ whenever $\|\eta - \eta_0\|_{L_2(\Omega)} < \eps$. Therefore, using a technique similar to that in the proof of Lemma \ref{lemma:taylor_expansion_functional}, 
 \begin{align}
 \br(\eta, \widehat\eta) 
 &=\btheta_\eta^* - \widetilde\btheta_\eta 
 =\btheta_\eta^*-\widehat\btheta_{L_2}-2\int_\Omega \left[\eta(\bx)-\widehat\eta(\bx)\right]\bV_{\widehat\eta}^{-1}\frac{\partial y^s}{\partial\btheta}(\bx,\widehat\btheta_{L_2})\mathrm{d}\bx\nonumber\\
 &=2\int_\Omega [\eta(\bx)-\widehat\eta(\bx)]\left[\bV_{\eta[u']}^{-1}\frac{\partial y^s}{\partial\btheta}(\bx,\btheta_{\eta[u']}^*) - \bV_{\widehat\eta}^{-1}\frac{\partial y^s}{\partial\btheta}(\bx,\widehat\btheta_{L_2})\right]\mathrm{d}\bx\nonumber\\
 &=2\int_\Omega [\eta(\bx)-\widehat\eta(\bx)]\left[(\bV_{\eta[u']}^{-1} - \bV_0^{-1} + \bV_0^{-1} - \bV_{\widehat\eta}^{-1})\frac{\partial y^s}{\partial\btheta}(\bx,\btheta_{\eta[u']}^*)\right]\mathrm{d}\bx\nonumber\\
 &\quad + 2\int_\Omega [\eta(\bx)-\widehat\eta(\bx)](\bV_{0}^{-1} + \bV_{\widehat\eta}^{-1} - \bV_0^{-1})\left[\frac{\partial y^s}{\partial\btheta}(\bx,\btheta_{\eta[u']}^*)-\frac{\partial y^s}{\partial\btheta}(\bx,\widehat\btheta_{L_2})\right]\mathrm{d}\bx\nonumber,
 \end{align}
 and hence,
 \begin{align}
 \|\br(\eta,\widehat\eta)\|
 &\leq 2\int_\Omega|\eta(\bx)-\widehat\eta(\bx)|\left[\left(\left\|\bV_{\eta[u']}^{-1} - \bV_0^{-1}\right\| + \left\|\bV_{\widehat\eta}^{-1} - \bV_0^{-1}\right\|\right)\sup_{(\bx,\btheta)\in\Omega\times\Theta}\left\|\frac{\partial y^s}{\partial\btheta}(\bx,\btheta)\right\|\right]\mathrm{d}\bx\nonumber\\
 &\quad + 2\int_\Omega|\eta(\bx)-\widehat\eta(\bx)|\left[\left(\left\|\bV_0^{-1}\right\| + \left\|\bV_{\widehat\eta}^{-1} - \bV_0^{-1}\right\|\right)\left\|\frac{\partial y^s}{\partial\btheta}(\bx,\btheta_{\eta[u']}^*)-\frac{\partial y^s}{\partial\btheta}(\bx,\widehat\btheta_{L_2})\right\|\right]\mathrm{d}\bx\nonumber\\
 &\lesssim \left(\|\eta[u'] - \eta_0\|_{L_2(\Omega)} + \|\widehat\eta - \eta_0\|_{L_2(\Omega)}\right)\int_\Omega|\eta(\bx) - \widehat\eta(\bx)|\mathrm{d}\bx\nonumber\\
 &\quad  + \left(\left\|\bV_0^{-1}\right\| + 2qC_{\eta_0}\eps\right)\int_\Omega|\eta(\bx) - \widehat\eta(\bx)|\mathrm{d}\bx\sup_{(\bx,\btheta)\in\Omega\times\Theta}\left\|\frac{\partial^2 y^s}{\partial\btheta\partial\btheta\transpose}(\bx, \btheta)\right\|\|\btheta_{\eta[u']}^* - \widehat\btheta_{L_2}\|\nonumber\\
 &\lesssim \left(\|\eta - \widehat\eta\|_{L_2(\Omega)}
 + \|\widehat\eta - \eta_0\|_{L_2(\Omega)}
 \right)\|\eta - \widehat\eta\|_{L_2(\Omega)} + \|\eta - \widehat\eta\|_{L_2(\Omega)}^2\nonumber\\
 &\lesssim \|\eta - \eta_0\|_{L_2(\Omega)}^2 + \|\widehat\eta - \eta_0\|_{L_2(\Omega)}^2\nonumber.
 \end{align}
Recall that we use $\calA_n = \{\|\eta - \eta_0\|_{L_2(\Omega)}\leq M_n\eps_n\}\cap\{\|\eta - \eta_0\|_{L_\infty(\Omega)}\}\cap\calB_n$ in the proof of Theorem \ref{thm:BvM_limit} for $M_n = \log n$ and $\eps_n = n^{-\alpha/(2\alpha + p)}$. Let $\calJ_n = \{\calD_n:\|\widehat\eta - \eta_0\|_{L_2(\Omega)}\leq M_n\eps_n\}$. 
Clearly, By the argument of the proof of Theorem \ref{thm:BvM_limit}, it suffices to show that
\[
\int_{\calA_n\cap\calC_n}\exp\left[\bt\transpose\sqrt{n}\left(\widetilde\btheta_\eta - \widehat\btheta_{L_2}\right)\right]\Pi(\mathrm{d}\eta\mid\calD_n)\to\exp\left[\frac{1}{2}\bt\transpose\left(4\sigma^2\bV_0^{-1}\bW\bV_0^{-1}\right)\bt\right]
\]
in $\mathbb{P}_0$-probability for any fixed vector $\bt\in\mathbb{R}^q$. 
First observe that by the previous derivation, for any $\calD_n \in \calJ_n$, 
\begin{align*}
\sup_{\eta\in\calA_n\cap\calC_n}\left|\bt\transpose\sqrt{n}(\widetilde\btheta_\eta - \btheta_\eta^*)\right|
&\leq \sqrt{n}\|\bt\|\sup_{\eta\in\calA_n}\left(\left\|\widetilde\btheta_\eta - \widehat\btheta_{L_2} - \btheta_\eta^* + \widehat\btheta_{L_2}\right\|\right)
= \sqrt{n}\|\bt\| \sup_{\eta\in\calA_n}\|\br(\eta, \widehat\eta)\|\\
&\lesssim \sqrt{n}M_n^2\eps_n^2 = (\log n)^2n^{-(\alpha - p/2)/(2\alpha + p)}\to 0.
\end{align*}
Therefore, for any $\eps > 0$,
\begin{align*}
\mathbb{P}_0\left(\sup_{\eta\in\calA_n\cap\calC_n}\left|\bt\transpose\sqrt{n}(\widetilde\btheta_\eta - \btheta_\eta^*)\right| > \eps\right)
&\leq \mathbb{P}_0(\calJ_n^c) + \mathbb{P}_0\left(\sup_{\eta\in\calA_n}\left|\bt\transpose\sqrt{n}(\widetilde\btheta_\eta - \btheta_\eta^*)\right| > \eps, \calD_n\in \calJ_n\right)\to 0.
\end{align*}
Since
\[
\int_{\calA_n\cap\calC_n}\exp\left[\bt\transpose\sqrt{n}(\btheta_\eta^* - \widehat{\btheta})\right]\Pi(\mathrm{d}\eta\mid\calD_n) = \exp\left[\frac{1}{2}\bt\transpose(4\sigma^2\bV_0^{-1}\bW\bV_0^{-1})\bt\right] + o_{\mathbb{P}_0}(1) 
\]
by the proof of Theorem \ref{thm:BvM_limit}, it follows that
\begin{align*}
&\int_{\calA_n\cap\calC_n}\exp\left[\bt\transpose\sqrt{n}(\widetilde\btheta_\eta - \widehat{\btheta})\right]\Pi(\mathrm{d}\eta\mid\calD_n)\\
&\quad=
\int_{\calA_n\cap\calC_n}\exp\left\{\bt\transpose\sqrt{n}\left[(\widetilde\btheta_\eta - \btheta_\eta^*) + (\btheta_\eta^* - \widehat{\btheta}_{L_2})\right]\right\}\Pi(\mathrm{d}\eta\mid\calD_n)\\
&\quad = (1 + o_{\mathbb{P}_0}(1))\left\{\exp\left[\frac{1}{2}\bt\transpose(4\sigma^2\bV_0^{-1}\bW\bV_0^{-1})\bt\right] + o_{\mathbb{P}_0}(1) \right\}
\to \exp\left[\frac{1}{2}\bt\transpose(4\sigma^2\bV_0^{-1}\bW\bV_0^{-1})\bt\right]
\end{align*}
in $\mathbb{P}_0$-probability. This completes the proof.


\section{Additional Numerical Results on KO Calibration} 
\label{sec:more_on_ko_calibration_method}

In this section we provide additional results regarding the computation issue of the classical KO approach for calibrating computer models. Recall that \cite{kennedy2001bayesian} formulate the computer model calibration problem as the following statistical model:
\[
\eta(\bx) = y^s(\bx, \btheta) + \delta(\bx),
\]
where $\eta$ is the physical system, $y^s$ is the computer model involving the calibration parameter $\btheta$, and $\delta$ is the discrepancy between them. Classical KO approach and the variations thereof are built on the assumption that $\delta$ follows a Gaussian process prior $\delta\sim\mathrm{GP}(\mu, \Psi_\psi)$ for some mean function $\mu:\Omega\to\mathbb{R}$ and some covariance function $\Psi(\cdot,\cdot\mid\psi):\Omega\times\Omega\to\mathbb{R}_+$ that is typically governed by a range parameter $\psi$, and $\btheta$ follows some prior $\pi(\btheta)$ based on certain expert knowledge. 
It is routine in the Bayes literature to further impose a hyperprior distribution $\pi(\psi)$ on the range parameter $\psi$. For example, in Section \ref{sec:numerical_examples} of the manuscript we take $\pi(\psi)$ to be the inverse-Gamma distribution.  
For simplicity we assume that $\mu$ is zero. 
After collecting noisy physical observations $y_i = \eta(\bx_i) + e_i$, $e_i\iidsim\mathrm{N}(0, \sigma^2)$, the joint posterior density of $\btheta$ and $\psi$ is
\begin{align}
\pi(\btheta, \psi\mid\calD_n)&\propto \frac{\pi(\btheta)\pi(\psi)}{\det(\boldsymbol{\Psi}(\bx_{1:n},\bx_{1:n}\mid\psi) + \sigma^2\eye_n)^{1/2}}\nonumber\\
\label{eqn:KO_density}
&\quad\times \exp\left[-\frac{1}{2}(\by - \by^s_\btheta)\transpose(\boldsymbol{\Psi}(\bx_{1:n},\bx_{1:n}\mid\psi) + \sigma^2\eye_n)^{-1}(\by - \by^s_\btheta)\right],
\end{align}
where $\by^s_\btheta = [y^s(\bx_1,\btheta),\ldots,y^s(\bx_n,\btheta)]\transpose$ and $\boldsymbol{\Psi}(\bx_{1:n},\bx_{1:n}\mid\psi) = [\Psi_\psi(\bx_i,\bx_j\mid\psi)]_{n\times n}$.

In principle, posterior computation can be directly carried out by routinely drawing samples of $\btheta$ and $\psi$ using Metropolis-Hastings algorithm. This could be cumbersome when $n$ is large, since each iteration of the algorithm requires inverting an $n\times n$ matrix. Here we present an alternative strategy to reduce the computational complexity. Rather than drawing $\psi$ from the Markov chain, we propose to directly estimate $\psi$ by maximizing the posterior density \eqref{eqn:KO_density}, \emph{i.e.}, we seek to find the maximum \emph{a posteriori} (MAP) estimate of $\btheta$ and $\psi$. 
In order for the MAP estimation to be valid, the hyperprior $\pi(\psi)$ for the range parameter needs to be carefully selected. We follow the suggestion of \cite{gu2018} and take $\pi(\psi)$ to be of the form 
\begin{align}\label{eqn:robust_prior}
\pi(\psi)\propto \left(\psi + \sigma^2\right)^{a_\psi}\exp\left[-b_\psi(\psi + \sigma^2)\right]
\end{align}
for some $a_\psi > -(p + 1)$ and $b_\psi > 0$. Eq. \eqref{eqn:robust_prior} is the one-dimensional version of the jointly robust prior proposed in \cite{gu2018}, and has been shown to yield valid MAP estimate of $\psi$ for Mat\'ern covariance function. 

Having  an estimate $\widehat\psi$ of $\psi$ by maximizing \eqref{eqn:KO_density} with $\pi(\psi)$ in \eqref{eqn:robust_prior}, the posterior inference regarding $\btheta$ can be conveniently carried out by Metropolis-Hastings scheme, and the precision matrix $(\boldsymbol{\Psi}(\bx_{1:n},\bx_{1:n}\mid\widehat\psi) + \sigma^2\eye_n)^{-1}$ can be computed before the MCMC. As pointed out by one of the reviewers, besides MCMC sampling, the normalizing constant in $\pi(\btheta\mid\calD_n)$ can also be computed by numerical integration method when $\Theta$ is low-dimensional. Namely, one first computes
\[
Z(\widehat\psi) = \int_\Theta\pi(\theta,\widehat\psi\mid\calD_n)\mathrm{d}\btheta
\]
using numerical integration methods (\emph{e.g.}, quadrature method), and then obtain the exact posterior density of $\btheta$ via $\pi(\btheta\mid\calD_n) = \pi(\btheta,\widehat\psi\mid\calD_n)/Z(\widehat\psi)$. The posterior density of $\btheta$ obtained via numerical integration can serve as an auxiliary result to check the accuracy of MCMC samples. 
In what follows we provide an illustrative numerical example. 

\begin{figure}[htbp]
  \centerline{\includegraphics[width=.7\textwidth]{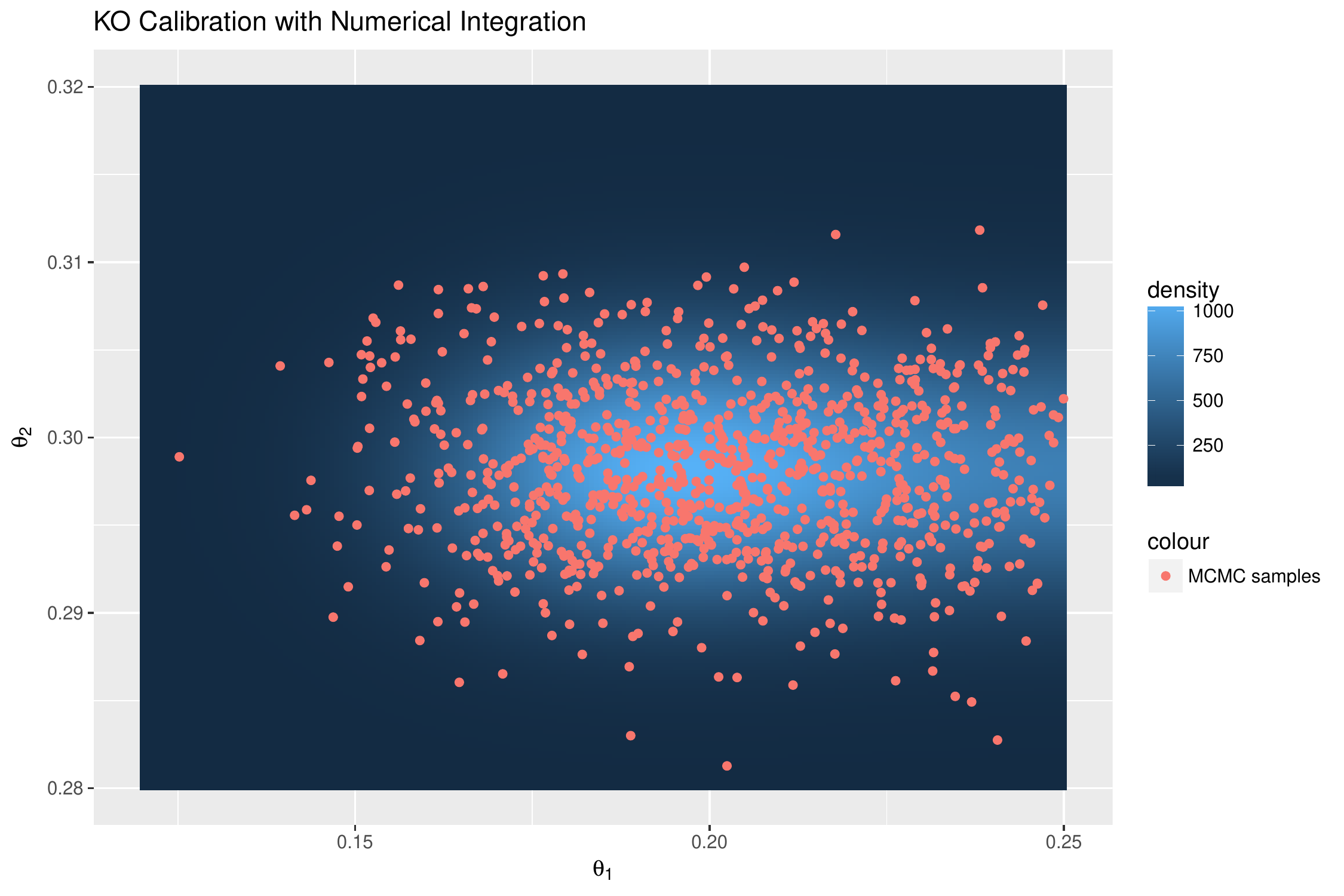}}
  \caption{Visualization of the comparison of MCMC sampling and numerical integration for posterior inference in KO method for configuration 1 in Section \ref{sub:simulation_studies}. The heatmap is the posterior density of $\btheta$ in KO method, the normalizing constant of which is computed using the cubature numerical integration method; The orange scatter points are the samples drawn from MCMC. }
  \label{fig:Projected_calibration_heatmap_configuration1}
\end{figure}

We adopt the same simulation setup as that of configuration 1 in Section \ref{sub:simulation_studies}, and is included here for readers' convenience. The computer model is 
\[
y^s(x,\btheta) = 7[\sin(2\pi\theta_1-\pi)]^2+2[(2\pi\theta_2-\pi)^2\sin(2\pi x-\pi)],
\]
and the physical system coincides with the computer model when $\btheta_0^* = [0.2, 0.3]\transpose{}$, \emph{i.e.}, $\eta_0(x) = y^s(x,\btheta_0^*)$. The design space $\Omega$ is $[0,1]$, and the parameter space $\Theta$ for $\btheta$ is $[0,0.25]\times[0,0.5]$.  We simulate $n = 50$ observations from the randomly perturbed physical system $y_i = \eta_0(x_i)+e_i$, where $(x_i)_{i=1}^n$ are uniformly sampled from $\Omega$, and the variance for the noises $(e_i)_{i=1}^n$ is set to $0.2^2$. We follow the aforementioned strategy to compute $\widehat
\psi$ and draw $1000$ posterior samples from the MCMC after $1000$ burn-in iterations. These post-burn-in samples are collected every $10$ iterations during the Markov chain. The comparison between the posterior samples and the exact posterior density obtained via numerical integration is visualized in Figure \ref{fig:Projected_calibration_heatmap_configuration1}. It can be seen that the distribution of these MCMC samples are in high accordance with the exact posterior density. 

Furthermore, we compute the means, standard deviations, and covariance matrices of $\btheta$ using the drawn MCMC samples and the exact posterior density, respectively, and tabulate them in Table \ref{table:Numerical_integration_config1}. It can be seen that the results computed using MCMC samples are close to their exact values, and there is no sign of non-accuracy occurring in these MCMC samples. 
\begin{table}[htbp]
  \centering
  \caption{Summary statistics comparison of MCMC sampling and numerical integration for posterior inference in KO method for configuration 1 in Section \ref{sub:simulation_studies}.}
  \begin{tabular}{c|c c|c c}
    \hline\hline
    &\multicolumn{2}{c|}{MCMC Sampling}&\multicolumn{2}{c}{Numerical Integration}
    \\
    \hline
    $\btheta$& $\theta_1$ & $\theta_2$ & $\theta_1$ & $\theta_2$ \\
    \hline
    Mean&$0.2013$& $0.2982$& 
         $0.2037$ & $0.2984$ \\
    Standard Deviation& $0.0244$ & $0.0048$ & 
                        $0.0255$ & $0.0052$ \\
        \hline
        Covariance Matrix 
        & \multicolumn{2}{c|}{$10^{-4}\times
        \begin{bmatrix*}
        5.91  & -0.0354\\
        -0.0354 & 0.23
        \end{bmatrix*}
        $}
        & \multicolumn{2}{c}{$10^{-4}\times 
        \begin{bmatrix*}
        6.48 & -0.0024\\
        -0.0024 & 0.27
        \end{bmatrix*}
        $} \\
    \hline\hline
  \end{tabular}%
  \label{table:Numerical_integration_config1}
\end{table}%


\end{document}